\documentclass[trackchanges]{aastex7}

\usepackage{graphicx}	
\usepackage{amsmath}
\usepackage{amssymb}	
\usepackage{longtable}
\usepackage{threeparttable}
\usepackage{multirow}
\usepackage{booktabs}
\usepackage{array}
\usepackage[figuresright]{rotating}
\usepackage[T1]{fontenc}   
\usepackage{subcaption}

\begin{document}

\title{Superoutbursts and Superhumps of Cataclysmic Variables observed with TESS}

\author[0000-0003-0516-404X]{Qi-Bin Sun}	
\email{sunqibin@ynu.edu.cn}
\affiliation{Yunnan Observatories, Chinese Academy of Sciences, Kunming 650216, China}
\affiliation{University of Chinese Academy of Sciences, No.19(A) Yuquan Road, Shijingshan District, Beijing, China}	

\author{Sheng-Bang Qian}
\affiliation{Department of Astronomy, School of Physics and Astronomy, Yunnan University, Kunming 650091, China}
\email[show]{qiansb@ynu.edu.cn}  

\author{Li-Ying Zhu}
\affiliation{Yunnan Observatories, Chinese Academy of Sciences, Kunming 650216, China}
\affiliation{University of Chinese Academy of Sciences, No.19(A) Yuquan Road, Shijingshan District, Beijing, China}	
\email{} 

\author{Qin-Mei Li}
\affiliation{Department of Astronomy, School of Physics and Astronomy, Yunnan University, Kunming 650091, China}
\email{} 

\author{Wen-Ping Liao}
\affiliation{Yunnan Observatories, Chinese Academy of Sciences, Kunming 650216, China}
\affiliation{University of Chinese Academy of Sciences, No.19(A) Yuquan Road, Shijingshan District, Beijing, China}	
\email{} 

\author{Fu-Xing Li}
\affiliation{Department of Astronomy, School of Physics and Astronomy, Yunnan University, Kunming 650091, China}
\email{} 

\author{Wen-Xu Lin}
\affiliation{Department of Astronomy, School of Physics and Astronomy, Yunnan University, Kunming 650091, China}
\email{} 

\author{Min-Yu Li}
\affiliation{Yunnan Observatories, Chinese Academy of Sciences, Kunming 650216, China}
\affiliation{University of Chinese Academy of Sciences, No.19(A) Yuquan Road, Shijingshan District, Beijing, China}	
\email{} 

\author{Ping Li}
\affiliation{Yunnan Observatories, Chinese Academy of Sciences, Kunming 650216, China}
\affiliation{University of Chinese Academy of Sciences, No.19(A) Yuquan Road, Shijingshan District, Beijing, China}	
\email{}

\author{Fang-Zhou Yan}
\affiliation{Department of Astronomy, School of Physics and Astronomy, Yunnan University, Kunming 650091, China}
\email{} 

\author{Zi-Mo Li}
\affiliation{Department of Astronomy, School of Physics and Astronomy, Yunnan University, Kunming 650091, China}
\email{} 

\author{Tao Wang}
\affiliation{Department of Astronomy, School of Physics and Astronomy, Yunnan University, Kunming 650091, China}
\email{} 

\author{Mao-Han Zhang}
\affiliation{Department of Astronomy, School of Physics and Astronomy, Yunnan University, Kunming 650091, China}
\email{}

\begin{abstract}

Superoutbursts and superhumps are characteristic signatures of SU UMa-type cataclysmic variables, yet the full diversity of superhump evolutionary behaviour remains poorly constrained owing to the limitations of ground-based photometric monitoring. Here we report a systematic analysis of 30 SU UMa-type dwarf novae observed with the Transiting Exoplanet Survey Satellite, covering 37 superoutburst events. We detect coherent superhump signals in 29 systems—including six first-time detections—and determine or refine orbital periods for 13 objects.  The high-cadence light curves clearly resolve the canonical three-stage (A–B–C) superhump period evolution. Using Stage~A periods, we derive dynamical mass ratios for 18 systems, consistent with established cataclysmic variable evolutionary tracks. The measured Stage~B period derivatives match the range of values from ground-based campaigns, with prominent positive drifts concentrated in short-period, low-mass-ratio systems. Repeated superoutbursts exhibit coherent evolutionary patterns, suggesting that superhump morphology may represent an intrinsic, repeatable property for some SU~UMa systems. Two systems break the standard template: RZ~LMi shows an inverted stage sequence inconsistent with classical precession theory, while ASASSN-14kj exhibits no stage evolution but a coherent long-period modulation. Persistent superhumps during normal outbursts in three systems provide direct evidence that eccentric disk structures can survive beyond their parent superoutbursts, revealing a decoupling between thermal accretion and tidal eccentricity cycles. These results demonstrate the power of space-borne photometry for probing accretion disk dynamics and testing precession models across the cataclysmic variable parameter space.

\end{abstract}

\section{Introduction} \label{sec:intro}

Cataclysmic variables (CVs) are compact interacting binaries where a white dwarf accretes material from a low-mass, Roche-lobe-filling companion through an accretion disk \citep{warner1995cat,ritter2003catalogue}. Dwarf novae (DNe), the largest subclass of CVs, show recurring 2--8\,mag outbursts triggered by episodes of enhanced mass accretion. Three main classes are widely recognized: U~Gem classical dwarf novae; Z~Cam stars, whose outburst decay can stall at intermediate brightness in so-called ``standstills''; and SU~UMa systems, which host superoutbursts. These superoutbursts reach roughly 0.5--1\,mag brighter than normal outbursts and last five to ten times longer, and they are accompanied by positive superhumps (PSHs): periodic photometric modulations whose periods are a few percent longer than the orbital period \citep{warner1995cat}.

Two distinct precessing disk configurations operate in CVs. An eccentric disk undergoing prograde precession experiences periodic tidal forcing from the companion, which generates PSHs. By contrast, a disk tilted out of the orbital plane precesses retrogradely. The mass stream then strikes opposite sides of the disk, producing negative superhumps (NSHs) with periods slightly shorter than the orbital period alongside super-orbital photometric modulations \citep{harvey1995superhumps,patterson1999permanent,wood2009sph,smak2009origin}. Recent space-survey analyses have greatly strengthened observational support for tilted disks and the stream-impact origin of NSHs \citep{Sun2023arXiv,2023arXiv230905891S,2023arXiv230911033S,Sun2024ApJ,2024arXiv240704913S}. This work focuses on eccentric, progradely precessing disks that power superoutbursts.

Ordinary outbursts are explained by the disk instability model (DIM). For a steady mass-transfer rate, hydrogen opacity varies with temperature and switches the disk between a cool, low-viscosity state and a hot, high-viscosity state \citep{1974MNRASBath,1974PASJOsaki,Meyer1983AA,Smak1983ApJ,1985MNRASLin,1983PASJ...35..377M,lasota2001disc,dubus2018testing}. Superoutbursts require the extra tidal physics introduced by the thermal-tidal instability (TTI) model of \citet{1989PASJ...41.1005O}. Material builds up within the disk over successive normal outbursts until the expanding disk reaches the 3:1 Lindblad resonance. There the companion’s tidal torque excites a global eccentric deformation that precesses in the prograde direction \citep{whitehurst1991superhumps,lubow1991simulations,hirose1990hydrodynamic,murray1996sph}. Enhanced tidal dissipation within this eccentric disk launches the superoutburst, while repeated tidal forcing of the outer disk generates PSHs \citep{vogt1982z,osaki1985irradiation,wood2011v344}.

Survey data have increasingly challenged this standard picture. Most notably, we recently identified Z~Cam dwarf novae with mass ratios $q>0.33$ — systems previously thought unable to sustain the 3:1 tidal resonance — that produce superoutbursts with PSHs during standstills, as seen in AT~Cancri \citep{2025ApJATCNC}. Standstills have also been reported in SU~UMa stars \citep{2019PASJ...71L...1K}, PSHs appear in nova-like variables and other high-$q$ binaries \citep{bruch2022tess}, PSH and NSH signals can coexist in single targets \citep{Sun2023arXiv}, and superoutbursts have been observed to transition directly into standstills \citep{2026ApJV803}. These findings demonstrate that eccentric disks can form and persist outside the classical TTI framework, and the full diversity of superoutburst behavior remains poorly understood.

Even for systems where TTI applies, the observed phenomenology of superhump period evolution is incompletely mapped. Long-running ground-based observing campaigns, most prominently the series led by Kato et al., (hereafter the Kato sequence; \citealp{2009PASJ...61S.395K,2010PASJ...62.1525K,2013PASJ...65...23K,2014PASJ...66...30K,2014PASJ...66...90K,2015PASJ...67..105K,2015PASJ...67....1K,2016PASJ...68...65K,2016PASJ...68...59K,2017PASJ...69...75K,2020PASJ...72...14K}) established that PSH periods follow a universal three-stage sequence across each superoutburst \citep{2009PASJ...61S.395K}. Stage~A occurs right at superoutburst onset and lasts only $\sim20$ superhump cycles; its period exceeds the later stage~B mean value by 1.0--1.5\%. This stage marks the initial excitation of eccentricity at the 3:1 resonance radius. The disk remains relatively cool, pressure terms are negligible, and the prograde precession rate is controlled solely by resonance-radius dynamics, set only by mass ratio $q$. Stage~A therefore yields a direct, calibration-free estimate of $q$. Stage~B is the longest, most stable phase: superhump amplitude slowly declines while the period evolves systematically. In short-period systems ($P_{\rm SH}\lesssim 0.08$\,d), the period rises with $\dot{P}/P\sim10^{-5}$. This behavior is interpreted as an eccentric disturbance wave propagating outward to larger radii, where precession is slower. Gas pressure inside the hot disk progressively counteracts this outward drift, and the stage~B period derivative depends mainly on $q$ rather than orbital period, peaking near $q\sim0.12$. Stage~B ends at the superoutburst plateau, giving way to Stage~C: the period drops abruptly by roughly 0.5\% and then plateaus. This change is attributed to tidal truncation removing the outer disk, stopping outward growth of the eccentric perturbation and re-establishing the resonance within the cooling inner disk. Later larger samples split this three-stage scheme into six sub-stages (A1/A2, B1/B2, C1/C2), without revising the underlying physical interpretation.

Superhump timing from ground networks of small telescopes suffers nightly phase gaps. Stage~A lasts at most $\sim1.5$\,d and is easily missed. Robust comparisons of stage evolution across repeated superoutbursts require continuous, homogeneous time series. The Transiting Exoplanet Survey Satellite (TESS; \citealt{Ricker2015journal}) delivers nearly continuous, high-precision photometry over $\sim27$\,d sectors. Its cadence and photometric stability fully sample complete superoutbursts, allowing us to measure hundreds of superhump maxima per outburst, build densely sampled $O$--$C$ diagrams, and resolve every evolutionary stage. Standard pipeline products \citep{2016SPIE.9913E..3EJ} make these light curves straightforward to retrieve. Following pioneering \textit{Kepler} work on SU~UMa stars \citep{wood2011v344,ramsay2012kepler}, TESS has now captured superoutbursts for a large sample of sources across the sky. Here we analyze TESS light curves for 30 dwarf novae classified as SU~UMa-type in the AAVSO Variable Star Index (VSX). We measure PSH maxima and construct $O$--$C$ diagrams to explore the diversity of superhump evolution at uniform precision: parabolic versus piecewise-linear trends in stage~B, anomalous or absent stage~A, targets with no clear evolutionary progression, and the repeatability of stage patterns between successive superoutbursts. 

The paper is organized as follows: Section~\ref{sec:data} describes the observations and analysis methods, Section~\ref{sec:Results} presents the results for each target, Section~\ref{sec:Discussions} discusses the implications for accretion disk physics, and Section~\ref{sec:Conclusions} summarizes our conclusions.

\begin{figure}
	\centering
	\begin{subfigure}{0.45\columnwidth}
		\includegraphics[width=\linewidth]{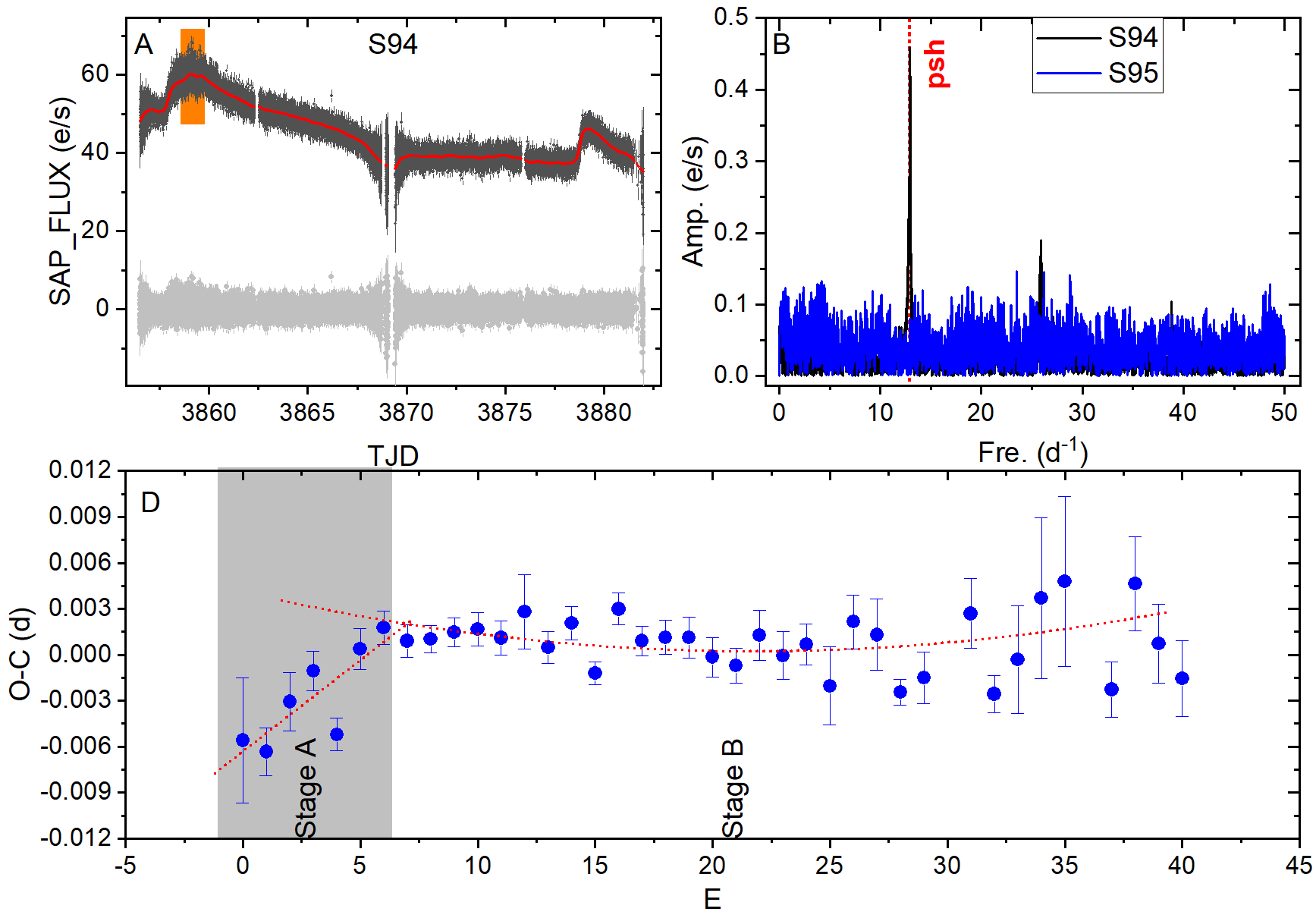}
		\caption{BE Oct}
	\end{subfigure}
	\begin{subfigure}{0.45\columnwidth}
		\includegraphics[width=\linewidth]{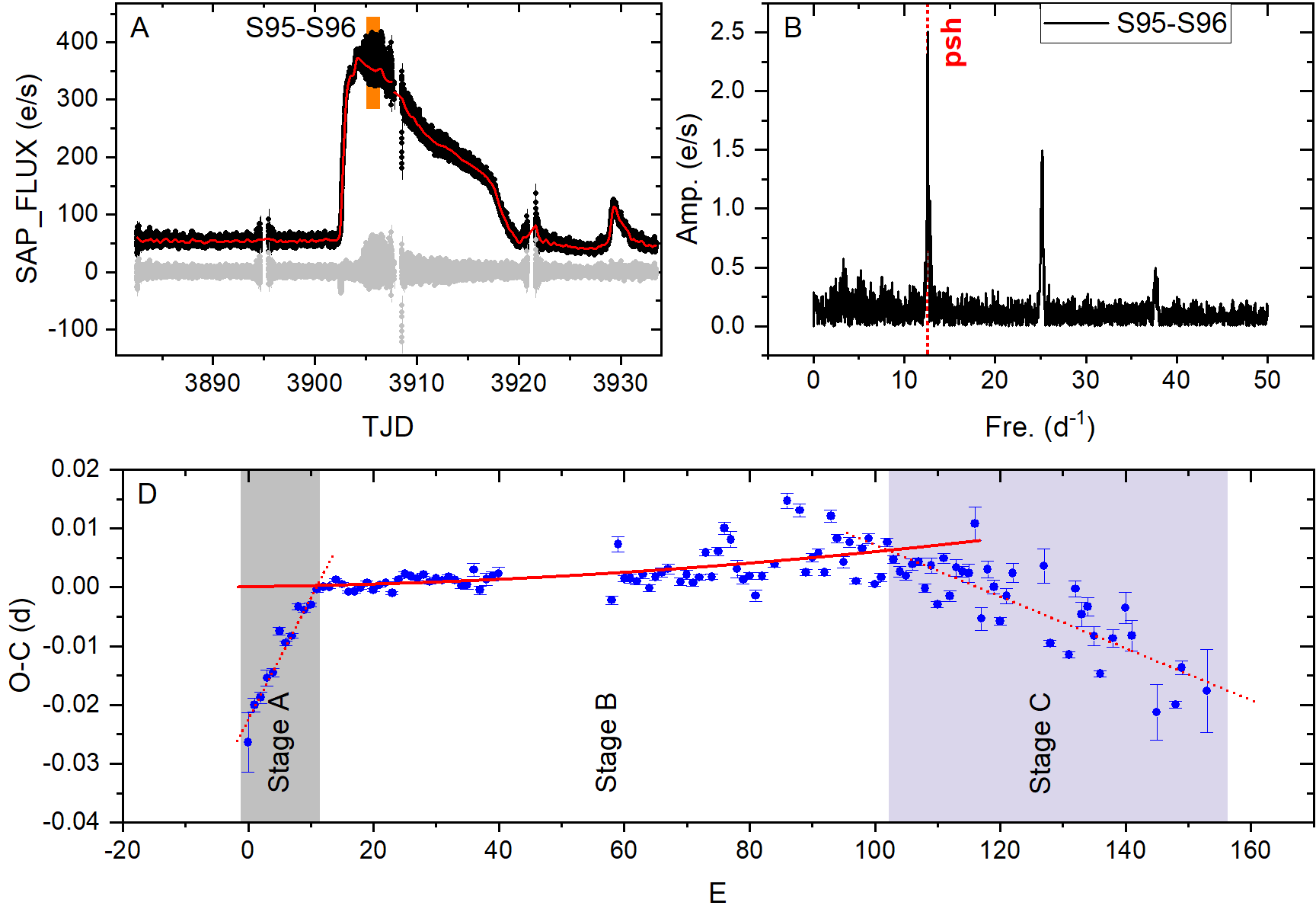}
		\caption{ASASSN-14eq}
	\end{subfigure}
	\begin{subfigure}{0.45\columnwidth}
		\includegraphics[width=\linewidth]{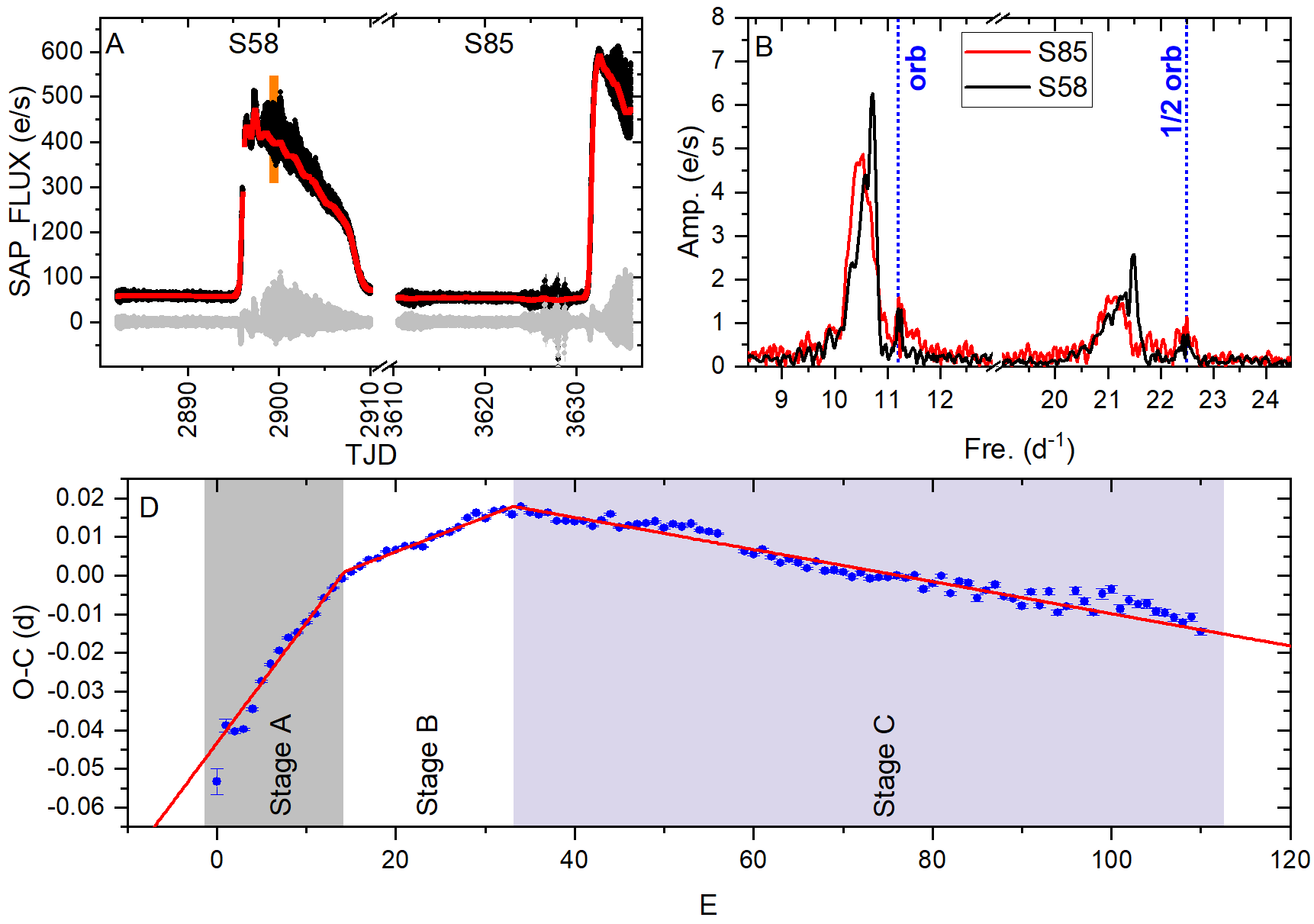}
		\caption{GX~Cas}
	\end{subfigure}
	\begin{subfigure}{0.45\columnwidth}
		\includegraphics[width=\linewidth]{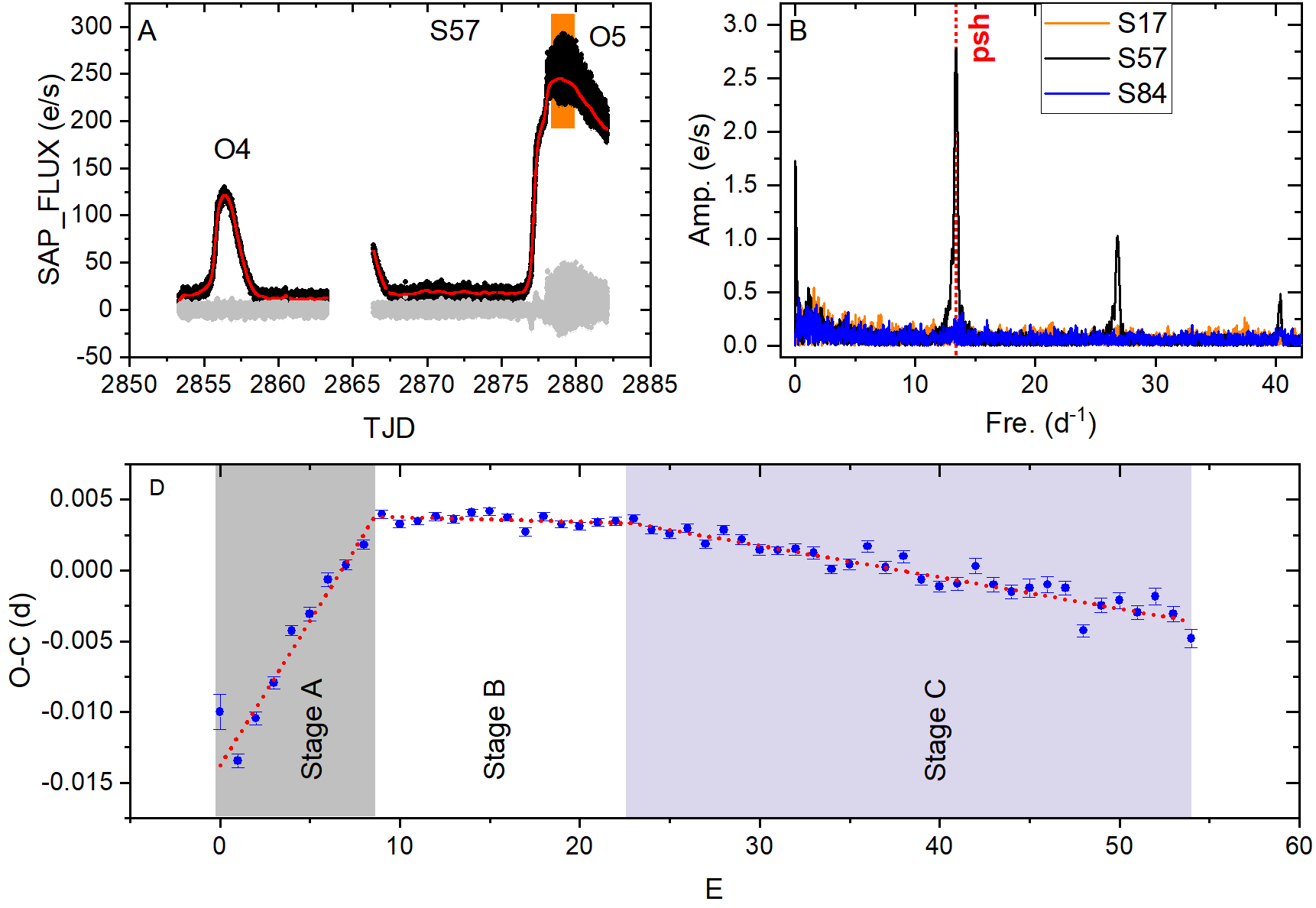}
		\caption{FO And}
	\end{subfigure}
	\begin{subfigure}{0.45\columnwidth}
		\includegraphics[width=\linewidth]{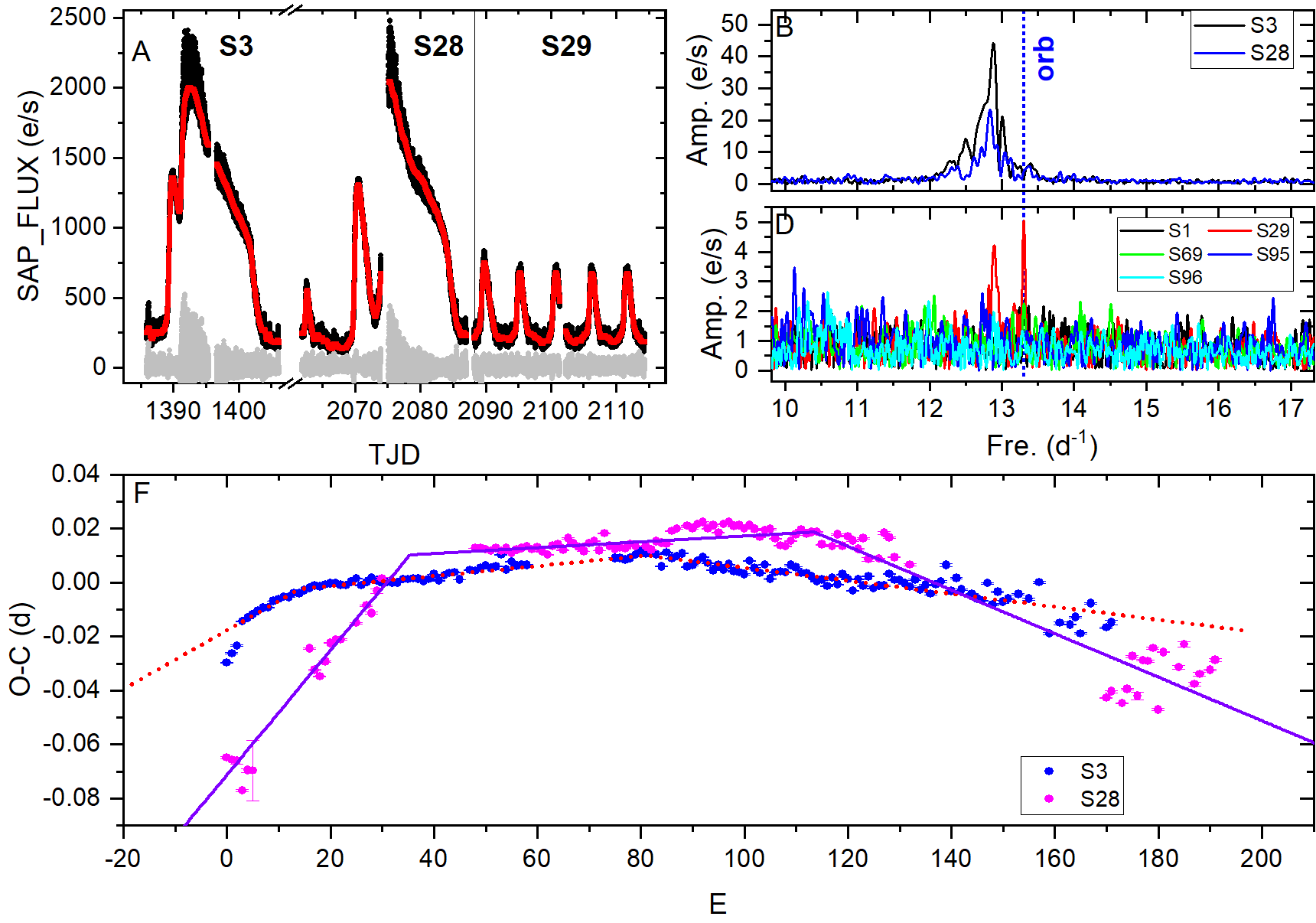}
		\caption{WX~Hyi}
	\end{subfigure}
	\begin{subfigure}{0.45\columnwidth}
		\includegraphics[width=\linewidth]{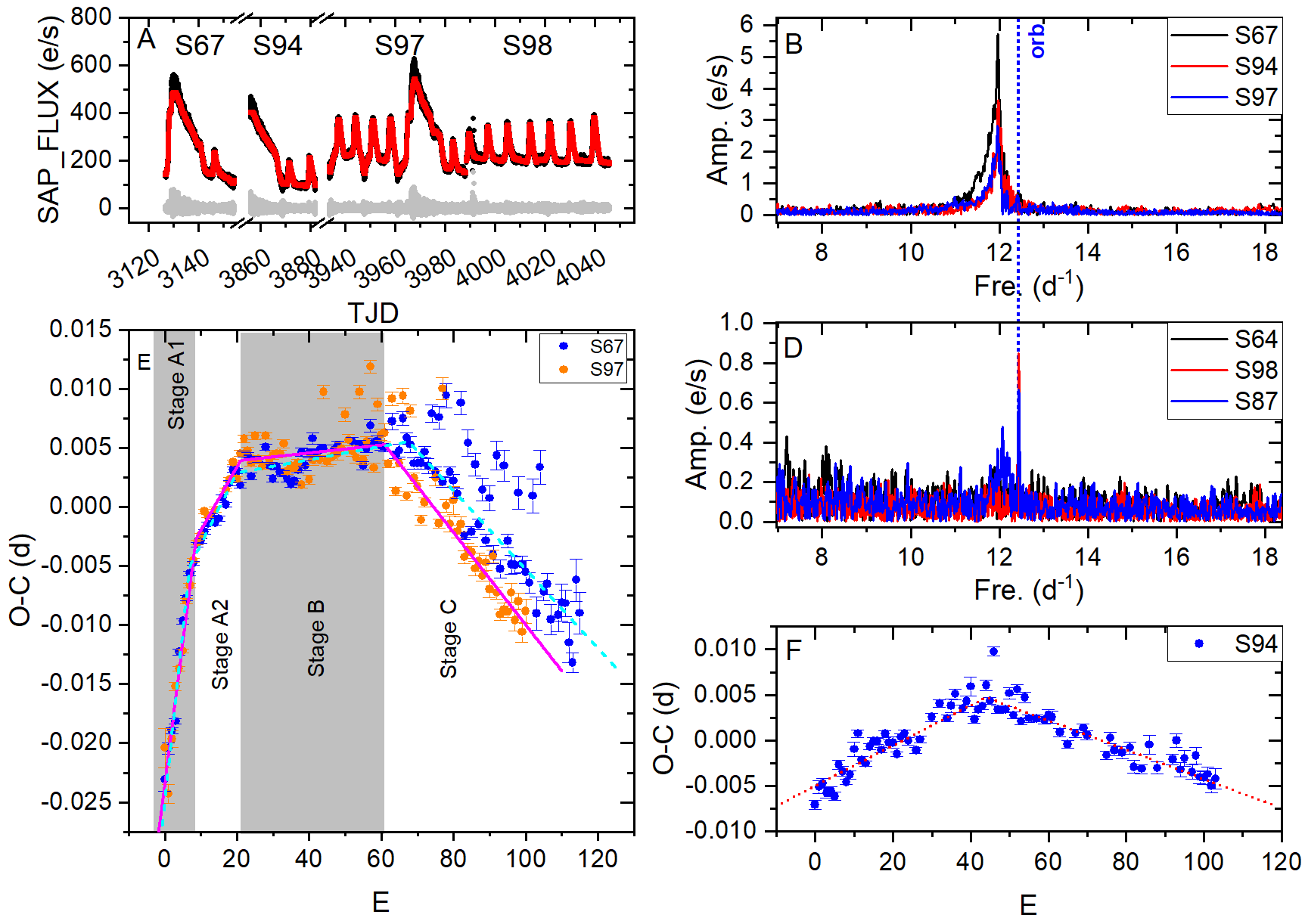}
		\caption{EC 05200-5205}
	\end{subfigure}
	\caption{The subpanels of Figure~\ref{fig:1-6}, namely panels (a)--(f) and their respective sub-panels, retain the identical numbering scheme as Figure~\ref{fig:1-6}. This figure, however, only presents TESS light curves of superoutbursts, periodograms, and $O$--$C$ diagrams. In all $O$--$C$ diagrams throughout the manuscript, grey shading corresponds to Stage~A, no fill to Stage~B, and light lavender to Stage~C. This color-coded shading scheme is not applied to $O$--$C$ diagrams containing multiple $O$--$C$ curves or those with an additional Stage~A2.}
	\label{fig:1-6-R}
\end{figure}

Detailed analysis results for each target in this paper, including TESS light curves, outburst details, periodograms, $O$--$C$ analyses, and folded light curves for selected signals, are presented as subpanels labelled A, B, C, etc. Each target occupies one panel ((a), (b), (c), etc.) that contains multiple subpanels of analysis outputs. These full results are shown in Appendix Figures~\ref{fig:1-6}--\ref{fig:25-30}. In the main-text Figures~\ref{fig:1-6-R}--\ref{fig:25-30-R}, we only display superoutburst light curves, periodograms, and $O$--$C$ diagrams.

\section{Data and Method}\label{sec:data}

The data analyzed in this work were obtained with the TESS \citep{Ricker2015journal}), a NASA mission conducting an all-sky survey for transiting exoplanets. TESS carries four 10-cm telescopes equipped with CCD cameras, each covering a $24^\circ \times 24^\circ$ field of view over a 600--1000~nm (``red'') bandpass. The sky is observed in contiguous \emph{sectors} of $\sim$27~days each, with brief interruptions near perigee for data downlink. Photometry is available as full-frame images (FFIs), recorded every 30~min during the prime mission and every 10~min from the extended mission onward, and as target-stamp light curves sampled at 2~min cadence. The standard pipeline produces Simple Aperture Photometry (SAP) and Pre-search Data Conditioning SAP (PDCSAP) light curves \citep{2016SPIE.9913E..3EJ}. Because the PDCSAP detrending can distort the large-amplitude outburst profiles of CVs \citep{2023arXiv230911033S,2024arXiv240903011S,2025AJLTEri,2026ApJV803}, we adopt the SAP light curves throughout this work. Although designed for exoplanet detection, TESS delivers nearly continuous, high-cadence time-series photometry for large samples of variable stars, including CVs. We selected 30 dwarf novae classified as SU~UMa-type in the AAVSO Variable Star Index (VSX; \citealt{2006SASSWatson}) and retrieved all TESS light curves covering their superoutbursts from the Mikulski Archive for Space Telescopes (MAST)(see Tab. \ref{tab:psh_param}).

Outburst peak, onset, and end times are determined interactively from the \texttt{SAP\_FLUX} light curves by means of a custom Python graphical interface. We first apply locally weighted regression (\textsc{LOWESS}; \citealp{cleveland1979robust}) smoothing to suppress the photometric noise and emphasize the outburst-related variations. Within a user-selected time window, the outburst peak is located by fitting a quadratic polynomial around the maximum of the LOWESS-smoothed light curve. The rise and decay boundaries are placed at the epochs where the LOWESS-smoothed flux crosses the quiescent baseline plus $1\sigma$; the baseline itself is obtained from a linear fit to manually selected quiescent segments (see Fig. \ref{fig:out-cal-examp1}). The timing uncertainty of each boundary is estimated from the interval between the crossings of the baseline level and of the baseline-plus-$1\sigma$ level.

Following \citet{2024arXiv240903011S,2025ApJATCNC,2026arXiv260910966S}, the epochs of superhump maxima are measured with a Gaussian-plus-linear model (see Fig.~\ref{fig:max-example}):
\begin{equation}
	y = A + Bx + \frac{\mathrm{amp}}{\mathrm{wid}\sqrt{\pi/(4\ln 2)}}
	\exp\left(-\frac{4\ln 2\,(x-\mathrm{cen})^2}{\mathrm{wid}^2}\right),
	\label{eq:1}
\end{equation}
where $A$ is the baseline magnitude, $B$ the linear drift term, $\mathrm{amp}$ the Gaussian peak amplitude, $\mathrm{cen}$ the central peak maxima, and $\mathrm{wid}$ the full width at half maximum (FWHM) of the Gaussian component.
LOWESS smoothing is first applied to remove the outburst trend, and preliminary candidate maxima are identified from the residuals. A Gaussian-plus-linear function is then fitted by least squares within a window of $\pm 0.3\,P_{\rm sh}$ around each candidate to determine the precise maximum epoch. Candidate superhump signals are searched with the Period04 software \citep{Period04}, which is used for all frequency spectra and period determinations presented in this work.

The $O$--$C$ (observed minus calculated) diagram is constructed by comparing the measured maximum epochs with those predicted from a linear ephemeris,
\begin{equation}
	T_{\rm MAX}(E) = T_0 + P_0 E,
	\label{eq:eph_init}
\end{equation}
where $E$ is the cycle number, $T_0$ is the reference epoch (taken as the time of the first measured superhump maximum), and $P_0$ is the superhump period derived from the Period04 analysis. Because the initial choice of $T_0$ and $P_0$ can bias the apparent $O$--$C$ morphology, we fit the residuals with a straight line, $O\!-\!C(E)=\alpha+\beta E$, and absorb it into the ephemeris. All subsequent analysis is based on the revised linear ephemeris
\begin{equation}
	T_{\rm MAX}(E) = T_0' + P_0'E, \qquad T_0' = T_0+\alpha, \quad P_0' = P_0+\beta,
	\label{eq:eph_rev}
\end{equation}
from which the evolutionary stages are identified. Following the notation of \citet{2009PASJ...61S.395K}, we define Stage~A as the initial growth stage with a relatively long period, Stage~B as the main stage with a systematically varying period, and Stage~C as the terminal stage with a shorter, nearly constant period.

To quantify the period change during stage~B, the $O$--$C$ data are fitted with either a parabolic model,
\begin{equation}
	O\!-\!C = a + bE + cE^2,
	\label{eq:oc_quad}
\end{equation}
or a two- to three-segment piecewise-linear model,
\begin{equation}
	O\!-\!C =
	\begin{cases}
		a_1 + b_1 E, & E < E_1, \\
		a_2 + b_2 E, & E_1 \leq E < E_2, \\
		a_3 + b_3 E, & E \geq E_2,
	\end{cases}
	\label{eq:oc_pwl}
\end{equation}
where $E_1$ and $E_2$ are the transition cycle numbers between adjacent stages; the two-segment variant contains a single break at $E_1$. For the parabolic fit, the stage~B period derivative is
\begin{equation}
	\dot{P}_{\rm sh} = \frac{2c}{P_0} \quad [{\rm d\,d^{-1}}].
	\label{eq:pdot}
\end{equation}
For the piecewise-linear fits, the mean period within segment~$i$ is $P_0+b_i$, so that the differences between segment slopes directly measure the period changes across the stage transitions. The choice between the parabolic and piecewise-linear models is guided by the fit residuals and by the morphology of the $O$--$C$ diagram: when clear breaks indicating discrete stage transitions are present, the piecewise-linear fit is preferred.

\begin{deluxetable}{ccccccccccc}
	\tablecaption{Superhump parameters of SU~UMa-type dwarf novae from TESS light curves.\label{tab:psh_param}}
	\tablewidth{1pt}
	\tabletypesize{\scriptsize}
	\tablehead{
		\colhead{Name} & \colhead{Coordinates} & \colhead{TESS} & \colhead{$P_{\rm SH}$} & \colhead{error} & \colhead{Sector} & \colhead{Cycle} & \colhead{$T_0'$} & \colhead{error} & \colhead{$P_{\rm 0}'$} & \colhead{error}\\
		\colhead{} & \colhead{(J2000)} & \colhead{sectors} & \colhead{(d)} & \colhead{(d)} & \colhead{} & \colhead{$E$ range} & \colhead{(TJD)\tablenotemark{a}} & \colhead{(d)} & \colhead{(d)} & \colhead{(d)}}
	\startdata
	BE Oct & 00 00 49.02 $-$77 18 57.8 & S94--S95 & 0.077346 & 0.000005 & S94 & E(0--40) & 3857.87627 & 0.00477 & 0.077345 & 0.000042 \\
	ASASSN-14eq & 00 21 30.93 $-$57 19 22.2 & S28--S29, 68, 69, 95--96, 102 & 0.079935 & 0.000004 & S95 & E(0--153) & 3904.35536 & 0.00559 & 0.079017 & 0.000015 \\
	GX Cas & 00 49 01.49 $+$56 52 43.6 & S58, S85 & 0.093310 & 0.000004 & S58 & E(0--110) & 2898.22660 & 0.00419 & 0.093361 & 0.000046 \\
	& & & 0.095211 & 0.000006 & S85 & & & & & \\
	FO And & 01 15 32.16 $+$37 37 35.6 & S17, S57, S84 & 0.074605 & 0.000003 & S57 & E(0--54) & 2878.07320 & 0.00227 & 0.074626 & 0.000036 \\
	WX Hyi & 02 09 50.82 $-$63 18 39.8 & S1, S3, S28--S29, S69, S95--S96 & 0.077610 & 0.000002 & S3 & E(0--171) & 1390.90270 & 0.00143 & 0.077575 & 0.000015 \\
	& & & 0.077882 & 0.000003 & S28 & E(0--191) & 2071.49106 & 0.00542 & 0.077716 & 0.000052 \\
	& & & 0.077519 & 0.000013 & S29 & & & & & \\
	EC 05200-5205 & 05 21 13.83 $-$52 02 59.0 & 64, 67, 87, 94, 97--98 & 0.083591 & 0.000003 & S67 & E(0--115) & 3129.02705 & 0.00224 & 0.083523 & 0.000021 \\
	& & & 0.083521 & 0.000004 & S94 & E(0--103) & 3856.34301 & 0.00120 & 0.083334 & 0.000016 \\
	& & & 0.083626 & 0.000004 & S97 & E(0--100) & 3966.28962 & 0.00312 & 0.083576 & 0.000029 \\
	& & & 0.082864 & 0.000015 & S87 & & & & & \\
	V1434 Tau & 05 29 58.81 $+$18 48 09.9 & S43--S45, S71 & 0.069959 & 0.000003 & S71 & E(0--120) & 3244.93253 & 0.00657 & 0.069895 & 0.000011 \\
	MASTER OT J055845.55+391533.4 & 05 58 45.50 $+$39 15 33.3 & S73 & 0.055006 & 0.000001 & S73 & E(0--116) & 3292.54359 & 0.00414 & 0.054988 & 0.000007 \\
	ASASSN-14je & 06 32 27.68 $-$56 38 52.5 & S87--S90, S93--S98 & 0.069507 & 0.000002 & S89 & E(0--162) & 3723.14265 & 0.00196 & 0.069453 & 0.000012 \\
	& & & 0.069473 & 0.000002 & S96 & E(0--148) & 3922.19621 & 0.00383 & 0.069437 & 0.000014 \\
	UV Gem & 06 38 44.17 $+$18 16 11.4 & S71--S72 & 0.092943 & 0.000003 & S71--S72 & E(0--99) & 3254.43400 & 0.00652 & 0.092790 & 0.000045 \\
	IR Gem & 06 47 34.68 $+$28 06 22.3 & S44--S45, S71--S72 & 0.071039 & 0.000003 & S71 & E(0--151) & 3256.80341 & 0.00199 & 0.070975 & 0.000018 \\
	PNV J06501960+3002449 & 06 50 19.50 $+$30 02 43.5 & S44--S45, S71--S72 & 0.067504 & 0.000002 & S45 & E(0--169) & 2526.03270 & 0.00234 & 0.067357 & 0.000012 \\
	SDSS J075107.50+300628.4 & 07 51 07.51 $+$30 06 28.5 & S72 & 0.058005 & 0.000004 & S72 & E(0--84) & 3261.47071 & 0.00434 & 0.057988 & 0.000023 \\
	ASASSN-14kj & 08 02 34.39 $-$00 09 40.2 & S61, S88 & 0.095167 & 0.000054 & S88 & E(0--51) & 3691.84013 & 0.00388 & 0.095167 & 0.000054 \\
	SDSS J080303.90+251627.0 & 08 03 03.90 $+$25 16 27.0 & S71--S72 & 0.090572 & 0.000007 & S72 & E(0--92) & 3264.48707 & 0.00700 & 0.090812 & 0.000064 \\
	YZ Cnc & 08 10 56.65 $+$28 08 33.2 & S44--S47, S71--S72 & 0.090728 & 0.000003 & S47 & E(0--86) & 2585.01534 & 0.00258 & 0.090921 & 0.000045 \\
	& & & 0.090604 & 0.000004 & S72 & E(0--125) & 3272.52304 & 0.00298 & 0.090502 & 0.000031 \\
	RZ LMi & 09 51 48.91 $+$34 07 23.8 & S21, S48 & 0.059581 & 0.000003 & S48 & E(0--99) & 2614.50563 & 0.00942 & 0.059395 & 0.000013 \\
	& & & 0.059600 & 0.000003 & S21 & & & & & \\
	KS UMa & 10 20 26.52 $+$53 04 33.1 & S48, S75 & 0.070249 & 0.000002 & S75 & E(0--66) & 3362.81429 & 0.00087 & 0.070174 & 0.000013 \\
	V0748 Hya & 10 25 22.24 $-$15 42 21.7 & S89 & 0.063347 & 0.000003 & S89 & E(0--151) & 3736.07109 & 0.00307 & 0.063311 & 0.000014 \\
	SX LMi & 10 54 30.42 $+$30 06 10.4 & S22, S48 & 0.069522 & 0.000003 & S48 & E(0--107) & 2628.56177 & 0.00151 & 0.069420 & 0.000011 \\
	MASTER OT J172758.09+380021.5 & 17 27 58.14 $+$38 00 22.4 & S79--S80 & 0.057850 & 0.000001 & S80 & E(0--144) & 3486.31900 & 0.00397 & 0.057878 & 0.000012 \\
	ASASSN-18hr & 18 57 54.66 $+$34 46 24.7 & S74, S80--S81 & 0.078450 & 0.000007 & S74 & E(0--51) & 3314.93003 & 0.01273 & 0.078522 & 0.000054 \\
	& & & 0.078462 & 0.000010 & S80 & E(0--61) & 3487.96906 & 0.00788 & 0.078568 & 0.000081 \\
	V0419 Lyr & 19 10 14.02 $+$29 06 13.9 & S74, S80--S81 & 0.089912 & 0.000006 & S74 & E(0--115) & 3321.38570 & 0.01003 & 0.090394 & 0.000063 \\
	V1113 Cyg & 19 22 41.98 $+$52 43 59.2 & S74--S76, S82--S83 & 0.079058 & 0.000003 & S75 & E(0--144) & 3345.74787 & 0.00324 & 0.079090 & 0.000020 \\
	& & & 0.078309 & 0.000009 & S82 & & & & & \\
	Gaia17cuh & 19 34 36.96 $+$10 02 13.4 & S81 & & & & & & & \\
	ASASSN-15qr & 19 50 39.55 $+$05 56 20.3 & S81 & 0.079681 & 0.000010 & S81 & E(0--74) & 3519.84201 & 0.00285 & 0.079543 & 0.000041 \\
	ASASSN-18xz & 20 27 45.96 $+$25 43 26.4 & S81--S82 & 0.065755 & 0.000010 & S81 & & & & & \\
	V0774 Peg & 21 26 25.08 $+$20 19 46.4 & S82 & 0.088660 & 0.000010 & S82 & E(0--99) & 3546.12112 & 0.01320 & 0.088327 & 0.000058 \\
	MGAB-V233 & 21 34 19.72 $+$51 38 57.0 & S77, S83 & 0.096562 & 0.000009 & S83 & & & & & \\
	V0630 Cyg & 21 34 59.22 $+$40 40 18.8 & S76, S82--S83 & 0.078709 & 0.000007 & S76 & E(0--44) & 3367.71905 & 0.00196 & 0.078756 & 0.000041 \\
	\enddata
	\begin{threeparttable}
		\begin{tablenotes}
			\item [a] Names and equatorial coordinates (J2000) are taken from the AAVSO Variable Star Index (VSX).
			\item [b] ``TESS sectors'' lists all sectors in which the target was observed; ``Sector '' gives the sector from which the superhump measurement in that row was derived. 
			\item [c] $P_{\rm SH}$ is the mean superhump period determined from the Period04 frequency analysis.
			\item [d] $T_0'$ and $P_0'$ are the parameters of the revised linear ephemeris after the $O$--$C$ linear correction (see Section \ref{sec:data}), $T_{\rm MAX}(E) = T_0' + P_0'E,$, where $E$ is the cycle number. $T_0$ is given in TJD = BJD $-$ 2457000. The cycle column lists the range of timed superhump maxima used in the $O$--$C$ fit.
		\end{tablenotes}
	\end{threeparttable}
\end{deluxetable}

\section{Results}\label{sec:Results}

\subsection{BE Oct}

The nature of BE~Oct was first suggested by \citet{1970C&T....86..229P}, who detected prolonged outburst activity and proposed its classification as an SU~UMa-type dwarf nova candidate. \citet{1981MitVS...9...16G} subsequently listed this source as a tentative U~Gem-type variable in a southern visual survey of the Hydri field, and the GCVS ultimately classified BE~Oct as SU UMa-type with a magnitude range of 15.4--$>$17.5. The International Variable Star Index (VSX) reports an orbital period of $P_{\rm orb}=0.0747$~d.
Kemp \& Patterson detected superhump modulations with $P_{\rm sh}=0.07712(13)$~d during the August 1998 superoutburst (vsnet-obs 3461), confirming the SU~UMa classification. Quiescent spectroscopy by \citet{1991PASP..103..300H} revealed H$\alpha$, H$\beta$, He\,\textsc{i}, and Fe\,\textsc{ii} emission lines, with H$\alpha$ exhibiting a FWHM of $\sim$1300~km~s$^{-1}$, while the light curve showed typical 0.2~mag flickering without periodic modulations. \citet{2003A&A...403..699M} obtained time-resolved spectroscopy confirming the emission-line spectrum characteristic of dwarf novae near minimum light. A superoutburst in 2017 July was monitored by ASAS-SN and ground-based observers; \citet{2020PASJ...72...14K} reanalyzed the light curve and derived a refined superhump period of $P_{\rm sh}=0.07715(132)$~d.

TESS observed BE~Oct in Sectors~94 and~95, capturing one superoutburst ($\gtrsim$11~d in duration) and three normal outbursts (Figs. \ref{fig:1-6}(a)A and E). The normal outbursts exhibited a mean duration of 2.46~d and a mean recurrence interval of 11.86~d. Fourier analysis of Sector~94 revealed a coherent signal at 0.077346(5)~d, identified as the PSH period (Fig. \ref{fig:1-6-R}(a)B). No significant periodic signal was detected in Sector~95. Strong PSH modulations were present throughout the superoutburst plateau phase.
Gaussian fitting of the PSH maxima yielded 39 independent measurements (cycles $E=0$--$40$). The following ephemeris was adopted for the $O$--$C$ analysis:
\begin{equation}
	T_{\rm MAX}={\rm TJD}\,3857.87626 + E\times 0.077345.
\end{equation}
The resulting $O$--$C$ diagram displays two distinct evolutionary stages (A and~B; Fig. \ref{fig:1-6-R}(a)D). The period derivative during stage~B is $\dot{P}=(21.26\pm12.21)\times10^{-5}$~d~d$^{-1}$.

\subsection{ASASSN-14eq}

ASASSN-14eq was discovered as a transient at $V=13.53$ on 2014 July~28 by the ASAS-SN survey \citep{2015PASJ...67..105K}. During the same outburst, \citet{2015PASJ...67..105K} detected superhumps and derived $P_{\rm sh}=0.079467(69)$~d, accompanied by a large negative $\dot{P}$ ($\dot{P}/P=-40.2\times10^{-5}$~cycle$^{-1}$) that they attributed to a stage transition. The shortest supercycle interval was $\sim$410~d \citep{2015PASJ...67..105K}. Notably, \citet{2019MNRAS.486.2422P} identified a 0.0813(3)~d signal during quiescence, which they interpreted as the orbital period, suggesting that the superhump-like signal reported by \citet{2015PASJ...67..105K} may instead correspond to negative superhumps.

TESS observed ASASSN-14eq across seven sectors (S28-S29, S68-S69, S95-S96, and~102). One superoutburst was captured in S95--S96, lasting $\sim$17.23~d, along with three normal outbursts of mean duration $\sim$3~d (Figs. \ref{fig:1-6}(b)A, F and E); these normal outbursts occurred in different TESS sectors, precluding determination of their recurrence period. Fourier analysis uncovered a strong PSH signal at 0.079935(4)~d during S95--S96, while no significant periodicity was found in the remaining sectors (Fig. \ref{fig:1-6-R}(b)B). Our measured PSH period is closer to the value reported by \citet{2015PASJ...67..105K} than to the quiescent signal of \citet{2019MNRAS.486.2422P}, arguing against the latter being the true orbital period. Prominent PSH signatures persisted throughout the superoutburst. A total of 117 PSH maxima (cycles $E=0$--$153$) were extracted via Gaussian fitting, and the following ephemeris was adopted:
\begin{equation}
	T_{\rm MAX}=3904.35536 + E\times 0.079017.
\end{equation}
The $O$--$C$ curve shows well-defined stages~A and~C (Fig. \ref{fig:1-6-R}(b)D), whereas stage~B is only weakly expressed. A parabolic fit yields $\dot{P}=(1.19\pm1.36)\times10^{-5}$~d~d$^{-1}$, with a correspondingly large uncertainty.

\subsection{GX~Cas}

GX\,Cas was the first dwarf nova found to host a superhump period exceeding two hours, establishing its location at the upper boundary of the period gap \citep{1998PASJ...50..411N}. \citet{2003PASP..115.1308P} confirmed both the superhump period ($P_{\rm sh}\simeq0\fd093$) and the orbital period through coordinated photometry and spectroscopy, detecting broad double-peaked H$\alpha$ emission with a peak separation of $\sim$1000~km~s$^{-1}$. The Kato sequence of superhump period variations \citep{2009PASJ...61S.395K} analysed four superoutbursts (1994, 1996, 1999, and~2006) and revealed a characteristic stage~A $\to$ stage~C evolution, with stage~A superhumps reaching $P_{\rm sh}>0\fd0964$~d and stage~C stabilizing at $P_{\rm sh}=0\fd092958(23)$~d. The system has a supercycle of $\sim$386~d and normal outbursts recurring every 50--70~d. Its superhump period remains the longest among confirmed SU~UMa-type dwarf novae, making GX~Cas a critical test for the upper period limit of tidal instability.

Two superoutbursts of GX~Cas were recorded by TESS in S58 and~S85. The S58 superoutburst lasted 14.16~d, whereas only a half-covered event was captured in S85 (Fig. \ref{fig:1-6}(c)A). Skewed PSH profiles are present at the superoutburst peak. The PSH periods derived from Fourier analysis are 0.093310(4)~d (S58) and 0.095211(6)~d (S85) (Fig. \ref{fig:1-6-R}(c)B); the longer period in S85 arises because the TESS data mainly sample the early short-period PSH stages (stages~A and~B). In addition, weak signals near 0.089206(17)~d together with their second harmonic are detected, possibly corresponding to the orbital period. Adopting this periodicity, the superhump excess ($\epsilon \equiv P_{\rm SH}/P_{\rm orb} - 1$) for S58 is 0.046. Since only a little more than ten PSH maxima are available during the S85 superoutburst, we extracted PSH maxima exclusively from S58 to construct a complete $O$--$C$ evolution. A total of 109 PSH maxima ($E=0$--$110$) were obtained via Gaussian fitting, giving the ephemeris:
\begin{equation}
	T_{\rm MAX}=2898.22660 + E\times 0.093361.
\end{equation}

As in V0419\,Lyr, V1113\,Cyg, and V0774\,Peg, the $O$--$C$ diagram exhibits linear rather than parabolic evolution in stage~B (Fig. \ref{fig:1-6-R}(c)D). A three-segment piecewise-linear function provides an excellent fit (Pearson correlation coefficient $r=0.98$, null‑hypothesis significance $p<2.31\times10^{-88}$), with mean periods decreasing sequentially: 0.096461~d, 0.094260~d, and 0.092945~d.

\subsection{FO And}

FO~And was discovered as a dwarf nova by \citet{1967AN....289..205H}. 
\citet{1984IBVS.2483....1M} identified short outburst intervals of 15--23~d 
and reported likely superoutbursts, classifying FO~And as a probable 
SU~UMa-type dwarf nova. \citet{1989A&AS...78..145B} obtained a quiescent 
spectrum showing Balmer and He\,{\sc ii} emission lines, confirming its 
cataclysmic nature. \citet{1989PASP..101..899S} reported time-resolved 
photometry in quiescence but did not detect a significant periodic signal.
The first solid detection of superhumps was published by \citet{1995IBVS.4242....1K}, 
who observed a superoutburst in 1994 August lasting at least 10~d. CCD photometry 
on Aug.\ 15 revealed fully developed superhumps with an amplitude of 0.17~mag. 
The derived superhump period was $P_{\mathrm{SH}} = 0.07411(5)$~d. 
\citet{1996PASP..108...73T} determined the orbital period of $P_{\mathrm{orb}} = 0.07161(18)$~d 
from a radial-velocity study, yielding a fractional superhump excess of 
$\varepsilon^{+} = 3.5 \pm 0.3$\%, consistent with the typical distribution of 
SU~UMa-type dwarf novae at this orbital period.
Reanalysis of the 1994 data by \citet{2009PASJ...61S.395K} revealed a stage 
transition: the mean superhump period was $0.07455(5)$~d for $E \leq 14$ 
(stage~B) and $0.07402(1)$~d for $13 \leq E \leq 27$ (stage~C), with the 
observation most likely capturing the stage B--C transition. 

FO And was observed by TESS in Sectors 17, 57 and 84, capturing one superoutburst (O5) and seven normal outbursts. Owing to missing data at the end of the superoutburst, its duration is constrained to be longer than $5.57$~d (Figs. \ref{fig:1-6}(d)A, E and F). Based on recurrence time and outburst amplitude, the normal outbursts can be divided into two categories. Type\,1 includes peaks O1, O2, O4, O6 and O7 with amplitudes below $150\ \mathrm{e\,s^{-1}}$, while Type\,2 consists of O3 and O8 with amplitudes exceeding $150\ \mathrm{e\,s^{-1}}$. These subtypes are also reflected in their recurrence intervals: the peak‑to‑peak interval between O1 and O2 is $5.66$~d, whereas the interval between O2 and O3 is $10.94$~d. In addition, the O6–O7 interval is $6.9$~d and O7–O8 is $9.47$~d. This behaviour suggests that the outburst recurrence period of FO And is unstable. Fourier analysis identifies a PSH signal with a period of $0.074605(3)$~d during the superoutburst in Sector~57 (Fig. \ref{fig:1-6-R}(d)B), which is consistent with previous studies. No orbital‑period signature is detected in any of the available sectors. A total of 55 PSH maxima ($E(0\text{--}54)$) were extracted via fitting. We adopted the following ephemeris for the $O$--$C$ analysis:
\begin{equation}
	T_{\rm MAX}=TJD2878.07320 + E\times 0.074626
\end{equation}
The $O$--$C$ diagram of FO\,And resembles that of GX\,Cas and can be well‑fitted by a three‑segment piecewise‑linear function ($r=0.96$, $p<1.86\times10^{-34}$; Fig. \ref{fig:1-6-R}(d)D).
The fitted periods for the three successive stages are $0.076666$\,d, $0.074593$\,d, and $0.074403$\,d, respectively.

\subsection{WX~Hyi}

WX~Hyi is a southern SU~UMa-type dwarf nova in which superhumps were among the first to be photometrically identified \citep{1979MNRAS.188..681B}. High-resolution spectroscopy by \citet{1981A&A....97..185S} yielded $K_1=67$~km~s$^{-1}$ and the precise orbital period $P_{\rm orb}=0.0748134(2)$, implying a superhump period excess $\varepsilon^+\simeq3.7$\% from the stage~B superhump period $P_{\rm sh}\simeq0.07761$ measured by \citet{2014PASJ...66...90K}. The system shows normal outbursts ($V\sim12\farcm7$, $\sim$11~d interval) and superoutbursts ($V\sim11\farcm4$, $\sim$185~d cycle) \citep{1991A&A...242..401K}, who also detected QPOs of $\sim$19--26~min near outburst maximum and found no orbital modulation in quiescence, indicating a low inclination.

TESS observed WX~Hyi across seven sectors (S1, S3, S28--S29, S69, and S95--S96), yielding a rich set of outburst light curves. The S3 superoutburst lasted 15.20~d, while the S28 superoutburst had a duration of 16.60~d and exhibits a prominent precursor resembling a normal outburst (Figs. \ref{fig:1-6}(e)A and B). 
Eighteen normal outbursts are detected in the remaining sectors, and their light-curve morphologies indicate distinct subclasses. A small outburst ($\sim$1.28~d, peak count rate 549.27~e~s$^{-1}$) preceded the S28 superoutburst. S29 contains five similar outbursts with nearly identical durations (mean 2.52~d), a typical peak count rate of 688.14~e~s$^{-1}$, and a mean recurrence interval of 5.47~d. Four normal outbursts in S1 have a mean duration of 3.44~d, a typical peak count rate of 1080.60~e~s$^{-1}$, and a mean recurrence of 8.10~d---both duration and interval are longer than those in S29. In S69, four outbursts are present: the first lasts 1.39~d with a modest peak count rate of 345.67~e~s$^{-1}$, while the other three last $\sim$3.05~d at $\sim$538.00~e~s$^{-1}$. The situation is more intricate in S95--S96, where two candidate outbursts with peak count rates near 1000~e~s$^{-1}$ and durations $>$3.6~d are identified, though these remain tentative owing to nearby data gaps; weaker intervening outbursts (1.99~d and 1.25~d) are also present. Such complex outburst behaviour implies an unstable accretion disc with intricate accretion dynamics.

Fourier analyses were performed for each sector. The PSH periods for S3 and S28 are 0.077610(2)~d and 0.077882(3)~d, respectively, with deviations from historical values appearing only at the fourth decimal place (Figs. \ref{fig:1-6-R}(e)B and D). Notably, PSHs at 0.077519(13)~d are also recovered in S29---analogous to the quiescent-state PSHs seen in Z~Cha \citep{2026arXiv260910966S}---and an orbital-period signal at 0.075160(11)~d is additionally detected.

The two superoutbursts provide abundant PSH timing data. Gaussian fitting yielded 145 PSH maxima ($E=0$--$171$) for S3 and 114 ($E=0$--$191$) for S28. The $O$--$C$ analyses adopted the following ephemerides:
\begin{eqnarray}
	T_{\rm MAX,S3} &=& 1390.90270 + E\times 0.077575, \\
	T_{\rm MAX,S28} &=& 2071.49106 + E\times 0.077716.
\end{eqnarray}
The $O$--$C$ diagrams resemble those of FO And and GX~Cas: stage~B exhibits linear rather than parabolic evolution (Fig. \ref{fig:1-6-R}(e)F). Both superoutbursts are well fitted by three-segment piecewise-linear functions (S3: $r=0.93$, $p<1.28\times10^{-78}$; S28: $r=0.94$, $p<1.05\times10^{-67}$). The mean periods decrease sequentially in each case: S3 gives 0.078665~d, 0.077760~d, and 0.077334~d; S28 gives 0.080026~d, 0.077825~d, and 0.076909~d.

\subsection{EC 05200-5205}

EC\,05200$-$5205 is an SU~UMa-type dwarf nova in Pictor, originally identified in the Edinburgh--Cape Blue Object Survey \citep{2013MNRAS.431..240O}.
The system is classified as a SU UMa-type variable and exhibits well-defined superoutbursts reaching $V\sim13\farcm4$--$13\farcm7$ with an average supercycle of $\sim$145~d, as revealed by the ASAS-SN light curve \citep{2014ApJ...788...48S, 2017PASP..129j4502K}; normal outbursts do not exceed $V\sim14\farcm0$--$14\farcm5$.
During the 2017~December superoutburst, time-resolved photometry recorded superhumps, confirming the SU~UMa classification; the superoutburst decline rate of $\sim$0\fd1~d$^{-1}$ over 10~d is consistent with typical SU~UMa decay time-scales.
No orbital period has yet been firmly established for this system.

TESS observed EC\,05200$-$5205 in S64, S67, S87, S94, S97, and~S98 (Figs. \ref{fig:1-6}(f)A and C). S67 captured one complete superoutburst (16.50~d) and one normal outburst (2.56~d). An incomplete outburst and two normal outbursts (mean 2.90~d) were recorded in S94. S97 contained a full superoutburst (16.40~d, very similar to that in S67) together with five normal outbursts (mean 3.49~d); the mean recurrence interval prior to this superoutburst was 8.88~d. Seven normal outbursts were detected in S98 (mean 3.77~d, mean recurrence 8.34~d). The outburst duration and recurrence interval show only small changes ($\sim$0.28~d and $\sim$0.54~d variations, respectively) across the epochs bracketing the S97--S98 superoutburst. Assuming no missing superoutburst occurred between S94 and~S97, the superoutburst cycle is $\sim$111~d. S64 recorded two outbursts of distinct durations (3.84~d and 5.91~d) and peak rates (343.78 and 496.49~e~s$^{-1}$). Two outbursts of comparable duration ($\sim$3.31~d) but a long recurrence of 12.59~d were observed in S87.

Fourier analysis yielded PSH periods of 0.083591(3)~d, 0.083521(4)~d, and 0.083626(4)~d for S67, S94, and~S97 (Figs. \ref{fig:1-6-R}(f)B and D), respectively, with variations confined to the fourth-to-fifth decimal place. An additional mean periodic signal at 0.080334(15)~d, likely the orbital period, was detected in S64, S87, and~S98, corresponding to a PSH excess of $\sim$0.04. Notably, a candidate PSH-related signal at 0.082864(15)~d---shorter than the superoutburst PSH period---is present in S87, possibly indicating that the eccentric precessing disc did not fully disappear after superoutburst termination, as seen in Z~Cha \citep{2026arXiv260910966S}.

The three superoutbursts provide rich $O$--$C$ data. We obtained 108 ($E=0$--$115$), 78 ($E=0$--$103$), and 84 ($E=0$--$100$) PSH maxima from S67, S94, and~S97, respectively, yielding the ephemerides:
\begin{align}
	T_{\rm MAX,S67} &= 3129.02706 + E\times 0.083523, \\
	T_{\rm MAX,S94} &= 3856.34301 + E\times 0.083334, \\
	T_{\rm MAX,S97} &= 3966.28962 + E\times 0.083576.
\end{align}
The $O$--$C$ diagrams of S67 and~S97 are highly similar and well reproduced by a four-segment piecewise-linear function ($r=0.85$ for S67; $r=0.89$ for S97) (Figs. \ref{fig:1-6-R}(f)E and F), with closely matching breakpoints and slopes. 
To avoid conflict with previous nomenclature, and given that the slopes of the first two linear segments in the $O$--$C$ diagram are both positive, we designate these two components as stage\,A1 and stage\,A2. The subsequent stages follow the classical scenario, with zero period derivative during stage\,B.
In contrast, the S94 $O$--$C$ curve deviates markedly from the other two and is adequately fitted by a two-segment piecewise-linear function ($r=0.81$). This discrepancy is likely caused by the incomplete superoutburst in S94, where the first two evolutionary stages (stage\,A1 and stage\,A2) are absent. Evidence for this comes from the mean period of the first linear segment in S94 ($0.083555$\,d), which is very close to the mean PSHs period of stage\,B in S67 ($0.083580$\,d) and S97 ($0.083609$\,d).

\begin{figure}
	\centering
	\begin{subfigure}{0.45\columnwidth}
	\includegraphics[width=\linewidth]{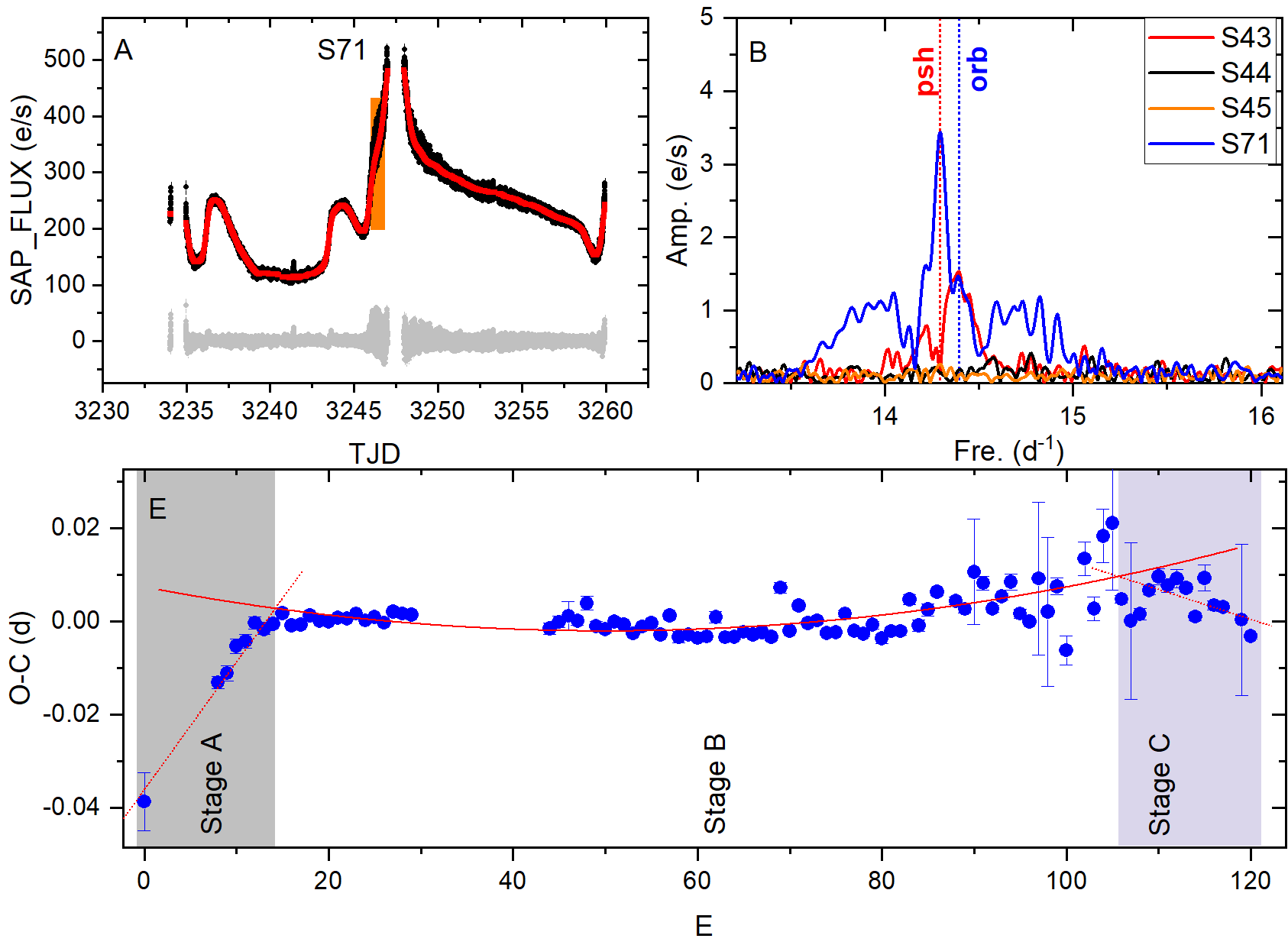}
	\caption{V1434 Tau}
\end{subfigure}	
	\begin{subfigure}{0.45\columnwidth}
		\includegraphics[width=\linewidth]{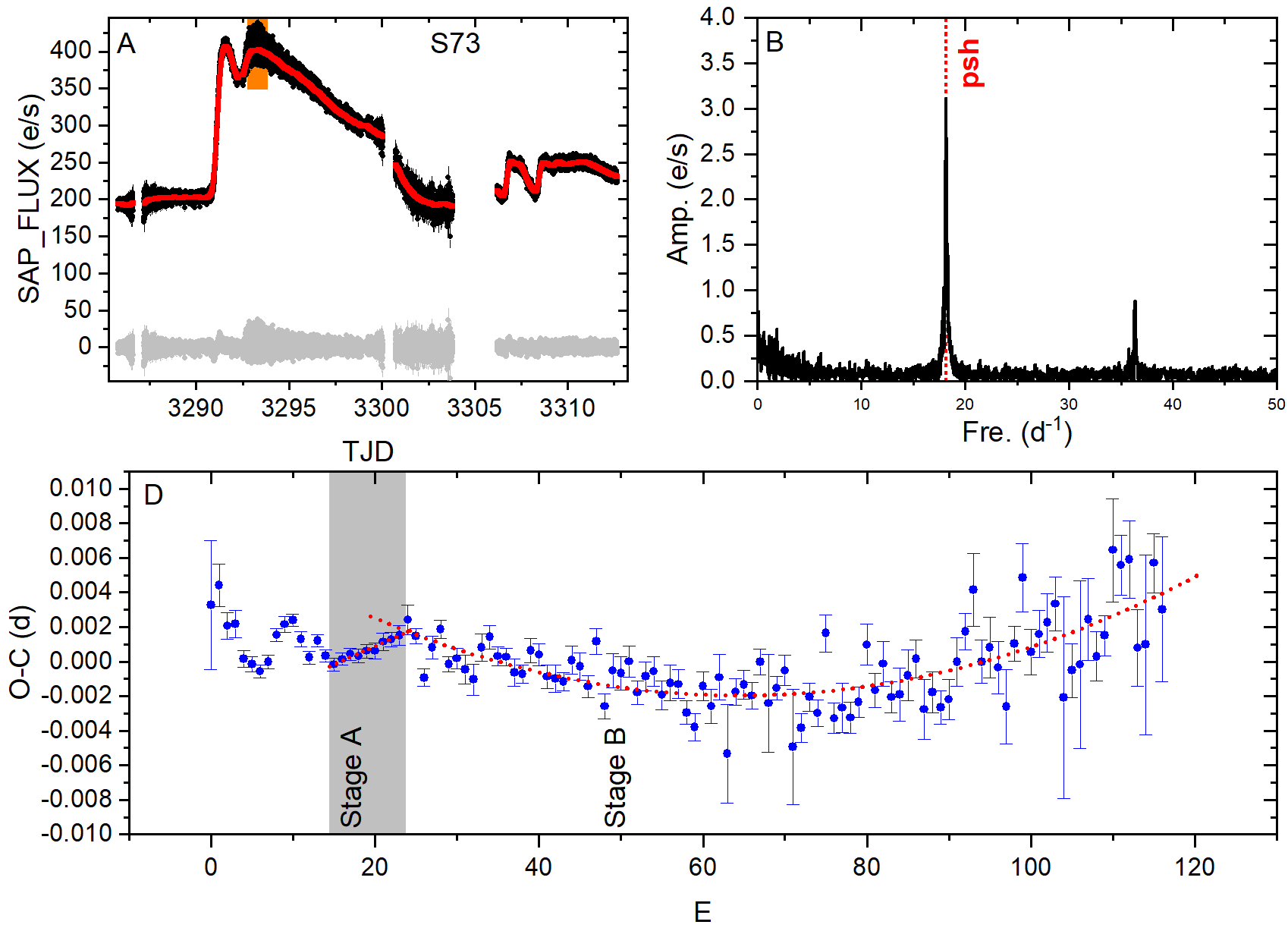}
		\caption{MASTER OT J055845.55+391533.4}
	\end{subfigure}

	\begin{subfigure}{0.45\columnwidth}
		\includegraphics[width=\linewidth]{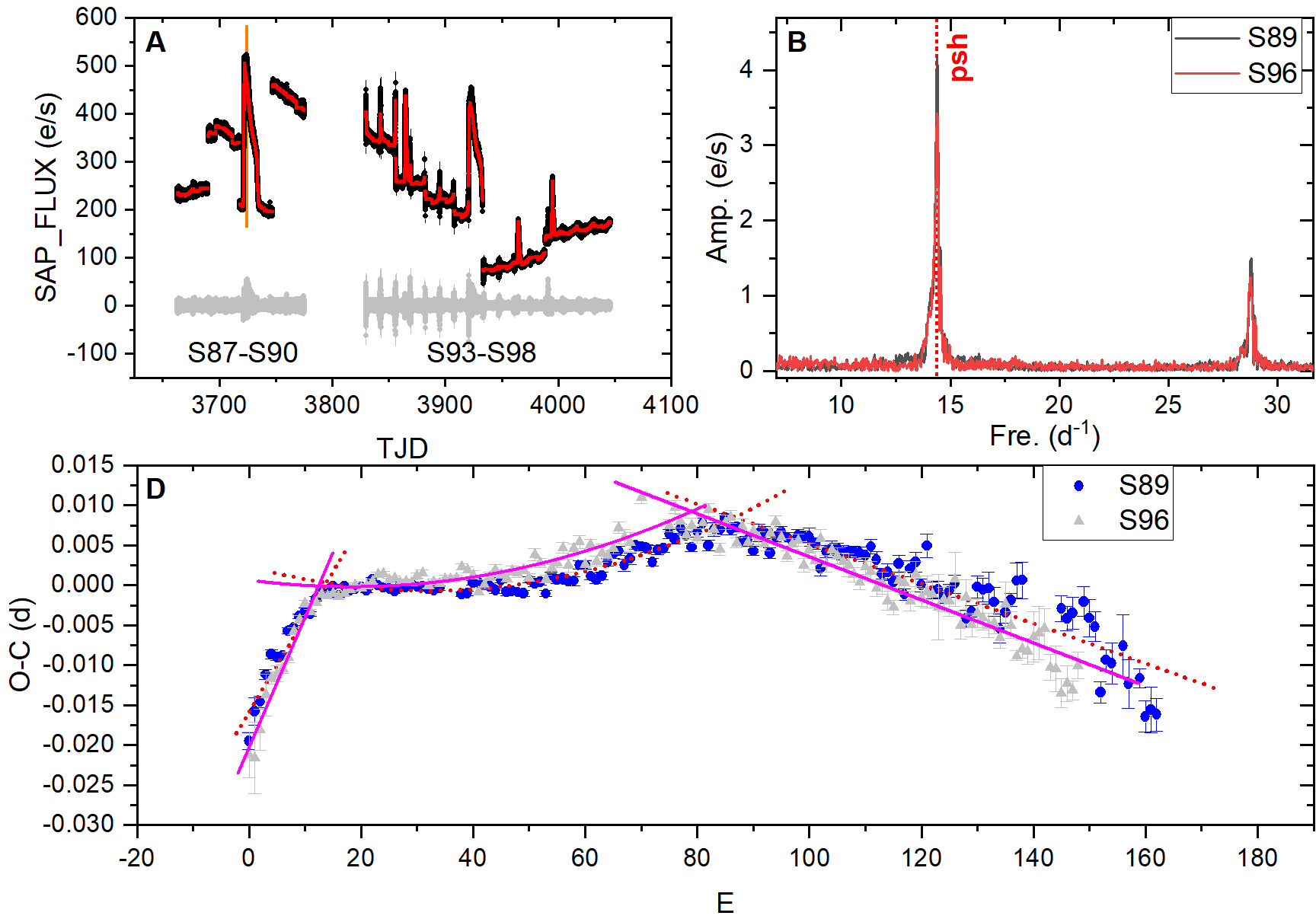}
		\caption{ASASSN-14je}
	\end{subfigure}
	\begin{subfigure}{0.45\columnwidth}
	\includegraphics[width=\linewidth]{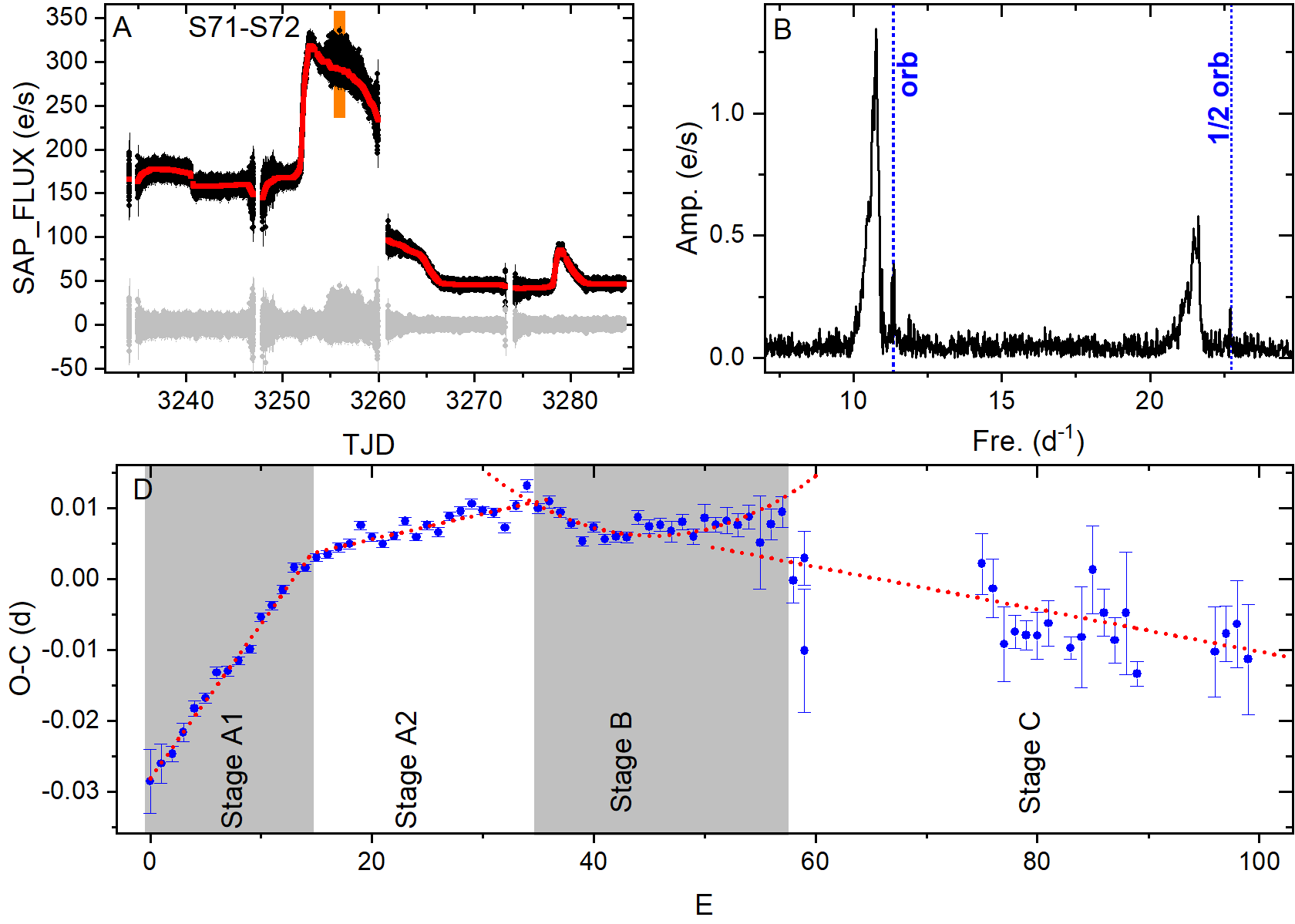}
	\caption{UV Gem}
\end{subfigure}

	\begin{subfigure}{0.45\columnwidth}
		\includegraphics[width=\linewidth]{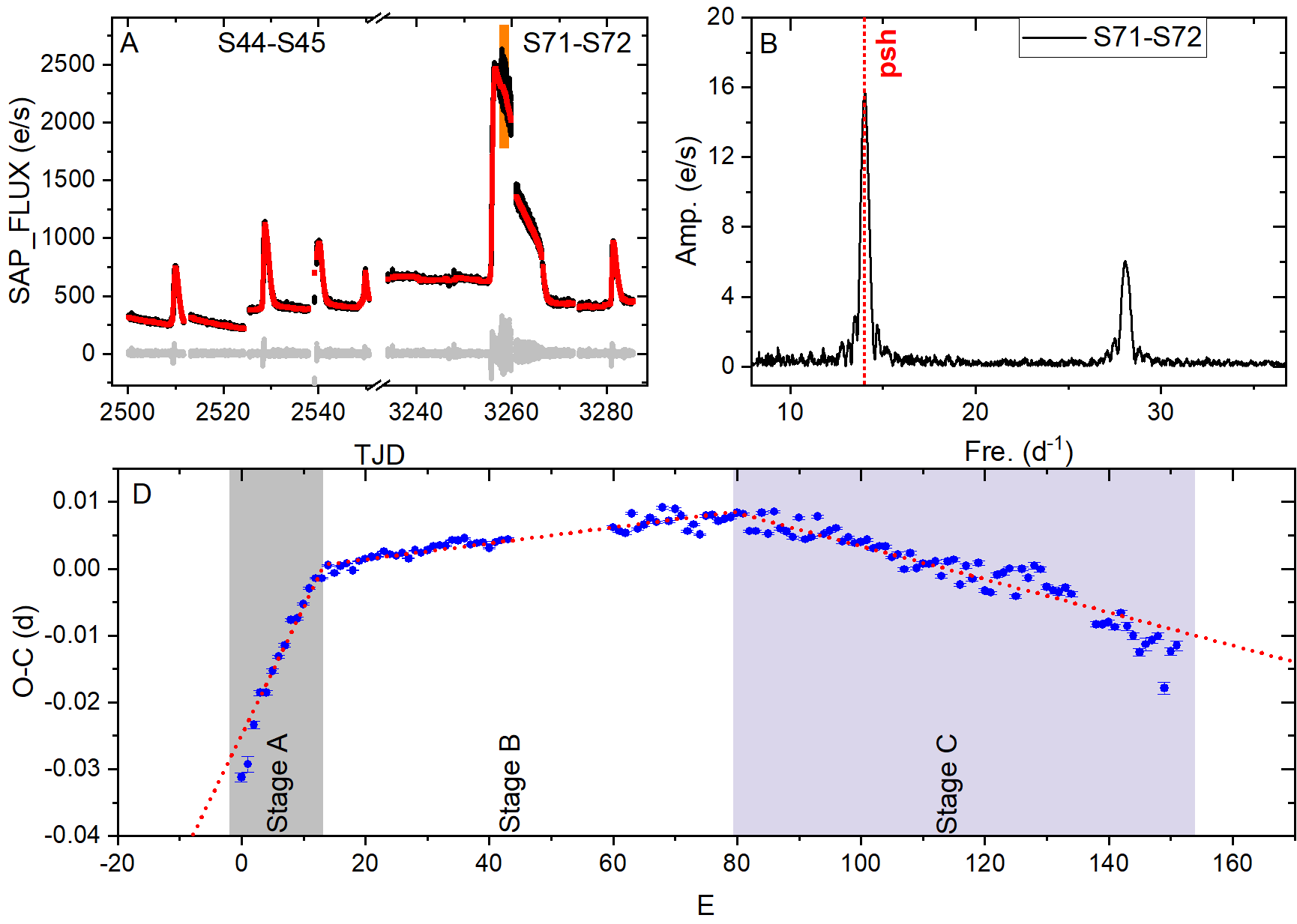}
		\caption{IR Gem}
	\end{subfigure}
	\begin{subfigure}{0.45\columnwidth}
		\includegraphics[width=\linewidth]{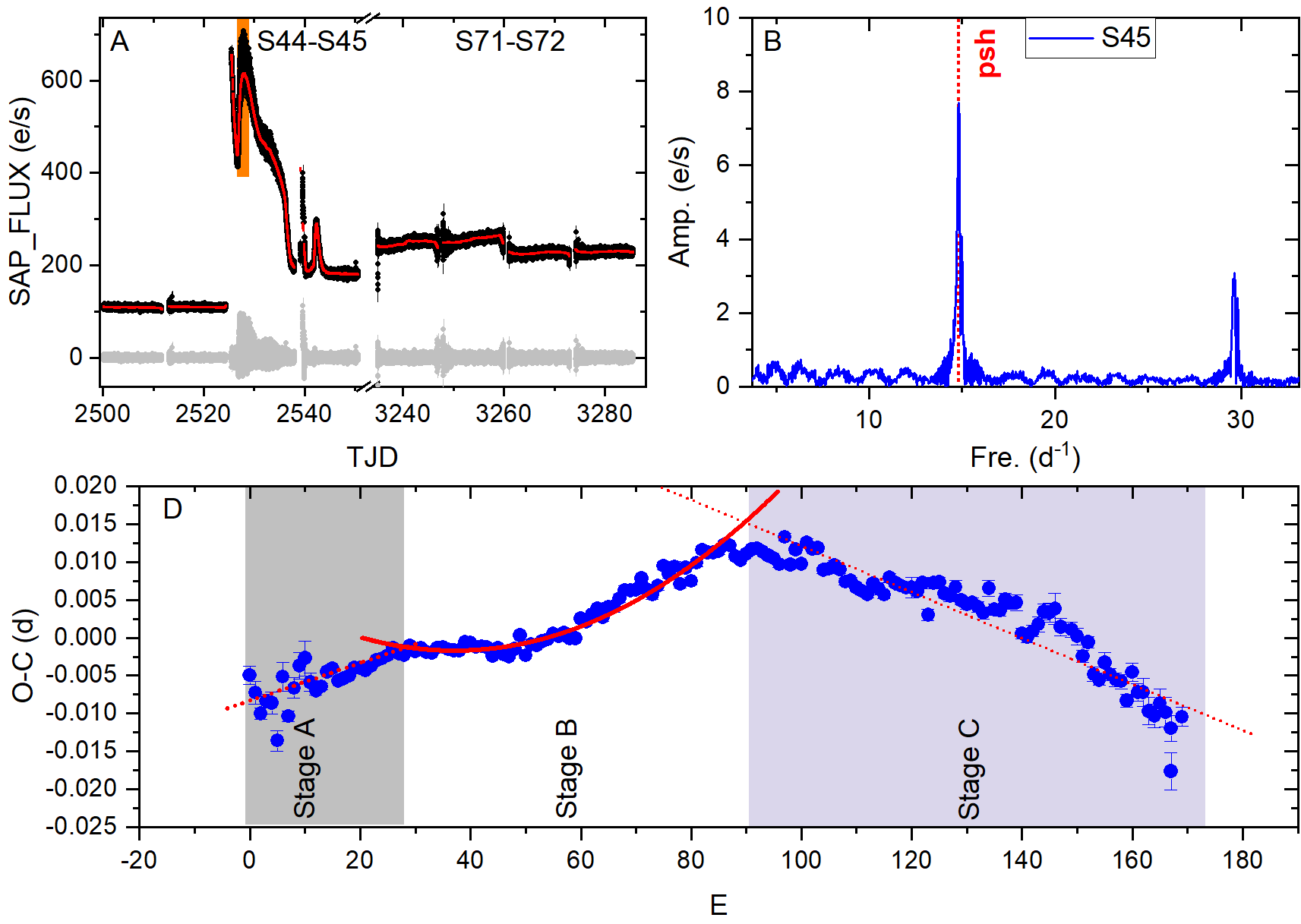}
		\caption{PNV J06501960+3002449}
	\end{subfigure}
\caption{Panels (a)--(f) and their subpanels follow the same numbering convention as Figure~\ref{fig:7-12}. This figure only displays TESS superoutburst light curves, periodograms, and $O$--$C$ diagrams.}
	\label{fig:7-12-R}
\end{figure}

\subsection{V1434 Tau}

V1434\,Tau (NSV\,02026) was first identified as a variable star (HV\,6907) by
\citet{1935BHarO.901...20H}, with precise coordinates later provided by
\citet{2002IBVS.5298....1W}.
The object was recognized as a dwarf nova following outburst detections by the
CRTS Mount Lemmon Survey (vsnet-alert~12503; \citealt{2016PASJ...68...65K}).
Its SU~UMa nature was confirmed during the 2015~November superoutburst,
which yielded $P_{\rm sh}=0.06978(2)$~d
(vsnet-alert~19258). \citet{2016PASJ...68...65K} obtained $P_{\rm sh}\approx0.06980$~d for stage~B, while the orbital period remains undetermined.
AAVSO monitoring reveals a regular supercycle of $\sim$95~d with normal
outbursts recurring every 6--14~d.
Exploratory spectroscopy obtained during outburst shows H$\beta$ in narrow
emission, H$\alpha$ in emission, and a blue-sloped continuum, consistent with
a typical SU~UMa-type dwarf nova \citep{2020AJ....159..114O}.

TESS monitored V1434\,Tau in Sectors~43--45 and~71, revealing one superoutburst ($\sim$16.95~d) and ten normal outbursts (Figs. \ref{fig:7-12}(a)A and D). Among the latter, O7 exhibits a distinctly smaller amplitude and shorter separations from its neighbors; excluding O7, the remaining peak-to-peak intervals range from 6.70~d to 12.99~d. Fourier analysis detects a PSH signal at 0.069959(3)~d during the S71 superoutburst (Fig. \ref{fig:7-12-R}(a)B), with prominent PSH signatures present throughout. A separate periodic signal at 0.069488(6)~d is found in S43, tentatively identified as the orbital period. A total of 97 PSH maxima ($E=0$--$120$) were extracted via Gaussian fitting, yielding the ephemeris:
\begin{equation}
	T_{\rm MAX}=3244.93253 + E\times 0.069895.
\end{equation}
The $O$--$C$ diagram reveals three distinct stages (Fig. \ref{fig:7-12-R}(a)E). The period derivative for stage~B is $(10.79\pm1.84)\times10^{-5}$~d~d$^{-1}$, with mean periods of 0.072635~d (stage~A), 0.069518~d (stage~B), and 0.069254~d (stage~C).

\subsection{MASTER\,OT\,J055845.55+391533.4}

MASTER\,OT\,J055845.55+391533.4 (hereafter MASTER\,J055845) was discovered as an optical transient by the MASTER-Amur auto-detection system \citep{2014ATel.5905....1Y}. Retrospective examination of the MASTER data revealed two prior outbursts (2011 November 28 and 2012 November 19), and the 2014 February superoutburst was already in progress at 13.9\,mag on February\,13, reaching 14.4\,mag on February\,19. Time-series photometry during the 2014 superoutburst detected superhumps with a period of $P_{\rm sh} = 0.0563$\,(4)\,d , confirming its SU\,UMa-type nature \citep{2015PASJ...67..105K}. If all three outbursts were superoutbursts, the supercycle would be of the order of 360--450\,d \citep{2015PASJ...67..105K}. A second superoutburst was detected by ASAS-SN in 2016 September; however, \citet{2017PASJ...69...75K} noted an unexplained discrepancy in the superhump periods between the 2014 and 2016 events, requiring further observations.

MASTER\,OT\,J055845.55+391533.4 was observed only in Sector~73, hosting one superoutburst lasting $\sim 11.80$~d and one normal outburst with a duration of approximately $1.47$~d. After this normal outburst, another putative similar outburst may have occurred (Fig. \ref{fig:7-12}(b)A), but its reality cannot be confirmed owing to data gaps. The frequency spectrum reveals a PSH signal at $0.055006(1)$~d (Fig. \ref{fig:7-12-R}(b)B), with prominent PSH signatures present around the peak of the superoutburst. A total of 117 PSH maxima ($E(0\text{--}116)$) were extracted via fitting. We adopted the following ephemeris for the $O$--$C$ analysis:
\begin{equation}
	T_{\rm MAX}=3292.54359 + E\times 0.054988
\end{equation}
The $O$--$C$ diagram shows a relatively short Stage\,A and a well‑defined Stage\,B, while Stage\,C is absent (Fig. \ref{fig:7-12-R}(b)D). Irregular variations are also seen prior to Stage\,A. A parabolic fit to Stage\,B yields a period derivative of $(8.14\pm0.96)\times 10^{-5}\ \mathrm{d\,d^{-1}}$.

\subsection{ASASSN-14je}
ASASSN-14je was discovered as a transient by the All-Sky Automated Survey for SuperNovae on 2014~October~20 at $V=14.26$ (vsnet-alert~17880; \citealt{2015PASJ...67..105K}). Follow-up time-resolved photometry detected coherent superhumps with an amplitude of $\sim$0.25~mag, identifying the object as an SU~UMa-type dwarf nova (vsnet-alert~17883). The superoutburst declined unusually swiftly, with rapid fading setting in within five days of outburst onset; consequently, all recorded superhump maxima ($E=0$--74) belong to stage~C, and a linear ephemeris gives $P_{\rm sh}=0.069070(54)$~d ($\sim$99.5~min; \citealt{2015PASJ...67..105K}). Neither stage~B superhumps nor an orbital modulation has been measured, so the orbital period and superhump excess remain undetermined. ASAS-SN recorded a second bright outburst, again likely a superoutburst, on 2015~May~31 at $V\simeq13.9$ (vsnet-alert~18677), but no time-resolved observations were obtained.

TESS monitored ASASSN-14je in S87--S90 and S93--S98. Two superoutbursts were detected in S89 ($\sim$13.95~d) and S96 ($\sim$12.90~d) (Figs. \ref{fig:7-12}(c)A and E), separated by $\sim$199.18~d. Three normal outbursts were also recorded, with a mean duration of 3.23~d; two consecutive normal outbursts occurred in S97--S98 with a recurrence interval of $\sim$30.16~d. Fourier analysis yielded PSH periods of 0.069507(2)~d and 0.069473(2)~d for S89 and~S96 (Fig. \ref{fig:7-12-R}(c)B), respectively; no orbital periodic signals were detected in any sector.
A total of 150 ($E=0$--$162$) and 144 ($E=0$--$148$) PSH maxima were obtained from S89 and~S96, giving the ephemerides:
\begin{align}
	T_{\rm MAX,S89} &= 3723.14265 + E\times 0.069452, \\
	T_{\rm MAX,S96} &= 3922.19621 + E\times 0.069437.
\end{align}
The two $O$--$C$ sequences are highly similar and nearly overlapping, exhibiting clear three-stage evolution with closely matched transition points (Fig. \ref{fig:7-12-R}(c)D). The period-change rates during stage~B are $(8.84\pm0.76)\times10^{-5}$ (S89) and $(7.35\pm1.80)\times10^{-5}$ (S96), consistent within uncertainties. The mean periods in corresponding stages are also very close: stage~A (S89: 0.070623~d; S96: 0.071067~d), stage~B (S89: 0.069257~d; S96: 0.069344~d), and stage~C (S89: 0.069203~d; S96: 0.069167~d).

\subsection{UV Gem}
UV~Geminorum is a long-period SU~UMa-type dwarf nova whose
SU~UMa nature was established from a bimodal outburst pattern and
superhumps recorded during the 2003, 2008, and 2011 superoutbursts
\citep{2001IBVS.5158....1K, 2009PASJ...61S.395K, 2014PASJ...66...30K}.
The superhump period is $P_{\rm sh}\simeq0.0936$~d
($0.092822(94)$~d in 2011), and the $O$--$C$ curves of the three
superoutbursts follow an exceptionally steep downward quadratic
trend with $\dot{P}=-5.34\times10^{-4}$, characteristic of
long-$P_{\rm orb}$ systems. During {\it K2} Campaign~0 a complete
$\sim$5~d, $\sim$3~mag normal outburst was captured, and quiescent
periodograms yielded coherent signals at 0.086089~d (2.07~hr) and
0.088342~d (2.12~hr), giving the first plausible orbital period
$P_{\rm orb}\simeq2.07$~hr; the weaker 2.12~hr peak may be a
residual fading superhump \citep{2016AJ....152....5D}.

TESS observed UV~Gem in S71--S72, where one complete superoutburst (14.68~d) and one normal outburst (3.23~d) were recorded (Fig. \ref{fig:7-12}(d)A).
Fourier analysis yielded a PSH period of 0.092943(3)~d  (Fig. \ref{fig:7-12-R}(d)B), consistent with previous measurements.
An orbital-period signal at 0.088045(9)~d together with its second harmonic was additionally detected, from which the PSH excess is calculated to be 0.056.
A total of 78 PSH maxima ($E=0$--$99$) were obtained via Gaussian fitting, giving the ephemeris:
\begin{equation}
	T_{\rm MAX}=3254.43400 + E\times 0.092790.
\end{equation}
Similar to the case of EC\,05200-5205 (stage\,A1 and stage\,A2), the $O$--$C$ of the first stage is well reproduced by a two‑segment piecewise‑linear function ($r=0.98$)  (Fig. \ref{fig:7-12-R}(d)D). A significant period change is present in stage\,B, whose period‑change rate is $(85.09\pm17.83)\times10^{-5}$.

\subsection{IR Gem}

IR~Geminorum was first listed as a U~Gem-type dwarf nova,
until $\sim$102~min superhumps detected during a superoutburst
established its SU~UMa nature \citep{1984ApJ...282..236S}; optical and
UV spectroscopy subsequently yielded the orbital period
$P_{\rm orb}=0.0684$~d (98.5~min) and a rather small
$K_1\simeq30$~km~s$^{-1}$, suggestive of a massive white dwarf or an
undermassive secondary
\citep{1988AJ.....96.1702F, 1991AJ....101..196L, 1992MNRAS.255..237W}.
Superhumps recorded during the 1991--2017 superoutbursts give a
characteristic stage~B period $P_{\rm sh}\simeq0.0711$~d, whereas the
2017 stage~A superhumps yield $\epsilon^*=0.068(11)$ and $q=0.22(4)$
\citep{2001IBVS.5122....1K, 2009PASJ...61S.395K, 2010PASJ...62.1525K,
	2017PASJ...69...75K}. Quiescent $V$-band photometry reveals a weak
orbital modulation together with episodic signals 5\% longer and 3\%
shorter than $P_{\rm orb}$, attributed to apsidal and nodal disc
precession \citep{2004ChJAA...4...88F}; the system fades from
$V\simeq12.1$ at superoutburst maximum to $V\simeq16.6$ in quiescence,
with normal outbursts every $\sim$21~d and a $\sim$150~d supercycle.

TESS monitored IR~Gem in S44--S45 and S71--S72.
Two normal outbursts were recorded in S44--S45; two additional candidate outbursts occurred near data gaps and are therefore of low reliability (Fig. \ref{fig:7-12}(e)A).
One complete superoutburst (13.13~d) and one normal outburst were detected in S71--S72.
The normal outbursts have comparable durations, with a mean of 2.55~d, and a recurrence interval of 18.74~d.
Fourier analysis yielded a PSH period of 0.071039(3)~d (Fig. \ref{fig:7-12-R}(e)B), while no orbital-period signal was identified.
A total of 133 PSH maxima ($E=0$--$151$) were extracted from light-curve fitting, giving the ephemeris:
\begin{equation}
	T_{\rm MAX}=3256.80340 + E\times 0.070975.
\end{equation}
The $O$--$C$ diagram is well fitted by a three-segment piecewise-linear function ($r=0.92$) (Fig. \ref{fig:7-12-R}(e)D),
with mean periods of 0.072884~d, 0.071095~d, and 0.070727~d for the three successive stages.

\subsection{PNV J06501960+3002449}

PNV\,J06501960+3002449 is an SU~UMa-type dwarf nova identified through the detection of superhumps with a period of 0.068~d and a large amplitude of 0.2~mag (Isogai 2017; vsnet-alert~21718). Archival observations from the NEAT project revealed an earlier outburst of $\sim$2.5~mag on 2002~January~17, with the system brightening from 18.4~mag to 15.8~mag (Denisenko 2017; vsnet-alert~21713).

TESS monitored PNV\,J06501960+3002449 in S44--S45 and S71--S72. One superoutburst ($>$11.40~d) and one normal outburst (2.70~d) were detected in S45 (Fig. \ref{fig:7-12}(f)A). Fourier analysis yielded a PSH period of 0.067504(2)~d  (Fig. \ref{fig:7-12-R}(f)B). Gaussian fitting provided 169 PSH maxima ($E=0$--$169$), giving the ephemeris:
\begin{equation}
	T_{\rm MAX}=2526.03270 + E\times 0.067357.
\end{equation}
The $O$--$C$ diagram reveals the presence of stages~A--C  (Fig. \ref{fig:7-12-R}(f)C). The period-change rate during stage~B is $(18.13\pm1.28)\times10^{-5}$, with mean periods of 0.067606~d (stage~A), 0.066905~d (stage~B), and 0.067053~d (stage~C).


\begin{figure}
	\centering
	\begin{subfigure}{0.45\columnwidth}
		\includegraphics[width=\linewidth]{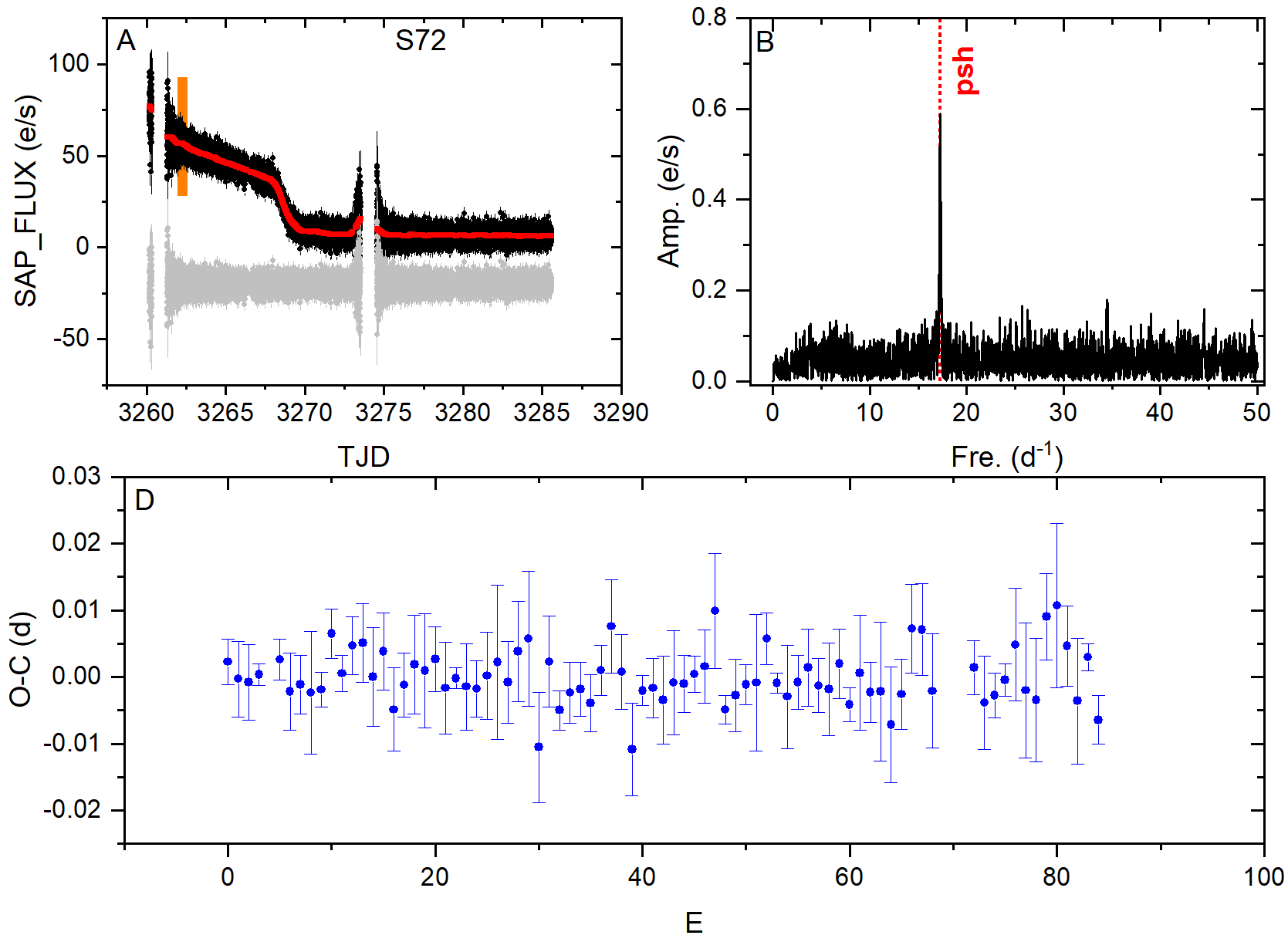}
		\caption{SDSS\,J075107.50+300628.4}
	\end{subfigure}	
	\begin{subfigure}{0.45\columnwidth}
		\includegraphics[width=\linewidth]{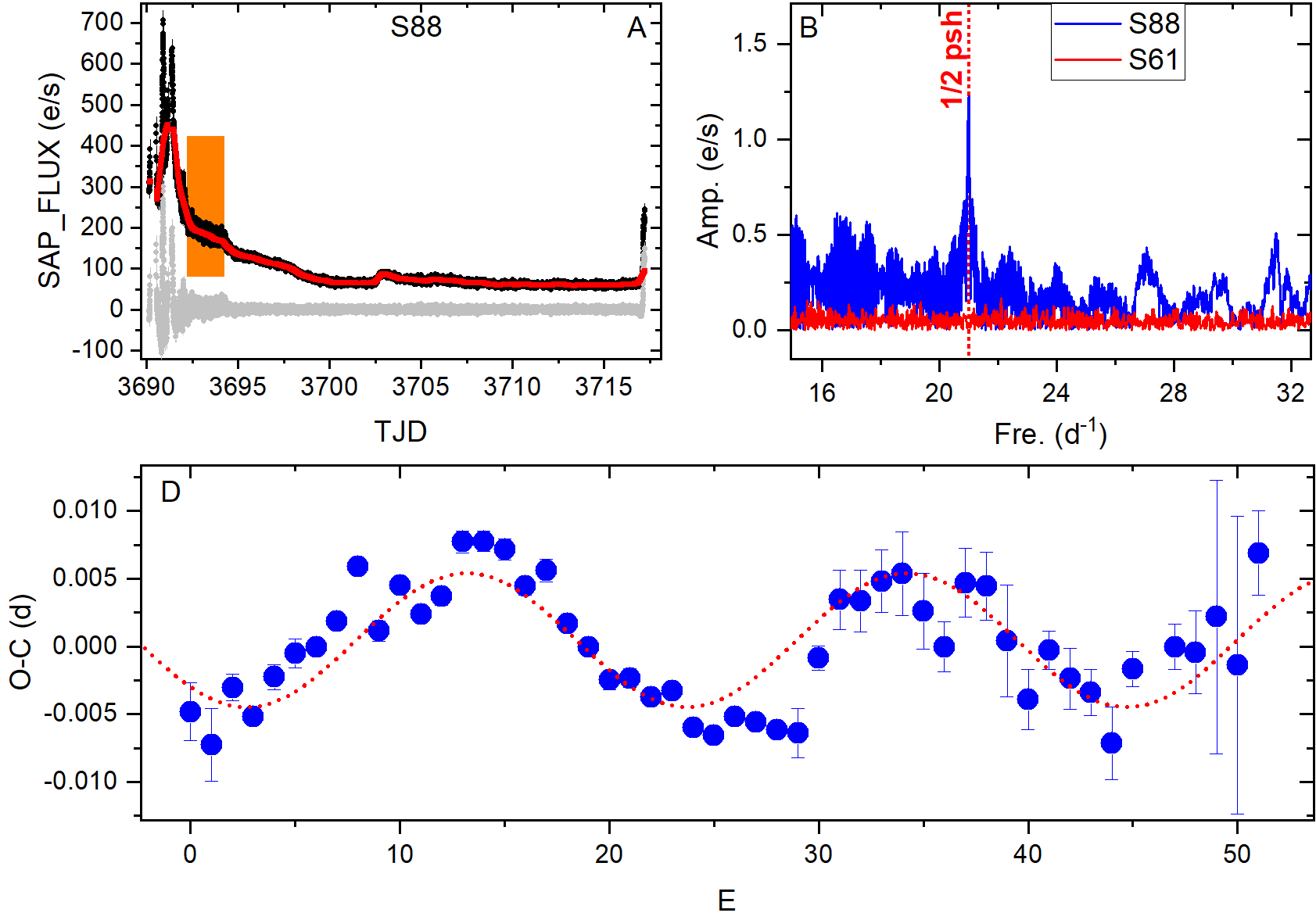}
		\caption{ASASSN-14kj}
	\end{subfigure}

	\begin{subfigure}{0.45\columnwidth}
		\includegraphics[width=\linewidth]{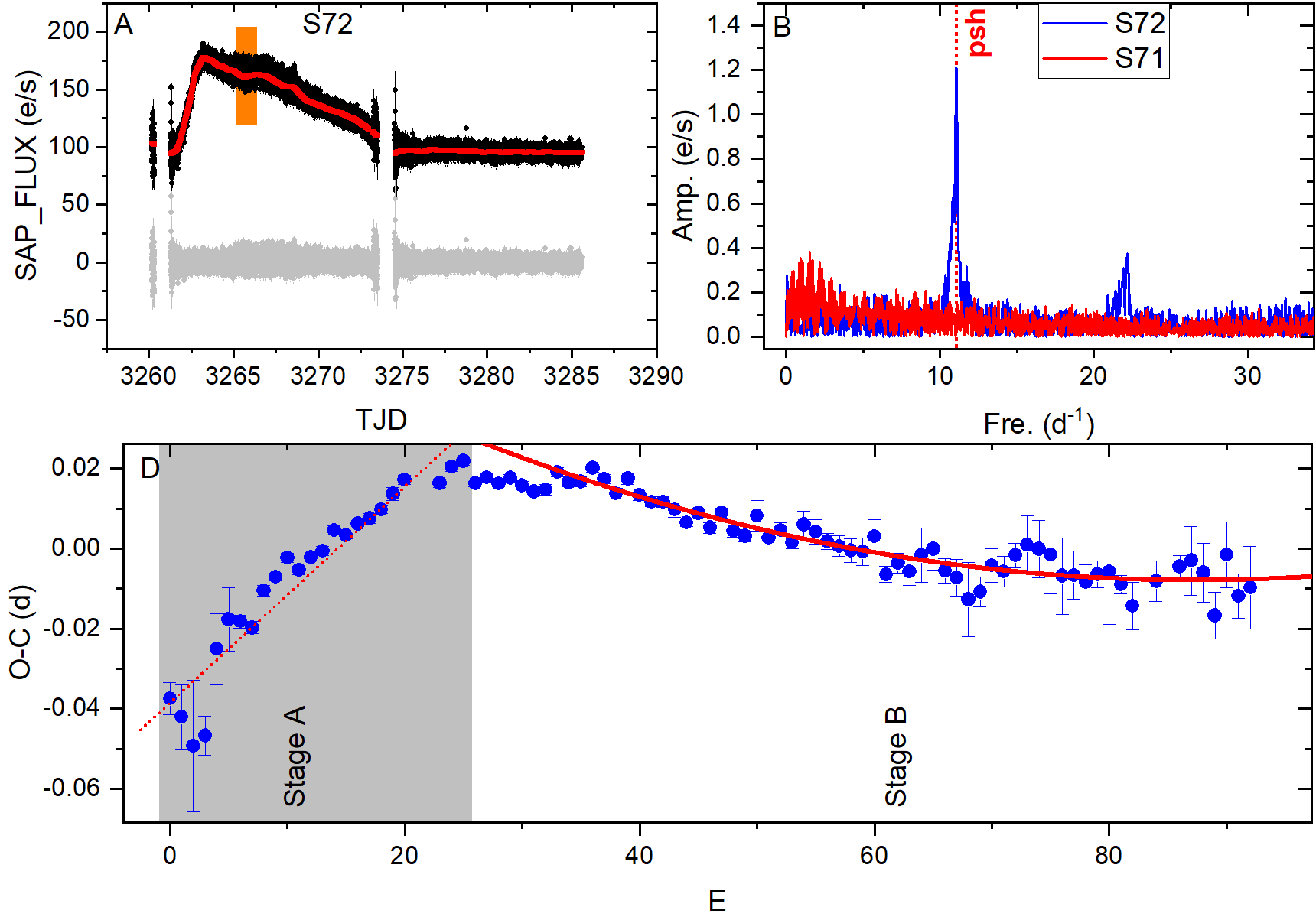}
		\caption{SDSS J080303.90+251627.0}
	\end{subfigure}
	\begin{subfigure}{0.45\columnwidth}
		\includegraphics[width=\linewidth]{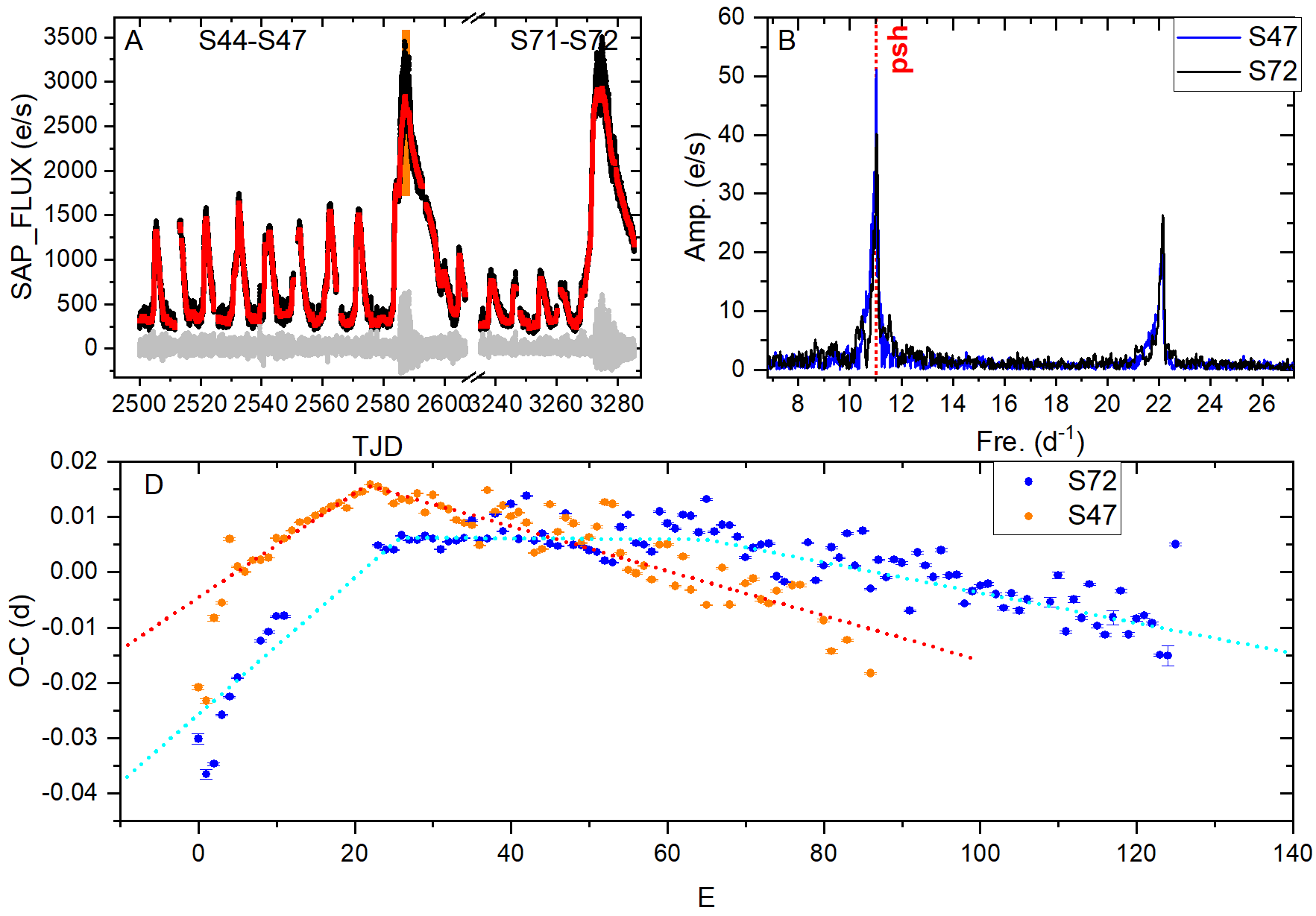}
		\caption{YZ Cnc}
	\end{subfigure}

	\begin{subfigure}{0.45\columnwidth}
		\includegraphics[width=\linewidth]{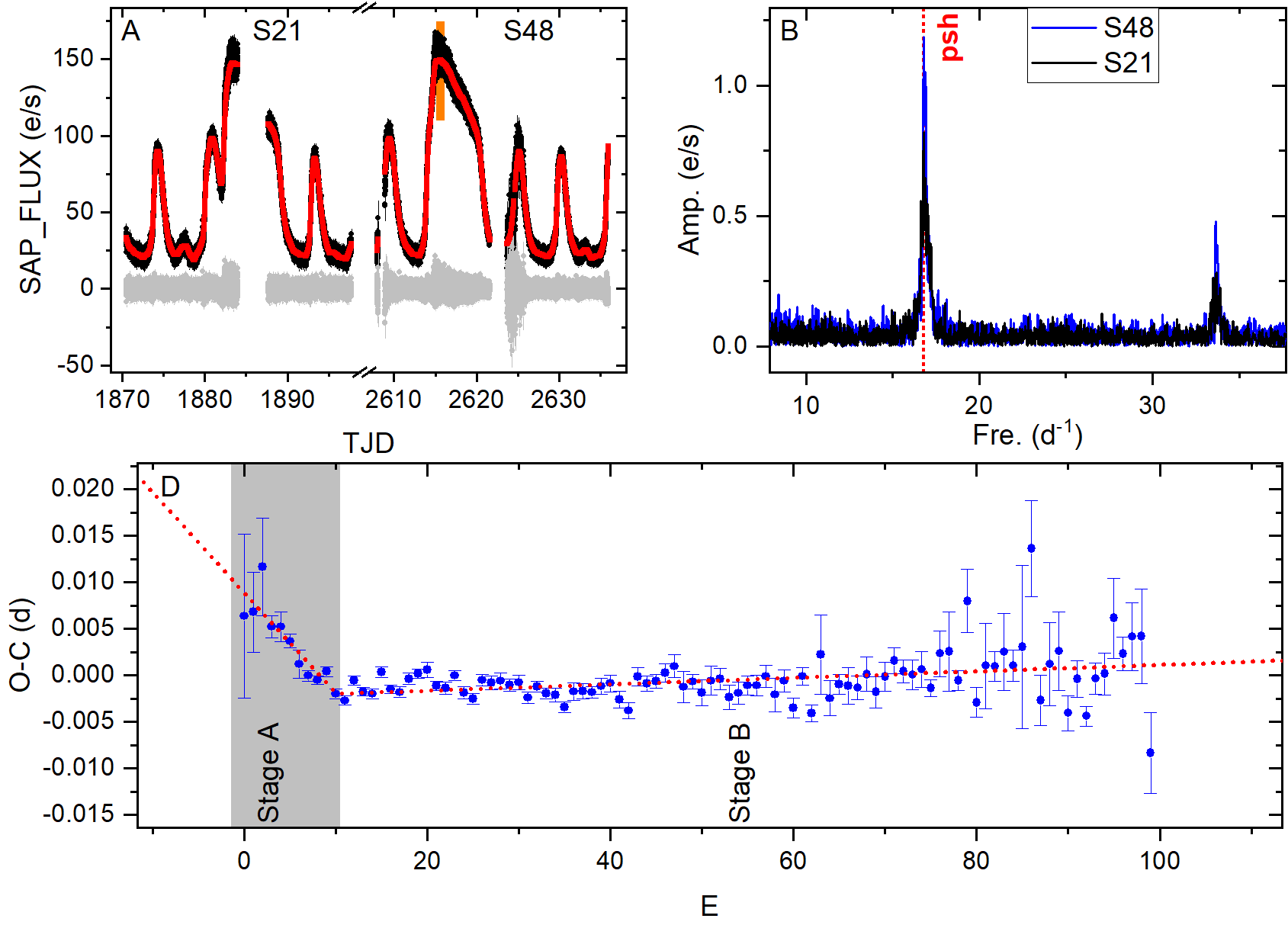}
		\caption{RZ LMi}
	\end{subfigure}
	\begin{subfigure}{0.45\columnwidth}
		\includegraphics[width=\linewidth]{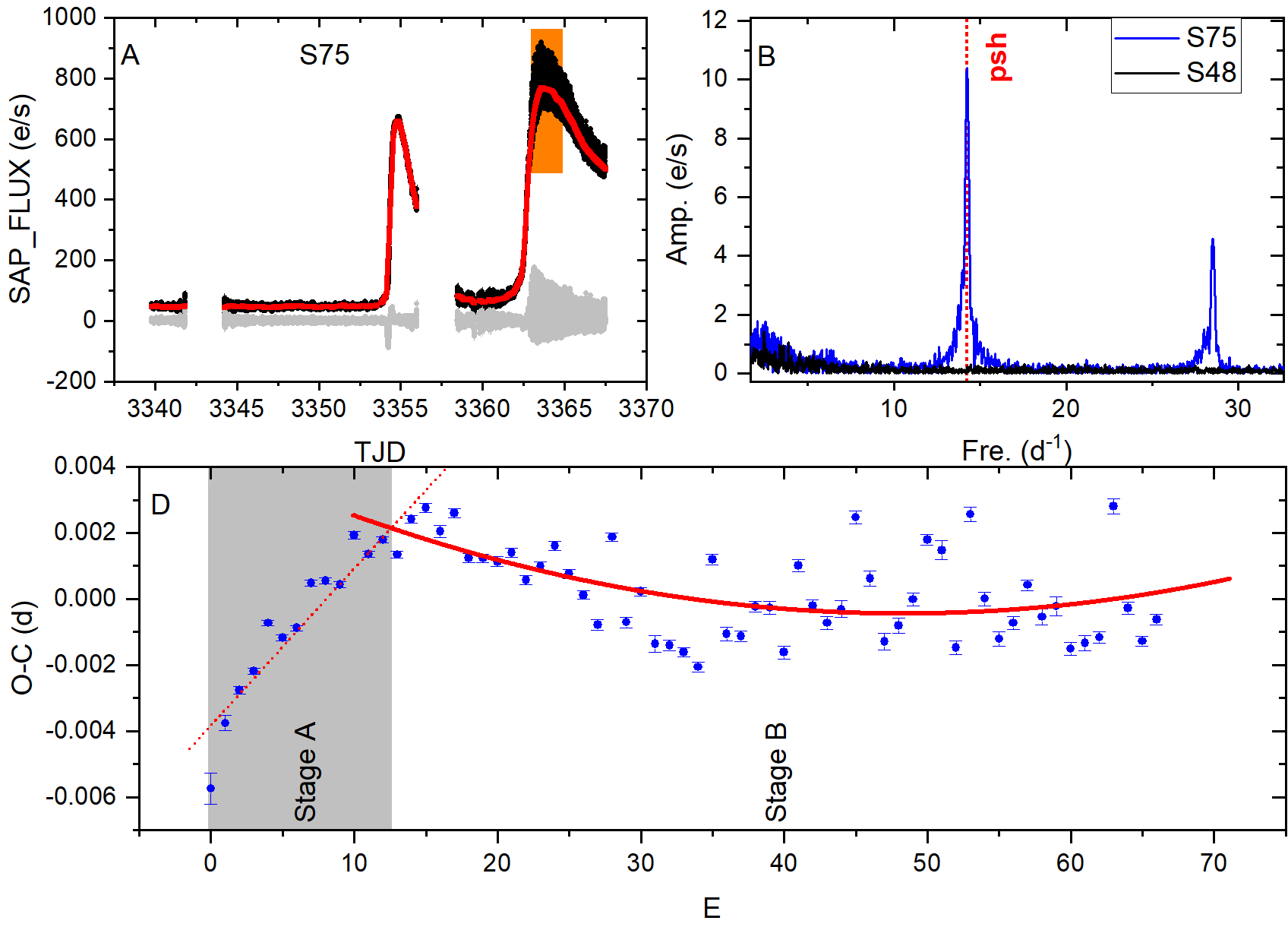}
		\caption{KS UMa}
	\end{subfigure}
\caption{Panels (a)--(f) and their subpanels follow the same numbering convention as Figure~\ref{fig:13-18}. This figure only displays TESS superoutburst light curves, periodograms, and $O$--$C$ diagrams.}
	\label{fig:13-18-R}
\end{figure}

\subsection{SDSS\,J075107.50+300628.4}

SDSS\,J075107.50+300628.4 (hereafter SDSS\,J075107) was identified as a dwarf
nova candidate through systematic data mining of SDSS, GALEX, and astrometric
catalogues by \citet{2010MNRAS.402..436W}, with SDSS colours suggesting an
orbital period of $\sim$0.067~d \citep{2014PASJ...66...30K}.
An outburst detected on 2013~February~12 by E.\,Muyllaert prompted time-series
photometry that revealed growing superhumps with $P_{\rm sh}\approx0.058$~d \citep{2014PASJ...66...30K}, confirming its
SU~UMa nature.
The object exhibits a large outburst amplitude of $\sim$4.7~mag
\citep{2010MNRAS.402..436W}.

TESS observed SDSS\,J075107 in Sector~72, capturing only the trailing half of a superoutburst (Fig. \ref{fig:13-18}(a)A). The power spectrum reveals a PSH signal at 0.058005(4)~d (Fig. \ref{fig:13-18-R}(a)B), broadly consistent with \citet{2014PASJ...66...30K}; this indicates that the previously reported $\sim$0.067~d signal is not the orbital period. Prominent PSH signatures are visible in the light curve. A total of 81 PSH maxima ($E=0$--$84$) were extracted, yielding the $O$--$C$ ephemeris:
\begin{equation}
	T_{\rm MAX}=3261.470701 + E\times 0.057988.
\end{equation}
No distinct staged evolution is seen in the resulting $O$--$C$ diagram.

\subsection{ASASSN-14kj}

ASASSN-14kj was discovered as a transient by the ASAS-SN survey
\citep{2014ApJ...788...48S}, matching a previously known SDSS source
($g=18.44$) with a GALEX ultraviolet counterpart.
Follow-up spectroscopy by \citet{2016AJ....152..226T} revealed a spectrum
typical of a dwarf nova at minimum light, with H$\alpha$ in emission
(EW $\sim$80~\AA, FWHM $\sim$1500~km~s$^{-1}$).
Time-resolved photometry yielded $P_{\rm orb}=0.09056(3)$~d (130.4(4)~min),
placing this system near the lower boundary of the $\sim$2--3~hr period gap
in the cataclysmic variable distribution \citep{2016AJ....152..226T}.

TESS monitored ASASSN-14kj in Sectors~61 and~88. Half of a superoutburst ($>$9.74~d) and one normal outburst ($\sim$1.32~d) were detected in S61 (Fig. \ref{fig:13-18}(b)A); no outbursts were identified in S88. The power spectrum shows no detectable orbital-period signal, whereas a prominent signal at $f=20.9934(29)$~d$^{-1}$ is present in S61 (Fig. \ref{fig:13-18-R}(b)B). Comparison with the published orbital period suggests that this signal is likely the second harmonic of the PSHs. Clear PSH signatures appear during the superoutburst. We extracted 51 PSH maxima ($E=0$--$51$) via Gaussian fitting. Taking the first maximum as the initial epoch and adopting twice the period of the detected signal, the $O$--$C$ analysis after linear correction yielded:
\begin{equation}
	T_{\rm MAX}=3691.84013 + E\times 0.095167.
\end{equation}
We therefore adopt $0.095167(54)$~d as the final PSH period, yielding a superhump excess of $0.048$. The $O$--$C$ diagram does not exhibit distinct staged behaviour, though periodic variations may be present; a sinusoidal fit suggests a modulation period of $1.00(2)$~d (Fig. \ref{fig:13-18-R}(b)D).

\subsection{SDSS\,J080303.90+251627.0}

SDSS\,J080303.90+251627.0 was catalogued as a cataclysmic variable
(GCVS designation Cnc) with $P_{\rm orb}\approx1.7$~hr by
\citet{2001PASP..113..764D}, and spectroscopically confirmed by the SDSS
DR12 survey, which detected H$\alpha$
(EW $\sim$46~\AA), H$\beta$, and He\,{\sc i}\,$\lambda$4471
emission lines \citep{2005AJ....129.2386S}.
Zwicky Transient Facility monitoring recorded outbursts reaching
$V\sim15.3$~mag from a quiescent level of $\sim$18.8~mag
\citep{2021AJ....162...94S}.

In TESS Sectors~71--72, SDSS\,J080303.90+251627.0 exhibited one relatively complete superoutburst lasting 11.70~d (Fig. \ref{fig:13-18}(c)A), with no normal outbursts detected. Prominent PSH signatures appear at the superoutburst peak. Fourier analysis yields a PSH period of 0.090572(7)~d (Fig. \ref{fig:13-18-R}(c)B), and no orbital-period signal is recovered. If the 1.7~h period were the true orbital period, the implied superhump excess would be 0.218---far exceeding the classical few-per-cent level---suggesting that 1.7~h may not represent the orbital period. A total of 89 PSH maxima ($E=0$--$92$) were extracted, giving the ephemeris:
\begin{equation}
	T_{\rm MAX}=3264.48707 + E\times 0.090812.
\end{equation}
The $O$--$C$ diagram shows stage~A plus an incomplete stage~B, whereas stage~C is absent (Fig. \ref{fig:13-18-R}(c)D). The period derivative of stage~B is $(8.30\pm0.14)\times10^{-5}$~d~d$^{-1}$.


\subsection{YZ Cnc}

YZ~Cnc is an eclipsing SU~UMa-type dwarf nova with an orbital period of 0.086924(7)~d \citep{1994MNRAS.267..465V,1961PZ.....13..428V}. \citet{1979AJ.....84..804P} established its SU~UMa nature by detecting superhumps with $P_{\rm sh}=0.09204$~d, corresponding to a superhump excess of $\sim$5.6\%. The system exhibits normal outbursts every $\sim$7--10~d and superoutbursts every $\sim$100--110~d, with large-amplitude flickering ($\sim$0.75~mag) in quiescence \citep{1974ApJ...194..141M}. X-ray observations with ROSAT revealed an unusual anti-correlation: the 0.1--2.4~keV flux during optical outburst is lower than in quiescence \citep{1999A&A...346..146V}, while XMM-Newton measurements yielded a quiescent X-ray luminosity of $\sim$1.4$\times$10$^{32}$~erg~s$^{-1}$ \citep{2004A&A...420..273H}. Spectroscopic analysis of the 2002 superoutburst gave a mass ratio $q\approx0.28$ and implied an eccentric-disc precession period of $\sim$1.5~d \citep{2005ChJAA...5..601Z}. Recent VLA observations finally detected radio emission at 17--27~$\mu$Jy during the 2014 superoutburst \citep{2016MNRAS.463.2229C}, in contrast to previous upper limits of 0.11--0.96~mJy \citep{1986A&A...154..377F}.

TESS observed YZ~Cnc in S44--S47 and S71--S72.
Two superoutbursts were detected in S47 ($\sim$15.70~d) and S92 ($>$18.60~d) (Fig. \ref{fig:13-18}(d)A).
In total, 13 normal outbursts were recorded.
For S44--S47, the normal outbursts have a mean duration of 5.46~d and a mean recurrence interval of 9.49~d; in S71--S72, the corresponding values are 3.63~d and 8.08~d.

Fourier analysis yielded PSH periods of 0.090728(3)~d and 0.090604(4)~d for S47 and~S72 (Fig. \ref{fig:13-18-R}(d)B), respectively; no orbital-period signal was found. Gaussian fitting yielded 78 ($E=0$--$86$) and 110 ($E=0$--$125$) PSH maxima for S47 and~S72, giving the ephemerides:
\begin{align}
	T_{\rm MAX,S47} &= 2585.01534 + E\times 0.090921, \\
	T_{\rm MAX,S72} &= 3272.52304 + E\times 0.090502.
\end{align}
The $O$--$C$ sequence of S47 exhibits only two-stage evolution (Fig. \ref{fig:13-18-R}(d)D), with mean periods of 0.091867~d and 0.090517~d. By contrast, the S72 $O$--$C$ diagram shows clear three-stage evolution, with corresponding mean periods of 0.091742~d, 0.090492~d, and 0.090228~d.

\subsection{RZ LMi}

RZ~LMi is an extreme ER~UMa-type dwarf nova, recognised as the most frequently outbursting SU~UMa system, with an orbital period of 0.05792~d \citep{1995PASP..107..615R, 2016PASJ...68...59K}. Its superhump period of $P_{\rm sh}=0.05944$~d corresponds to a period excess of $\varepsilon\sim2.5$\%, and Kato sequence observations have classified it as a stage~C object with a positive period derivative \citep{2013PASJ...65...23K}, while analysis of the 2005 superoutburst revealed a marginally positive period derivative $\dot{P}=+2.3(1.1)\times10^{-5}$ and stage-C superhumps at 0.05875(8)~d, along with phase-0.5-offset superhumps and candidate orbital humps not previously reported in this system \citep{2009PASJ...61S.395K}. The system exhibits the shortest known supercycle among ER~UMa stars, typically $\sim$19~d, with normal outbursts recurring every $\sim$4~d and superoutbursts lasting slightly over 10~d \citep{2008AcA....58..131O}. In 2016 the supercycle substantially lengthened to 35--60~d, suggesting a transition toward a novalike state with permanent superhumps \citep{2015PASJ...67....1K}; multicolour photometry further revealed the largest $(U-B)$ colour excess in quiescence and during outburst decline \citep{2018Ap&SS.363..100S}. Recent light-curve analysis confirmed the ER~UMa classification and derived a mass ratio $q=0.105(5)$ from negative superhumps \citep{2025MUPB...80S.236V}.

TESS monitored RZ~LMi across multiple sectors, with two superoutbursts detected in S21 ($\sim$10.92~d) and S48 ($\sim$8.47~d) (Fig. \ref{fig:13-18}(e)A), together with six normal outbursts (mean 2.77~d). Fourier analysis yielded PSH periods of 0.059600(3)~d and 0.059581(2)~d for S21 and~S48 (Fig. \ref{fig:13-18-R}(e)B). Only a little more than ten PSH maxima are available for S21; we therefore measured PSH maxima exclusively for S48, obtaining 100 ($E=0$--$99$) maxima. The following ephemeris was adopted:
\begin{equation}
	T_{\rm MAX,S48}=2614.50563 + E\times 0.059395.
\end{equation}
The $O$--$C$ diagram can be fitted by a two-segment piecewise-linear function and reveals unusual behaviour (Fig. \ref{fig:13-18-R}(e)D). Theoretically, stage~A should exhibit the longest period among all evolutionary stages; however, the linear-fit slope of stage~A in RZ~LMi is negative, and its mean period of 0.058315~d is shorter than that of the subsequent stage (0.059429~d).

\subsection{KS\,UMa}

KS\,UMa was first identified as an emission-line object
by \citet{1997Ap.....40..101B} and caught in outburst in 1998
\citep{2004AJ....128.1882S}.
Superhumps with $P_{\rm sh}\approx0.069$~d were detected during that
outburst, confirming its SU~UMa nature.
A combined analysis of the 2003 superoutburst traced a clear three-phase
period evolution: decreasing in stage~A ($E\lesssim15$), slightly
increasing ($\dot{P}_{\rm sh}=+2.2\times10^{-5}$) in stage~B, and
dropping to a shorter value after $E\sim95$ (stage~C)
\citep{2003AcA....53..175O, 2009PASJ...61S.395K}.
With $P_{\rm orb}=0.06796$~d ($\sim$97.9~min), this system lies below
the period gap \citep{2009MNRAS.397.2170G}.

TESS observed KS\,UMa in Sectors~48 and~75, detecting one incomplete superoutburst and two normal outbursts (Fig. \ref{fig:13-18}(f)A); one normal outburst is fully covered with a duration of 2.71~d. Fourier analysis yields a PSH period of 0.070249(2)~d (Fig. \ref{fig:13-18-R}(f)B). We extracted 67 PSH maxima ($E=0$--$66$), yielding the $O$--$C$ ephemeris:
\begin{equation}
	T_{\rm MAX}=3362.81429 + E\times 0.070174.
\end{equation}
The $O$--$C$ diagram displays clear stages~A and~B, whereas stage~C is absent (Fig. \ref{fig:13-18-R}(f)D). The period derivative for stage~B is $(5.77\pm2.39)\times10^{-5}$~d~d$^{-1}$.


\begin{figure}
	\centering
	\begin{subfigure}{0.45\columnwidth}
		\includegraphics[width=\linewidth]{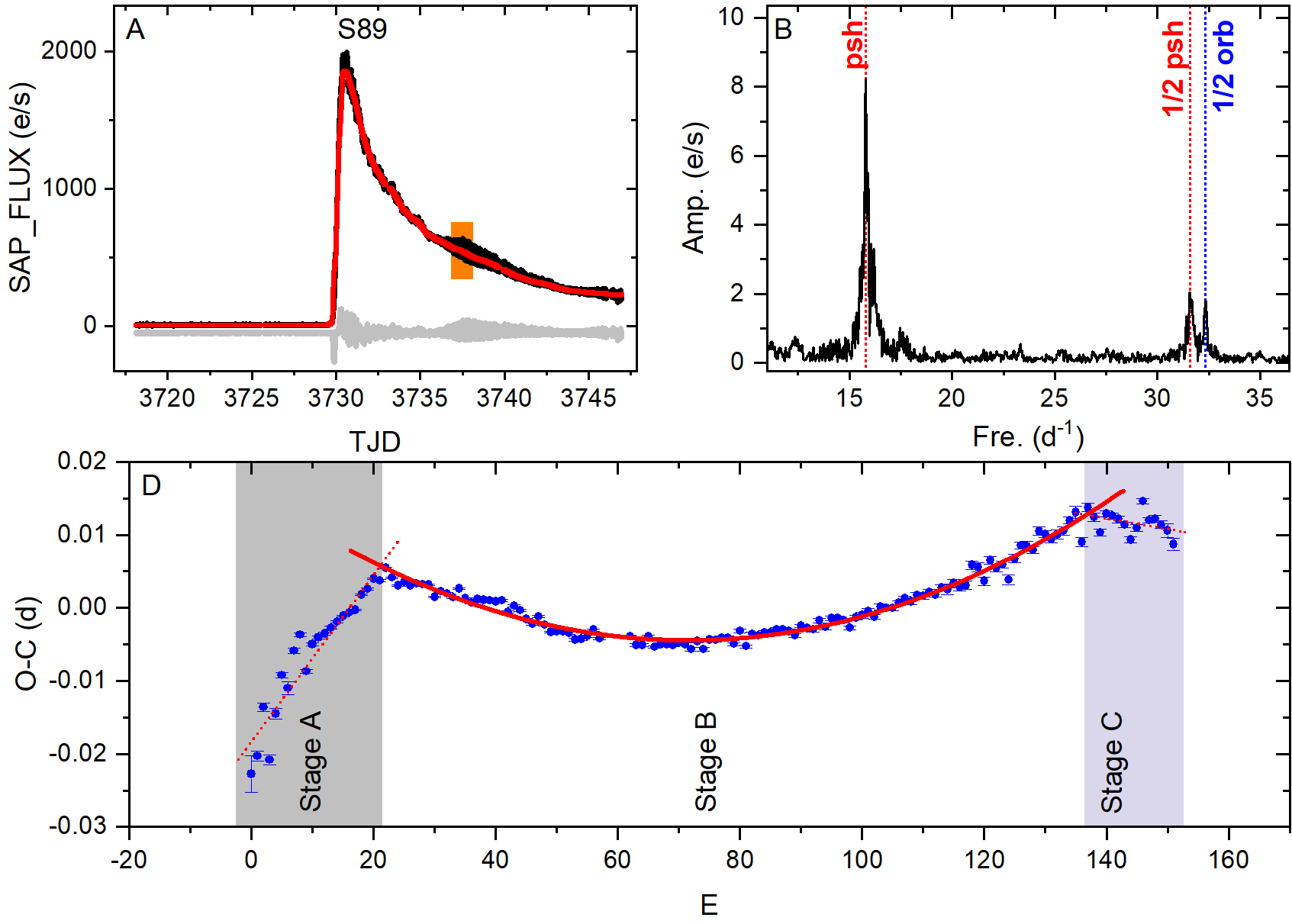}
		\caption{V0748 Hya}
	\end{subfigure}	
	\begin{subfigure}{0.45\columnwidth}
		\includegraphics[width=\linewidth]{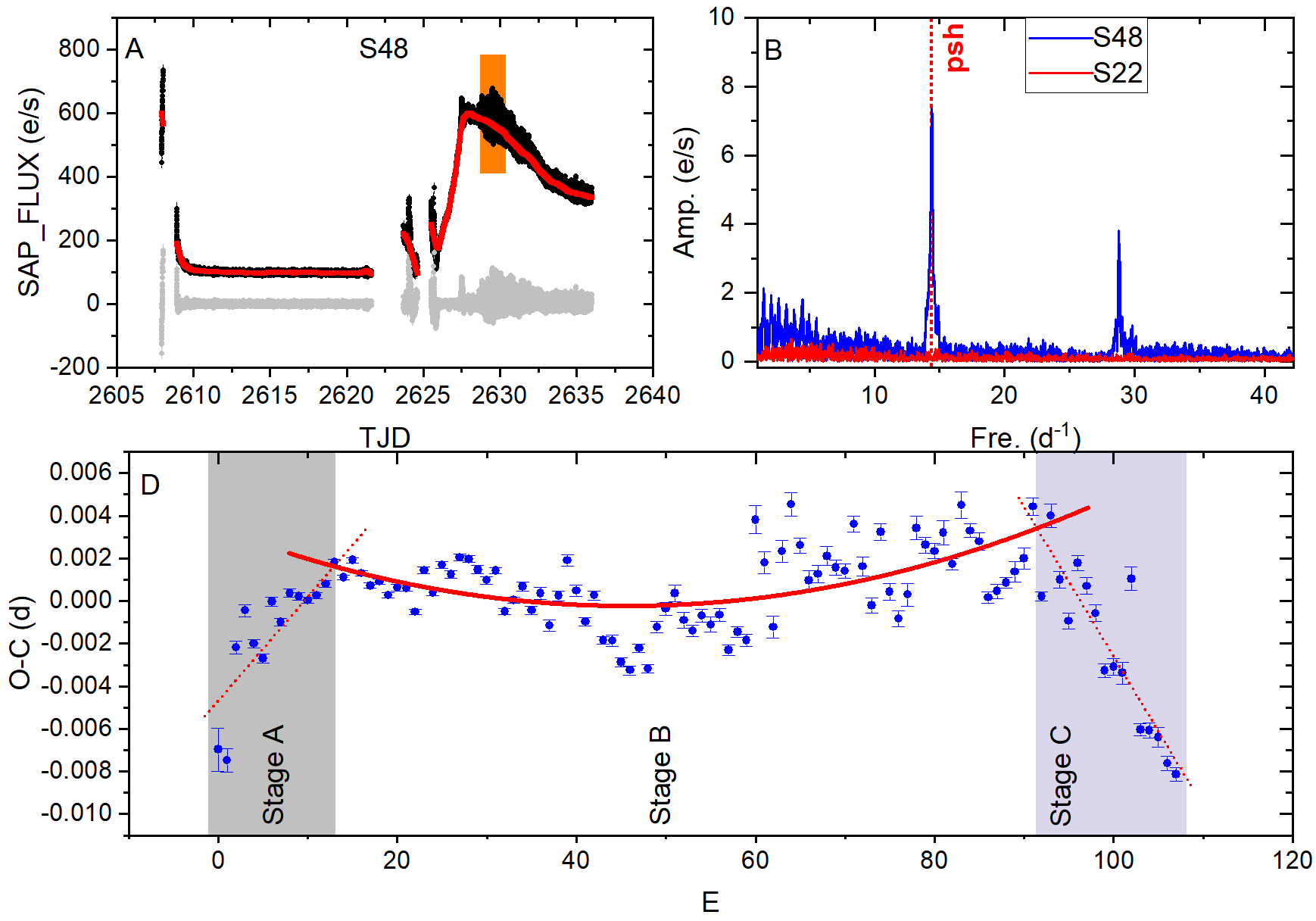}
		\caption{SX LMi}
	\end{subfigure}

	\begin{subfigure}{0.45\columnwidth}
		\includegraphics[width=\linewidth]{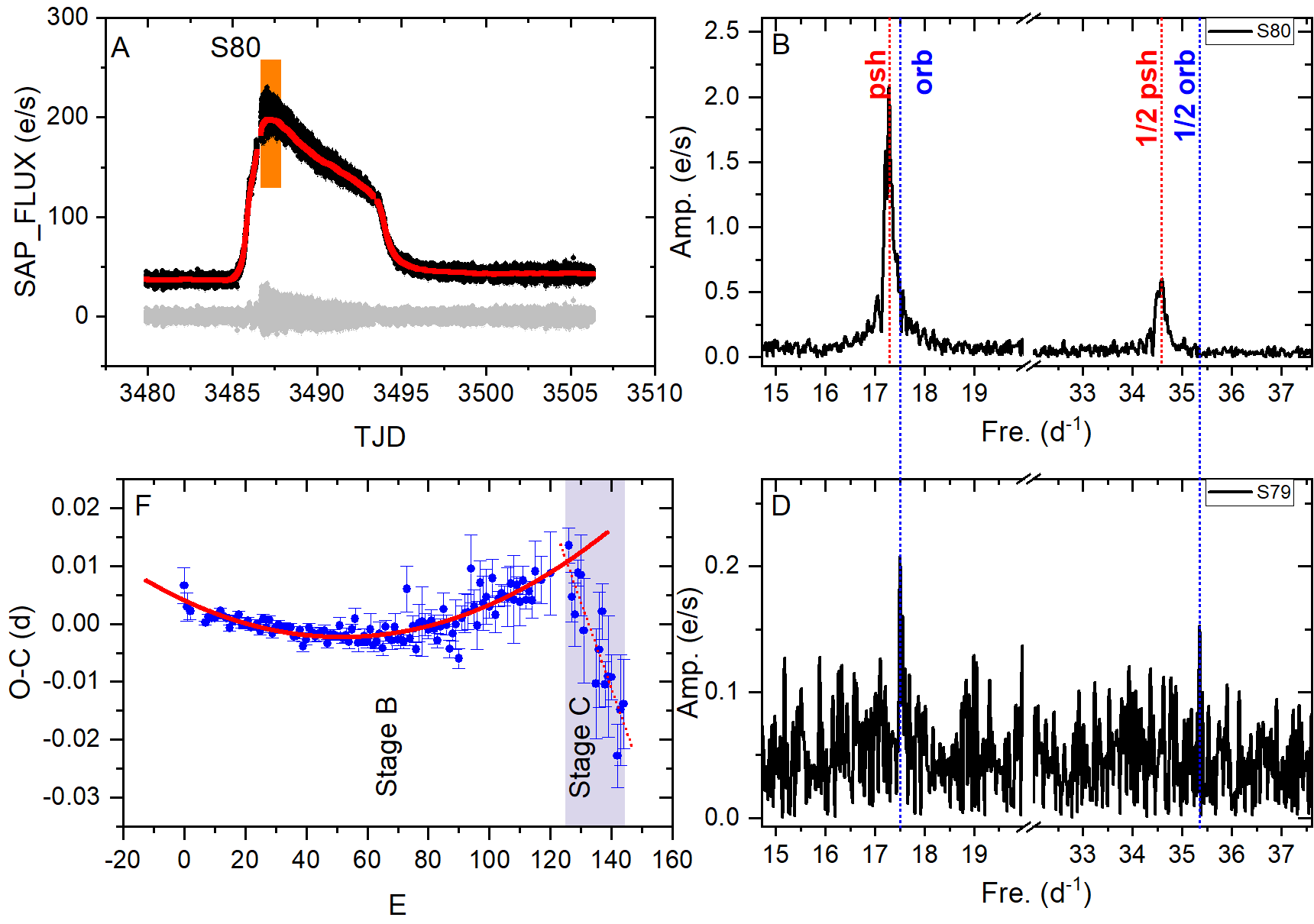}
		\caption{MASTER OT J172758.09+380021.5}
	\end{subfigure}
	\begin{subfigure}{0.45\columnwidth}
		\includegraphics[width=\linewidth]{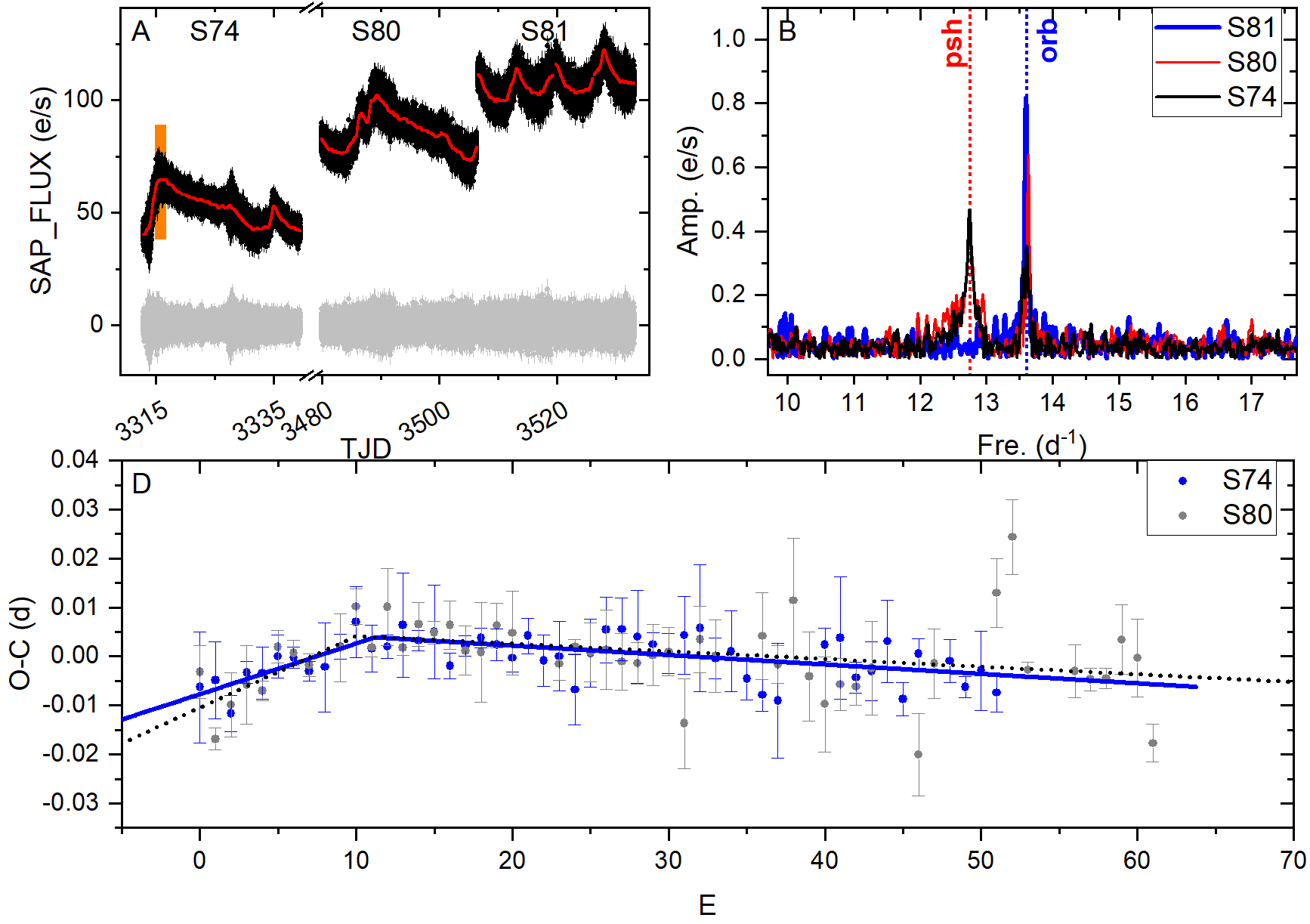}
		\caption{ASASSN-18hr}
	\end{subfigure}

	\begin{subfigure}{0.45\columnwidth}
		\includegraphics[width=\linewidth]{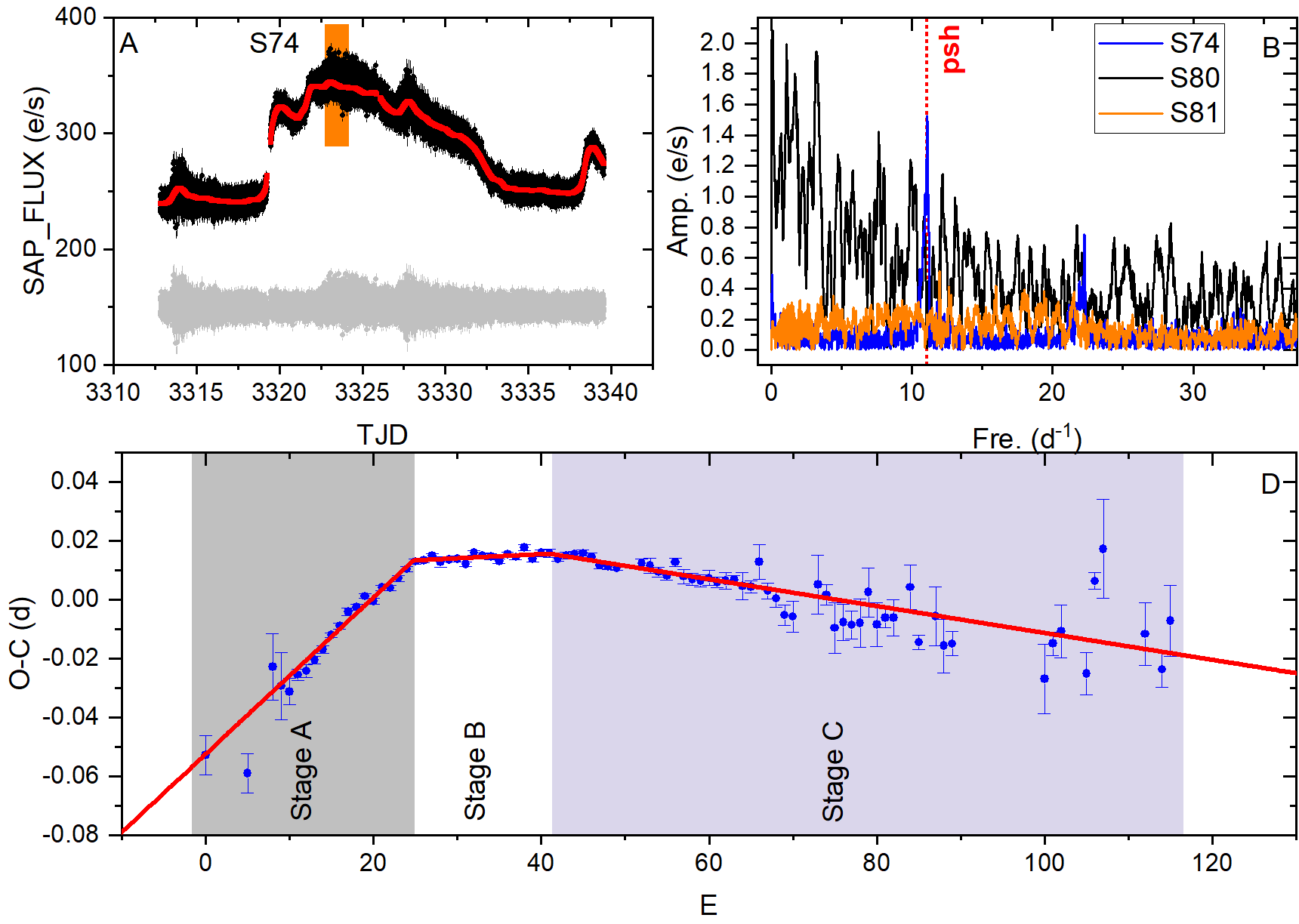}
		\caption{V0419 Lyr}
	\end{subfigure}
	\begin{subfigure}{0.45\columnwidth}
		\includegraphics[width=\linewidth]{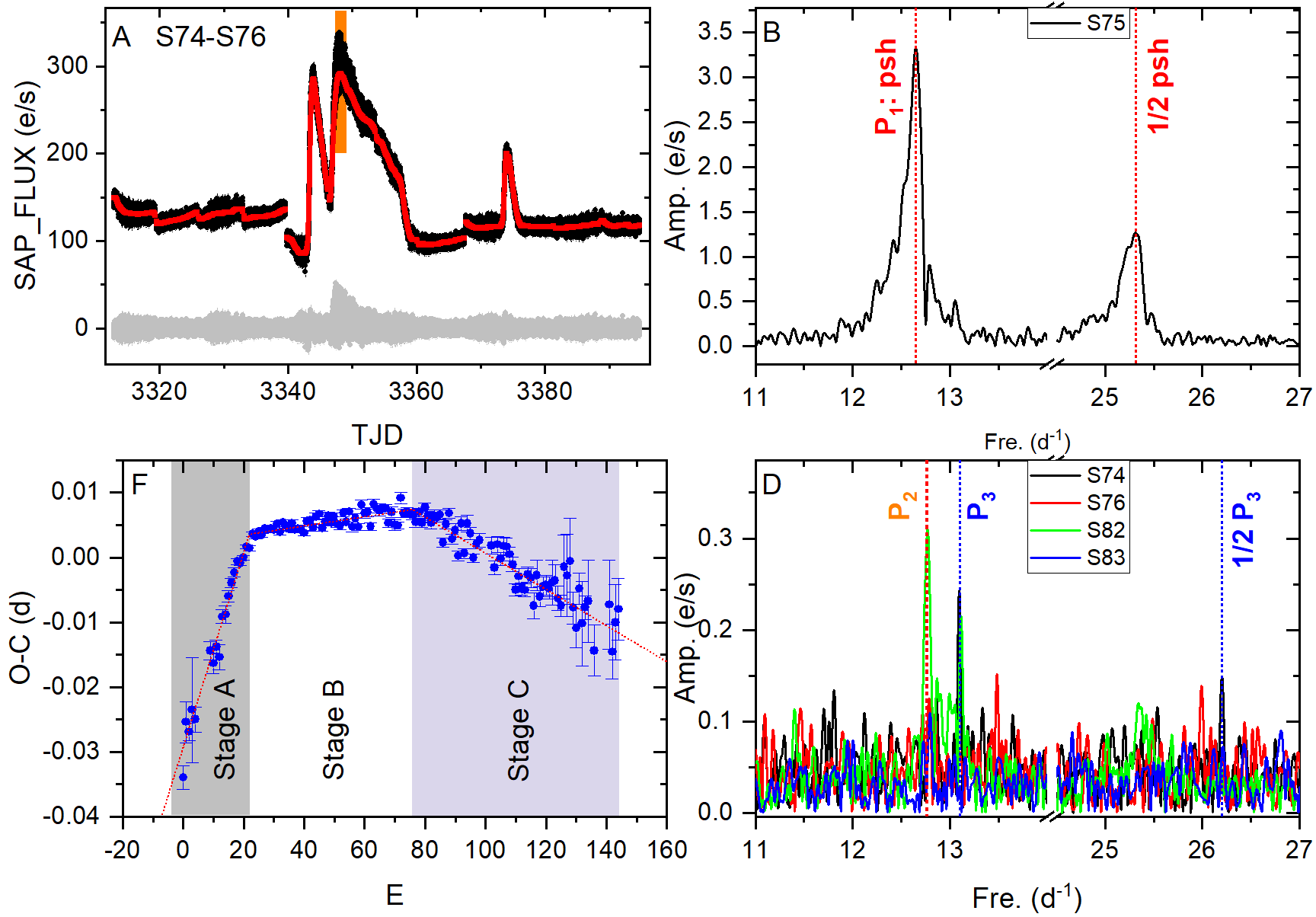}
		\caption{V1113 Cyg}
	\end{subfigure}
\caption{Panels (a)--(f) and their subpanels follow the same numbering convention as Figure~\ref{fig:19-24}. This figure only displays TESS superoutburst light curves, periodograms, and $O$--$C$ diagrams.}
	\label{fig:19-24-R}
\end{figure}

\subsection{V0748\,Hya}

V0748\,Hya ($=$\,ASAS\,J102522$-$1542.4) was detected in outburst by the
ASAS-3 survey on 2006~January~26 ($V=12.2$; \citealt{2009PASJ...61S.395K}).
The simultaneous presence of early superhumps
($P_{\rm esh}=0.06136(6)$~d)
and ordinary superhumps
($P_{\rm sh}=0.063314(5)$~d) led to its classification as a WZ\,Sge-type
dwarf nova (vsnet-alert~8824). The orbital period is $P_{\rm orb}=0.06136$~d.
A stage~A--C period evolution gives $\dot{P}_{\rm sh}=+10.9\times10^{-5}$~d~d$^{-1}$
during stage~B \citep{2009PASJ...61S.395K, 2015PASJ...67..108K}.

TESS monitored V0748\,Hya in Sector~89, revealing a relatively complete superoutburst lasting $>$17.44~d (Fig. \ref{fig:19-24}(a)A). The power spectrum shows three significant periodic signals: PSHs at 0.063347(3)~d, their second harmonic, and a third signal tentatively identified as the second harmonic of the orbital period (Fig. \ref{fig:19-24-R}(a)B). This yields an estimated $P_{\rm orb}=0.061838(5)$~d and a superhump excess of 0.0244. Prominent PSH modulations are present during the mid-decline of the superoutburst. A total of 148 PSH maxima ($E=0$--$151$) were extracted via Gaussian fitting, giving the ephemeris:
\begin{equation}
	T_{\rm MAX}=3736.07109 + E\times 0.063311.
\end{equation}
The $O$--$C$ diagram reveals well-defined stages~A, B, and a relatively short stage~C (Fig. \ref{fig:19-24-R}(a)D). The period derivative for stage~B is $(12.76\pm0.25)\times10^{-5}$~d~d$^{-1}$, with mean periods of 0.064451~d (stage~A), 0.062734~d (stage~B), and 0.063166~d (stage~C).


\subsection{SX LMi}

SX\,LMi ($=$\,CBS\,31 $=$\,Ton\,45) was first catalogued as a blue stellar object by \citet{1957BoOTT..16....3I} and independently by \citet{1984ApJS...55..517S}. \citet{1997PASP..109.1114N} confirmed its SU~UMa nature through the detection of superhumps ($P_{\rm sh}=0.06950$~d) during the 1994 superoutburst, with an overall decreasing period $\dot{P}=-8.2(1.1)\times10^{-5}$. The object was later independently identified as a cataclysmic variable in the SDSS spectroscopic survey \citep{2007AJ....134..185S}. Multi-epoch superoutburst observations revealed evolving $\dot{P}$ behaviour: the 2001 event showed a nearly constant superhump period ($\dot{P}\sim0$), while the 2002 event exhibited a two-stage pattern---a stable plateau ($\dot{P}=-0.7\times10^{-5}$, $14\leq E\leq115$) followed by a sudden transition to a shorter period between $E=116$ and $E=130$ \citep{2009PASJ...61S.395K}. With $P_{\rm orb}=0.06717$~d and a notably small outburst amplitude of $\sim$3.8~mag---atypical for ordinary SU~UMa stars but reminiscent of ER~UMa objects---SX~LMi has been proposed as a potential transitional system between the SU~UMa and ER~UMa populations \citep{1997PASP..109.1114N}.

TESS observed SX\,LMi in multiple sectors, revealing one superoutburst in Sector~48 and one normal outburst in Sector~22. The superoutburst lasted $>$10.00~d, while the normal outburst had a duration of 3.98~d (Fig. \ref{fig:19-24}(b)A). Prominent PSH signatures were present throughout the superoutburst. The power spectrum yields a PSH period of 0.069522(3)~d (Fig. \ref{fig:19-24-R}(b)B), very close to previously published values; no orbital-period signal is detected. A total of 108 PSH maxima ($E=0$--$107$) were extracted via Gaussian fitting, and the following ephemeris was adopted:
\begin{equation}
	T_{\rm MAX}=2628.56177 + E\times 0.069420.
\end{equation}
The $O$--$C$ diagram shows fully developed stages~A through~C (Fig. \ref{fig:19-24-R}(b)D). The period derivative for stage~B is $(5.01\pm1.10)\times10^{-5}$~d~d$^{-1}$, with mean periods of 0.069908~d (stage~A), 0.069262~d (stage~B), and 0.068718~d (stage~C).

\subsection{MASTER OT J172758.09+380021.5}

MASTER\,OT\,J172758.09+380021.5 was discovered as a bright transient by the MASTER-Amur survey \citep{2014ATel.5724....1D}, with archival CRTS data revealing at least nine outbursts between 2005 and~2013 and a quiescent magnitude of $\sim$18~mag. \citet{2016AJ....152..226T} derived an orbital period of 82.14(6)~min from quiescent spectroscopy---close to the CV period minimum---making it the shortest-period object in their sample. Superhumps were detected during the 2018 outburst, establishing the system as an SU~UMa-type dwarf nova with $P_{\rm sh}=0.0587(6)$~d (vsnet-alert~22335). \citet{2021CoSka..51..138P} classified it as an ER~UMa-type dwarf nova based on extensive photometry, revealing a highly variable supercycle of $\sim$50--100~d, unusually short superoutbursts of only $\sim$7~d, and a low mass ratio $q=0.08$ with $M_2\approx0.06\,M_\odot$. The inferred mass-transfer rate significantly exceeds that expected from gravitational radiation alone, suggesting a post-nova state and possibly providing the first observational evidence that nova eruptions can occur near the period minimum.

TESS observed this system in Sectors~79 and~80, capturing one superoutburst ($>$10.66~d in S80) and two normal outbursts in S79 (Fig. \ref{fig:19-24}(c)A). The two short outbursts have nearly identical durations of $\sim$2.87~d, separated by 13.79~d. The power spectrum yields a PSH period of 0.057850(1)~d. An orbital-period signal together with its second harmonic is detected in S79 (Figs. \ref{fig:19-24-R}(c)B and D), giving $P_{\rm orb}=0.057117(11)$~d. A total of 129 PSH maxima ($E=0$--$144$) were extracted, yielding the ephemeris:
\begin{equation}
	T_{\rm MAX}=3486.31900 + E\times 0.057878.
\end{equation}
The $O$--$C$ diagram shows no detectable stage~A, whereas stages~B and~C are present (Fig. \ref{fig:19-24-R}(c)D). The period derivative for stage~B is $(8.25\pm0.59)\times10^{-5}$~d~d$^{-1}$.

\subsection{ASASSN-18hr}

ASASSN-18hr is a ER UMa-type SU UMa dwarf nova discovered by ASAS-SN \citep{2017PASPKochanek}. Although its variability classification was established from ZTF photometric data \citep{2019ZTF}, no orbital period or superhump period has been published in the VSX catalogue or existing literature. 
ASASSN-18hr was recorded by TESS in S74, where one superoutburst ($\sim$17.09~d) and one normal outburst were detected.
Another superoutburst together with four normal outbursts (mean 3.12~d) were observed in S80--S81 (Fig. \ref{fig:19-24}(d)A).
Fourier transforms were performed for each of the three sectors. PSH periods of 0.078450(7)~d and 0.078462(10)~d were detected in S74 and~S80 (Fig. \ref{fig:19-24-R}(d)B), respectively. Significant orbital-period signals were present in all three sectors, yielding a mean orbital period of 0.073466(6)~d.

Gaussian fitting yielded 49 ($E=0$--$51$) and 50 ($E=0$--$61$) PSH maxima for S74 and~S80, giving the ephemerides:
\begin{align}
	T_{\rm MAX,S74} &= 3314.93003 + E\times 0.078522, \\
	T_{\rm MAX,S80} &= 3487.96906 + E\times 0.078568.
\end{align}
The two $O$--$C$ results are highly similar. Owing to the lack of sufficient minima, only two evolutionary stages are visible, well reproduced by a two-segment piecewise-linear function (Fig. \ref{fig:19-24-R}(d)D). The mean periods are 0.079552~d and 0.078329~d for S74, and 0.080028~d and 0.078412~d for S80.

\subsection{V0419 Lyr}

V0419\,Lyr was originally classified as a Z~Cam-type variable by \citet{1970PZ.....17..186K} with an outburst cycle of 8--14~d. \citet{1998PASJ...50..411N} identified superhumps during the 1995 superoutburst, revealing a superhump period near or exceeding two hours---making V0419\,Lyr one of the longest-$P_{\rm sh}$ SU~UMa systems known at that time. Follow-up observations of the 1999 superoutburst by \citet{2009PASJ...61S.395K} firmly established $P_{\rm sh}\approx0.090$~d and yielded a strongly negative period derivative $\dot{P}=-32.4(2.4)\times10^{-5}$, among the most negative values recorded in SU~UMa-type dwarf novae; \citet{2007AcA....57..267R} independently confirmed $\dot{P}=-24.8(2.2)\times10^{-5}$ during the 2006 superoutburst. This system belongs to a group of long-$P_{\rm sh}$ SU~UMa stars (including UV~Gem, MN~Dra, and NY~Ser) characterized by strongly negative $\dot{P}$ and frequent normal outbursts recurring every 9--12~d \citep{2001IBVS.5158....K, 1995IBVS.4208....I}.

TESS captured V0419\,Lyr in Sector~74 and Sectors~80--81. A superoutburst lasting 14.50~d occurred in S74 (Fig. \ref{fig:19-24}(e)A), with prominent superhump signatures at the outburst peak. The power spectrum yields a PSH period of 0.089918(6)~d (Fig. \ref{fig:19-24-R}(e)B). Two normal outbursts are detected in S80--81; one lasts $\sim$8.22~d. Owing to poor data quality, it cannot be reliably determined whether the large-amplitude variations correspond to a superoutburst or instrumental trends; neither PSH nor orbital-period signals are recovered.
Gaussian fitting of the S74 superoutburst yielded 87 PSH maxima ($E=0$--$115$), giving the ephemeris:
\begin{equation}
	T_{\rm MAX}=3321.38570 + E\times 0.090394.
\end{equation}
Unlike the parabolic stage~B evolution seen in many systems, the $O$--$C$ diagram exhibits only linear variations. The entire $O$--$C$ sequence is well described by a three-segment piecewise-linear function ($r=0.95$, $p<9.72\times10^{-51}$) (Fig. \ref{fig:19-24-R}(e)D), with mean periods decreasing sequentially: 0.093054~d, 0.090536~d, and 0.089938~d.

\subsection{V1113 Cyg}

V1113\,Cyg was originally catalogued as a dwarf nova by \citet{1966AN....289..139H}, who recorded four short outbursts with a recurrence interval of only $\sim$10~d. The SU~UMa nature was established by \citet{1996PASJ...48...45K}, who detected superhumps with $P_{\rm sh}=0.0792(1)$~d and noted the unusual coexistence of a short outburst recurrence time and a large ($\sim$6~mag) amplitude. Extensive monitoring over 1994--2001 revealed a mean supercycle of 189.8~d but an anomalously low number of normal outbursts---only $\sim$2 per supercycle \citep{2001IBVS.5110....K}. CCD photometry of the 2003 and 2005 superoutbursts by \citet{2010AcA....60..137B} refined $P_{\rm sh}$ to 0.07891(3)~d; the reported extreme period derivative $\dot{P}=-4.5(8)\times10^{-4}$ was later reinterpreted as a stage~B--C transition by \citet{2010PASJ...62.1525K}, a pattern consistently confirmed across multiple subsequent superoutbursts \citep{2009PASJ...61S.395K}. V1113\,Cyg thus stands out among short-supercycle SU~UMa stars for its suppressed normal outburst activity, pointing to a mechanism that preferentially excites the tidal instability over the thermal one \citep{2001IBVS.5110....K, 2010AcA....60..137B}.

TESS observed V1113\,Cyg in Sectors~74--76 and~82--83. A superoutburst lasting 16.09~d was recorded in S75 (Fig. \ref{fig:19-24}(f)A). Two well-defined normal outbursts were detected in S76 and~S82, both with durations of $\sim$2.76~d. The early portion of S82 suffers from data gaps, precluding confirmation of any outburst. Fourier analysis reveals PSHs at 0.079058(3)~d during the superoutburst. A periodic signal at $P_2=0.078309(9)$~d is recovered in S82, and another at $P_3=0.076306(13)$~d (with its second harmonic in S74) is found in S74 and~S82 (Figs.\ref{fig:19-24-R}(f)B and D). If $P_2$ represents the orbital period, the superhump excess is only $\sim$0.0095, far below the predicted value ($\sim$0.04); if instead $P_3$ is the orbital period, the superhump excess is 0.036, more consistent with empirical relations. Furthermore, the presence of a second harmonic and the absence of pre-superoutburst NSHs make $P_3$ more characteristic of an orbital signal. We therefore adopt $P_{\rm orb}=0.076306(13)$~d. Under this scenario, $P_2$ is interpreted as PSHs from an eccentric precessing disc that does not fully decay during quiescence, analogous to Z~Cha \citep{2026arXiv260910966S}.

A total of 133 PSH maxima ($E=0$--$144$) were extracted from the S75 superoutburst, yielding the ephemeris:
\begin{equation}
	T_{\rm MAX}=3345.74787 + E\times 0.079090.
\end{equation}
Similar to V0419\,Lyr, the $O$--$C$ diagram exhibits linear rather than parabolic evolution in stage~B. The data are well reproduced by a three-segment piecewise-linear function ($r=0.91$, $p<5.07\times10^{-67}$) (Fig. \ref{fig:19-24-R}(f)F), with mean periods of 0.080590~d, 0.079160~d, and 0.078812~d for the three successive stages.

\begin{figure}
	\centering
	\begin{subfigure}{0.45\columnwidth}
		\includegraphics[width=\linewidth]{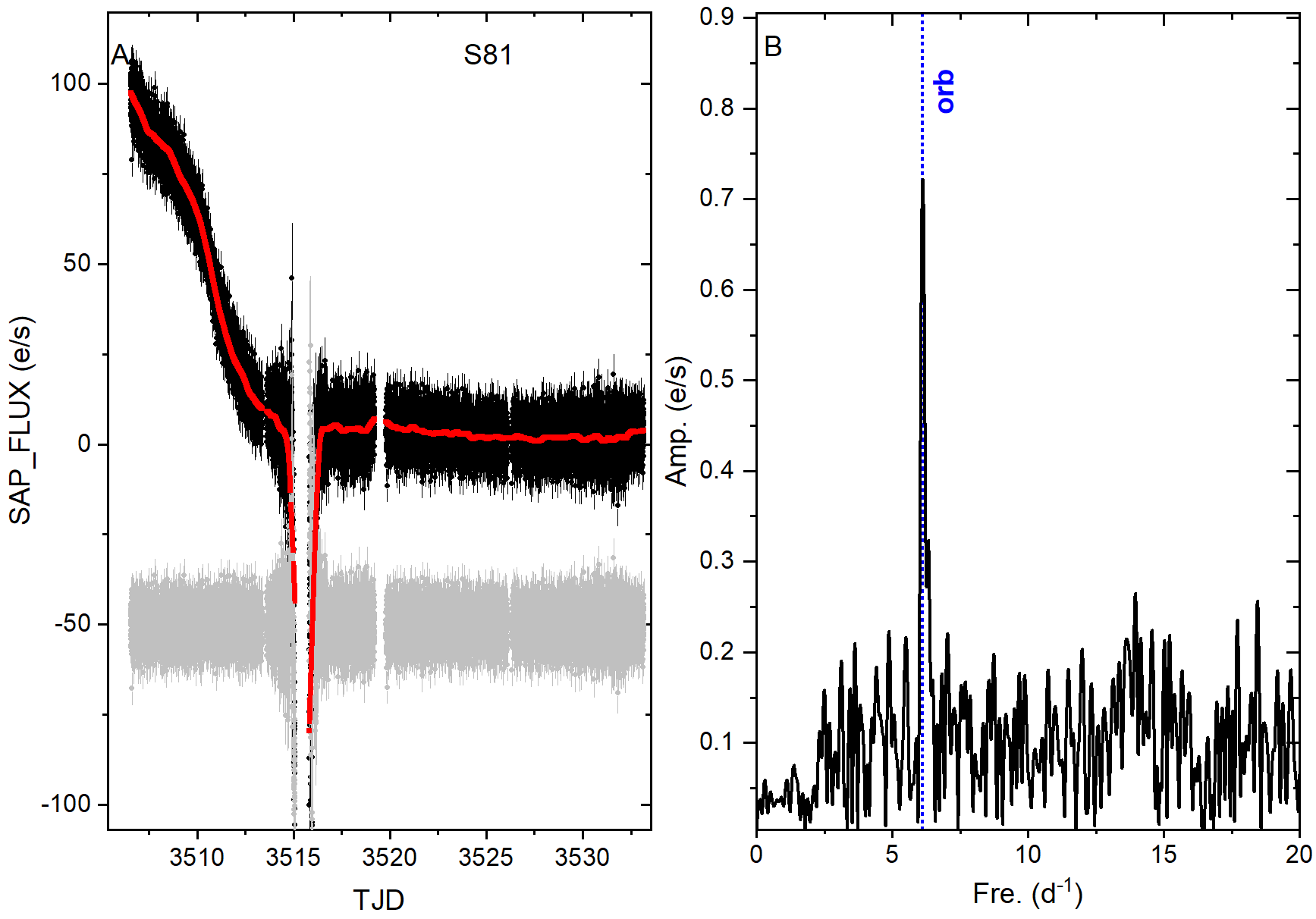}
		\caption{Gaia17cuh}
	\end{subfigure}	
	\begin{subfigure}{0.45\columnwidth}
		\includegraphics[width=\linewidth]{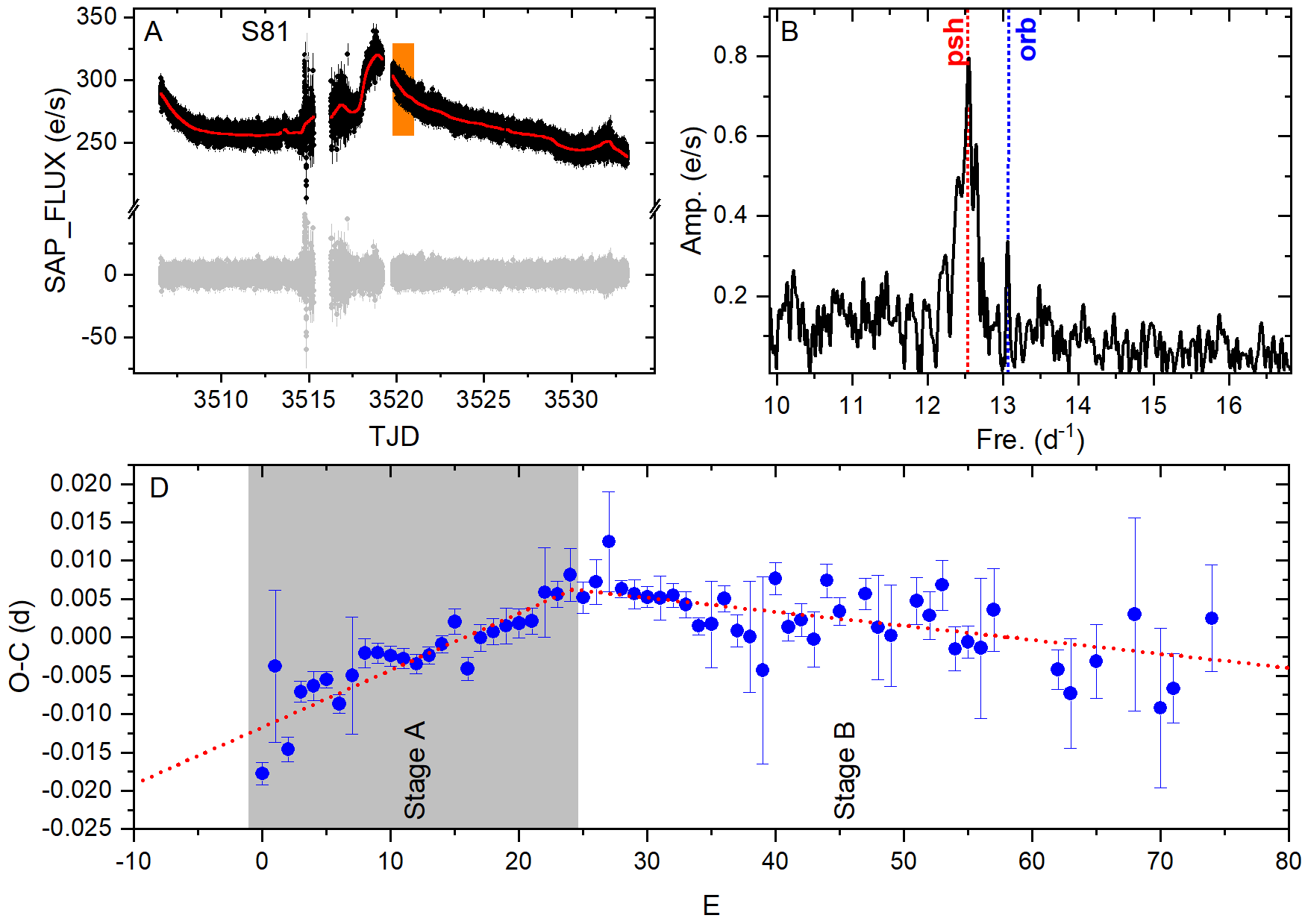}
		\caption{ASASSN-15qr}
	\end{subfigure}

	\begin{subfigure}{0.45\columnwidth}
		\includegraphics[width=\linewidth]{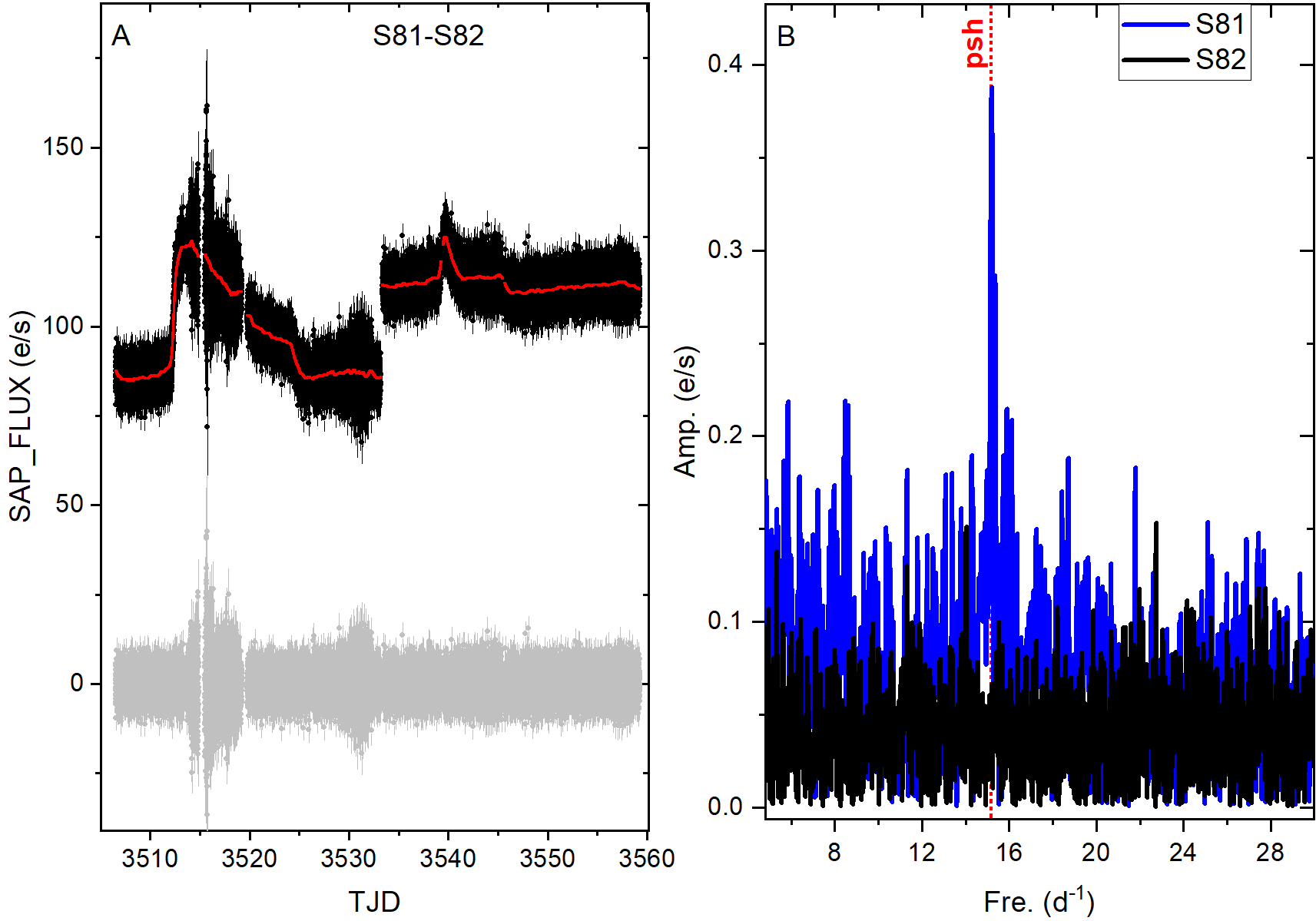}
		\caption{ASASSN-18xz}
	\end{subfigure}
	\begin{subfigure}{0.45\columnwidth}
		\includegraphics[width=\linewidth]{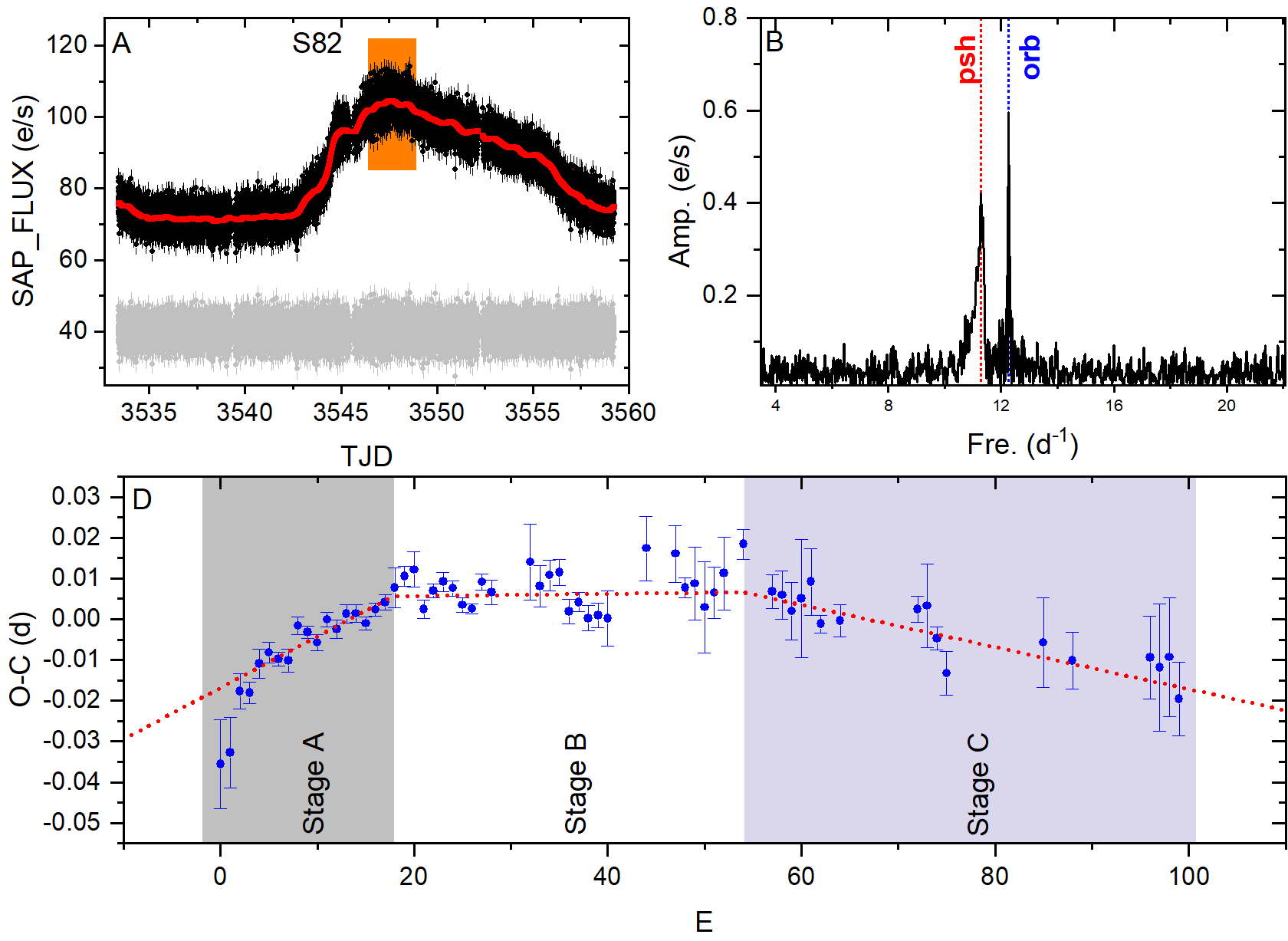}
		\caption{V0774 Peg}
	\end{subfigure}

	\begin{subfigure}{0.45\columnwidth}
		\includegraphics[width=\linewidth]{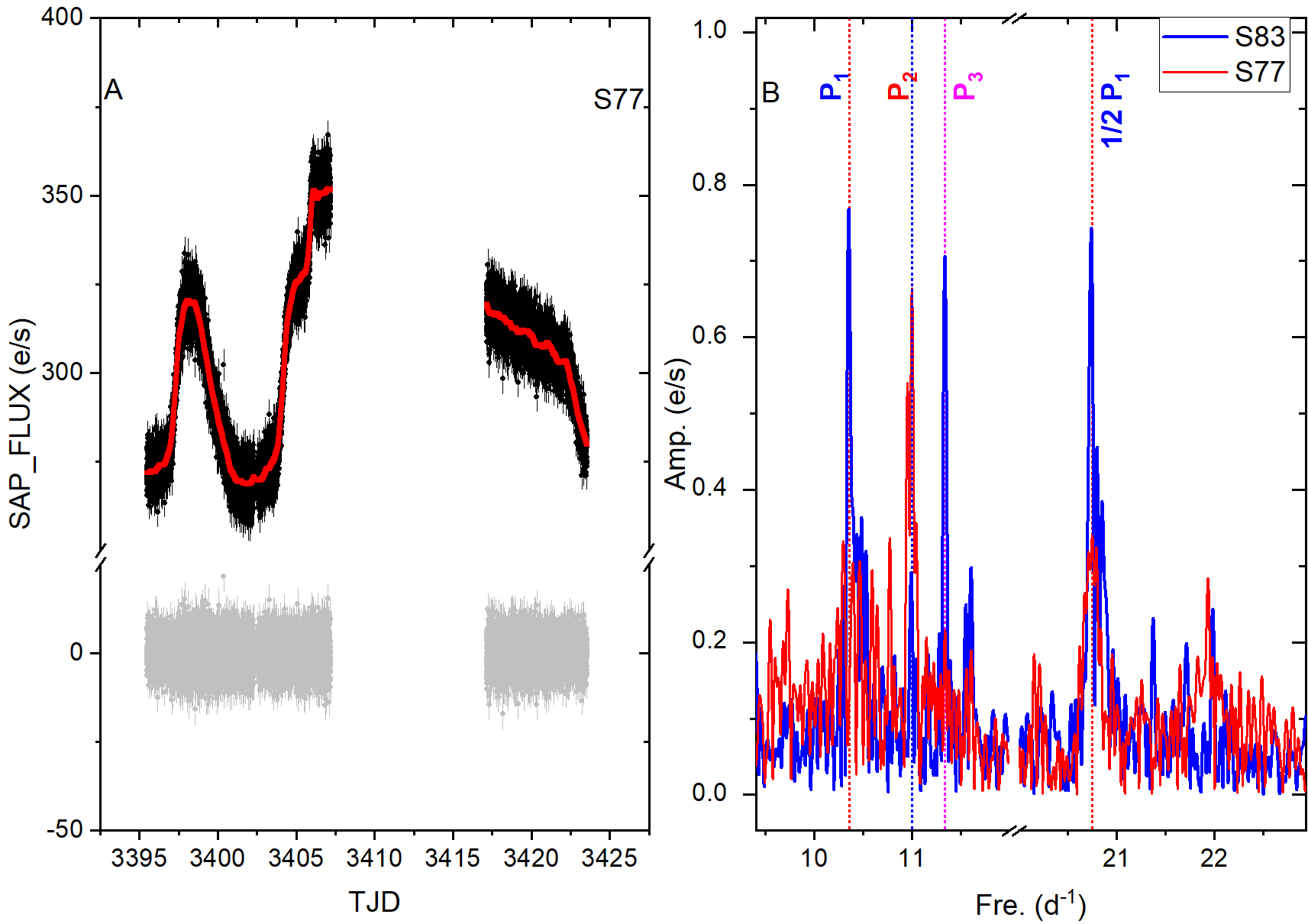}
		\caption{MGAB-V233}
	\end{subfigure}
	\begin{subfigure}{0.45\columnwidth}
		\includegraphics[width=\linewidth]{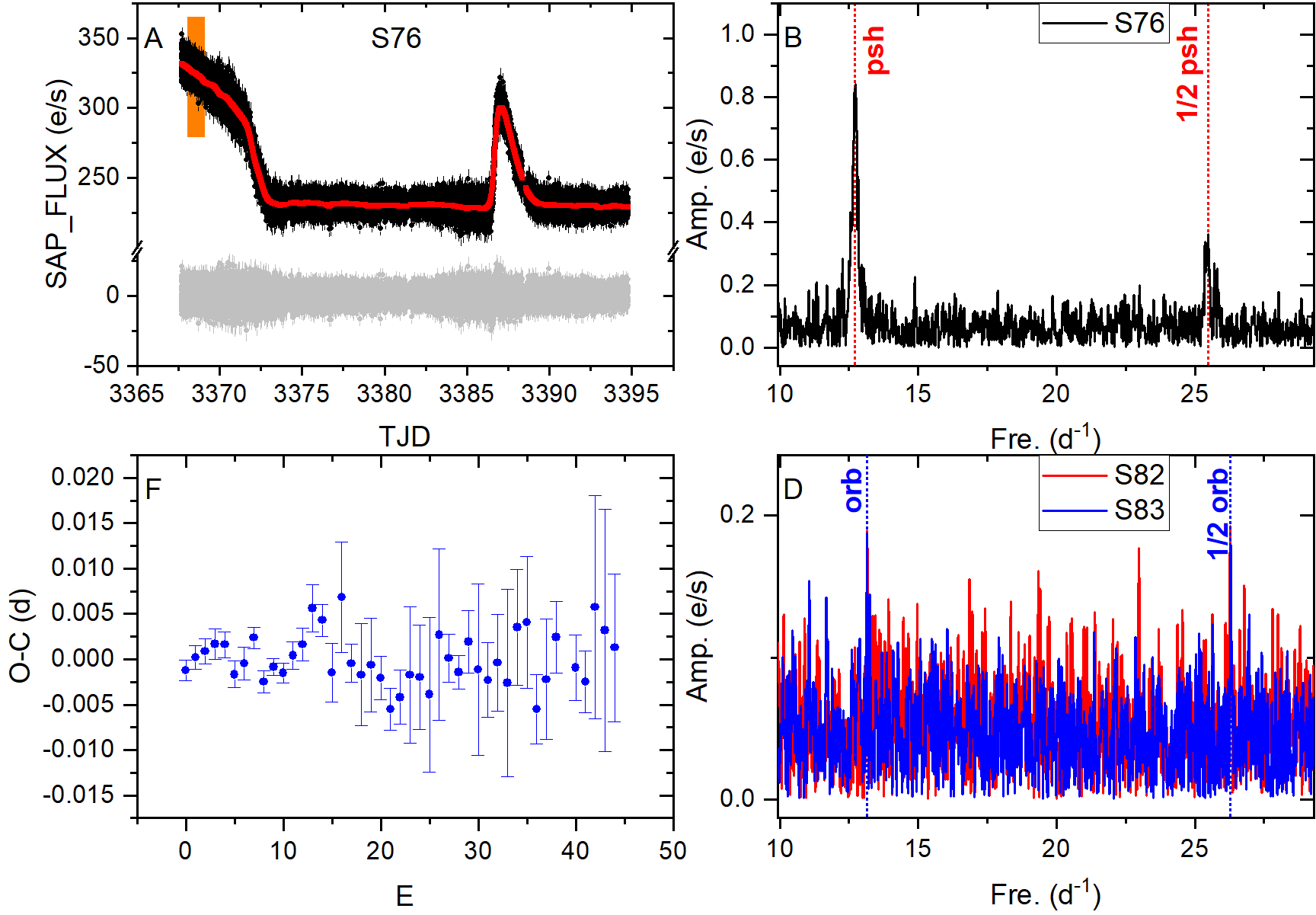}
		\caption{V0630 Cyg}
	\end{subfigure}
\caption{Panels (a)--(f) and their subpanels follow the same numbering convention as Figure~\ref{fig:25-30}. This figure only displays TESS superoutburst light curves, periodograms, and $O$--$C$ diagrams; no $O$--$C$ plots are provided for panels (a), (c), and (e).}
	\label{fig:25-30-R}
\end{figure}

\subsection{Gaia17cuh}

Gaia17cuh was identified as a bright blue transient originating from a faint Gaia source near the Galactic plane. The source brightened by more than 3~mag, reaching $G=15.48$ on 2017~November~2 \citep{2017ATel11006....1R}, and is catalogued as an SU~UMa-type dwarf nova in the VSX. Follow-up BVRI photometry obtained with the 2~m telescope at Terskol six days after the initial outburst revealed a fading source ($V=19.08$, $R=18.49$), consistent with the post-outburst decline of a cataclysmic variable \citep{2017ATel11006....1R}. A subsequent Gaia brightening in 2018~May led to the prediction and confirmation of a second outburst ($\Delta R\gtrsim3$~mag) starting on 2018~November~11. Tentative hourly variability with an amplitude of $\sim$0.1~mag was detected in the $R$-band during this event \citep{2018ATel12238....1S}.

In the TESS data, Gaia17cuh was observed exclusively in Sector~81, where only a half-covered outburst ($>$9~d) was recorded (Fig. \ref{fig:25-30-R}(a)A). This outburst is of uncertain nature and could correspond to either a superoutburst or a long normal outburst. A periodic signal at 0.16331(25)~d is extracted from the power spectrum and verified by the folded light curve (Fig. \ref{fig:25-30}(a)B). At 3.92~hr, this period substantially exceeds the typical range of both orbital periods and PSHs for known SU~UMa-type systems in the standard period-gap framework. This discrepancy strongly suggests that Gaia17cuh may not be a genuine SU~UMa-type dwarf nova; we conclude that 0.16331(25)~d most likely represents the true orbital period.

\subsection{ASASSN-15qr}
ASASSN-15qr is a newly identified SU~UMa-type dwarf nova.
Based on unfiltered CCD photometry obtained with the CBA Extremadura telescope,
Vanmunster (2020) reported clear superhump modulations in this system,
yielding preliminary parameters of $P_{\rm sh}=0.075$~d ($\simeq1.80$~hr)
and an amplitude of 0.28~mag.
The discovery was announced on the \texttt{vsnet-alert} mailing list (vsnet-alert~24642) on 2020~September~4.

TESS observed ASASSN-15qr in Sector~81, revealing one superoutburst lasting $\sim$19.55~d (Fig. \ref{fig:25-30}(b)A). Prominent PSH signals were present during the outburst. Fourier analysis yielded a PSH period of 0.079681(10)~d. Furthermore, we detect a weak candidate orbital periodic signal at $0.0765329(21)$\,d (Fig. \ref{fig:25-30-R}(b)B). Based on this value, the PSHs excess is calculated to be 0.041, consistent with expectations, which suggests that this period is likely the orbital period. A total of 63 PSH maxima ($E=0$--$74$) were extracted from the PSH light-curve modulations, giving the ephemeris:
\begin{equation}
	T_{\rm MAX}=3519.84201 + E\times 0.079543.
\end{equation}
The $O$--$C$ diagram shows only two-stage evolution, well fitted by a two-segment piecewise-linear function (Fig. \ref{fig:25-30-R}(b)D), with mean periods of 0.080286~d and 0.079360~d.

\subsection{ASASSN-18xz}

ASASSN-18xz was reported to host a candidate superoutburst from ZTF photometry (vsnet-alert~26147), yielding a supercycle length of about 170~d. A ZTF $zr$-band magnitude of 16.339 was recorded at 2021~August~4.2619 (UT) \citep{2019ZTF}.
TESS observed ASASSN-18xz, revealing one superoutburst in S81 ($\sim$12.56~d) and one normal outburst in S82 ($\sim$1.47~d) (Fig. \ref{fig:25-30-R}(c)A). The PSH period during the superoutburst was estimated to be 0.065755(10)~d (Fig. \ref{fig:25-30-R}(c)B). No orbital-period signal was detected in either sector. Owing to the low amplitude of the PSH modulations, PSH maxima could not be reliably measured, and therefore no $O$--$C$ analysis was performed for this object.

\subsection{V0774 Peg}

V0774\,Peg ($=$\,CSS\,121005:212625+201948) was identified as a candidate ER~UMa dwarf nova from its unusually frequent outbursts in the Catalina Sky Survey \citep{2015arXiv150507709S}. Intensive monitoring confirmed its SU~UMa nature, detecting superhumps with $P_{\rm sh}=0.08838(18)$~d during a 14-day superoutburst in 2014~November \citep{2015arXiv150507709S}. Its supercycle of 66.9(6)~d is among the shortest for a typical SU~UMa star, yet it falls short of the ER~UMa regime: $P_{\rm sh}>0.07$~d and only $\sim$20\% of its time is spent in superoutburst, compared with 30--45\% for true ER~UMa systems \citep{2015arXiv150507709S}. A 2021 ZTF-detected superoutburst suggested a lengthened supercycle of $\sim$80~d (vsnet-alert~26248), indicating possible long-term variability in the outburst frequency.

TESS monitored V0774\,Peg during a superoutburst lasting 14.53~d in Sector~81 (Fig. \ref{fig:25-30}(d)A). Prominent superhump signatures are present at the outburst peak. Fourier analysis yields a PSH period of 0.088660(10)~d. Another significant signal at 0.081566(6)~d is also detected, possibly representing the orbital period, with a corresponding superhump excess of 0.087. A total of 63 PSH maxima ($E=0$--$99$) were extracted, giving the ephemeris:
\begin{equation}
	T_{\rm MAX}=3546.12112 + E\times 0.088327.
\end{equation}
As seen in V0419\,Lyr and V1113\,Cyg, the $O$--$C$ diagram shows linear rather than parabolic evolution in stage~B (Fig. \ref{fig:25-30-R}(d)D). A three-segment piecewise-linear function provides a good fit ($r=0.74$, $p<1.00\times10^{-16}$), with mean periods of 0.089587~d, 0.088354~d, and 0.087810~d.

\subsection{MGAB-V233}

MGAB-V233 was classified as an ER~UMa-type dwarf nova through Zwicky Transient Facility survey observations, exhibiting a short supercycle of $\sim$70~d with 7--8 normal outbursts per cycle reaching $g=15.7$--$16.0$~mag \citep{2019ZTF}. Its broad magnitude range ($g=15.0$--$18.6$~mag) and near-infrared colours ($J-K=0.34$, $G_{\rm BP}-G=0.29$) are consistent with a high-inclination compact binary \citep{2019ZTF}.

TESS observed MGAB-V233 in Sectors~77 and~83. Sector~77 hosts one short outburst ($\sim$4.10~d) and a superoutburst lasting 19.86~d (Fig. \ref{fig:25-30}(e)A); no outburst activity is detected in Sector~83. Fourier analyses reveal four significant periodic signals: $P_1=0.096562(9)$~d (S83), $P_2=0.090909(11)$~d (S77), $P_3=0.088183(8)$~d (S83), and the second harmonic of $P_1$ (detected in both S77 and~S83) (Fig. \ref{fig:25-30-R}(e)B). All three independent signals are confirmed by the folded light curves (Figs. \ref{fig:25-30}(e)D-F).

The physical interpretation of these signals remains ambiguous. The 19.86-day outburst in S77 is most likely the superoutburst, making $P_2$ a candidate PSH; however, $P_1$ is 6.2\% longer than $P_2$, so $P_1$ could alternatively represent PSHs while $P_3$ (3.0\% shorter than $P_1$) might be the orbital period. Under yet another interpretation, $P_2$ could be the orbital period, with $P_1$ representing PSHs and $P_3$ NSHs. 
The PSHs are not significant during S77; only their second harmonic is detected, which may be caused by substantial data gaps during the superoutburst.
 Additional data from other TESS sectors or ground-based facilities are required to resolve their physical origins.

\subsection{V0630 Cyg}

V630\,Cyg was identified as an SU~UMa-type dwarf nova candidate through its distinctive outburst pattern \citep{1989IBVS.3405....1W}, and its nature was confirmed by the detection of superhumps with $P_{\rm sh}=0.0789(4)$~d during a 1996 superoutburst \citep{2001IBVS.5157....1N}. Subsequent surveys of superhump period variations \citep{2009PASJ...61S.395K, 2010PASJ...62.1525K, 2016PASJ...68...65K} characterised its three-stage superhump evolution and positive period derivative during stage~B. The refined orbital period is $P_{\rm orb}=0.07606(5)$~d \citep{2026RNAAS..10...67B}.

TESS observed V630\,Cyg in Sectors~76 and~82--83, detecting a half-covered superoutburst and three short outbursts with a typical duration of $\sim$3.27~d (Fig. \ref{fig:25-30}(f)A). The interval between two consecutive normal outbursts is 30.44~d. Fourier analysis yields a PSH period of 0.078709(7)~d in S76 (Fig. \ref{fig:25-30-R}(f)B), consistent with previous measurements. An orbital-period signal at 0.076025(22)~d (the mean from both sectors) together with its second harmonic is recovered in S82 and~S83 (Fig. \ref{fig:25-30-R}(f)D), in good agreement with \citet{2026RNAAS..10...67B}. Despite the incomplete superoutburst, PSH signatures remain clearly detectable. A total of 44 PSH maxima ($E=0$--$44$) were extracted, yielding the ephemeris:
\begin{equation}
	T_{\rm MAX}=3367.719048 + E\times 0.078756.
\end{equation}
Owing to the incomplete superoutburst coverage, no distinct staged evolutionary features are seen in the $O$--$C$ diagram.

\section{Discussions} \label{sec:Discussions}

\begin{figure}
	\centering
	\includegraphics[width=0.45\columnwidth]{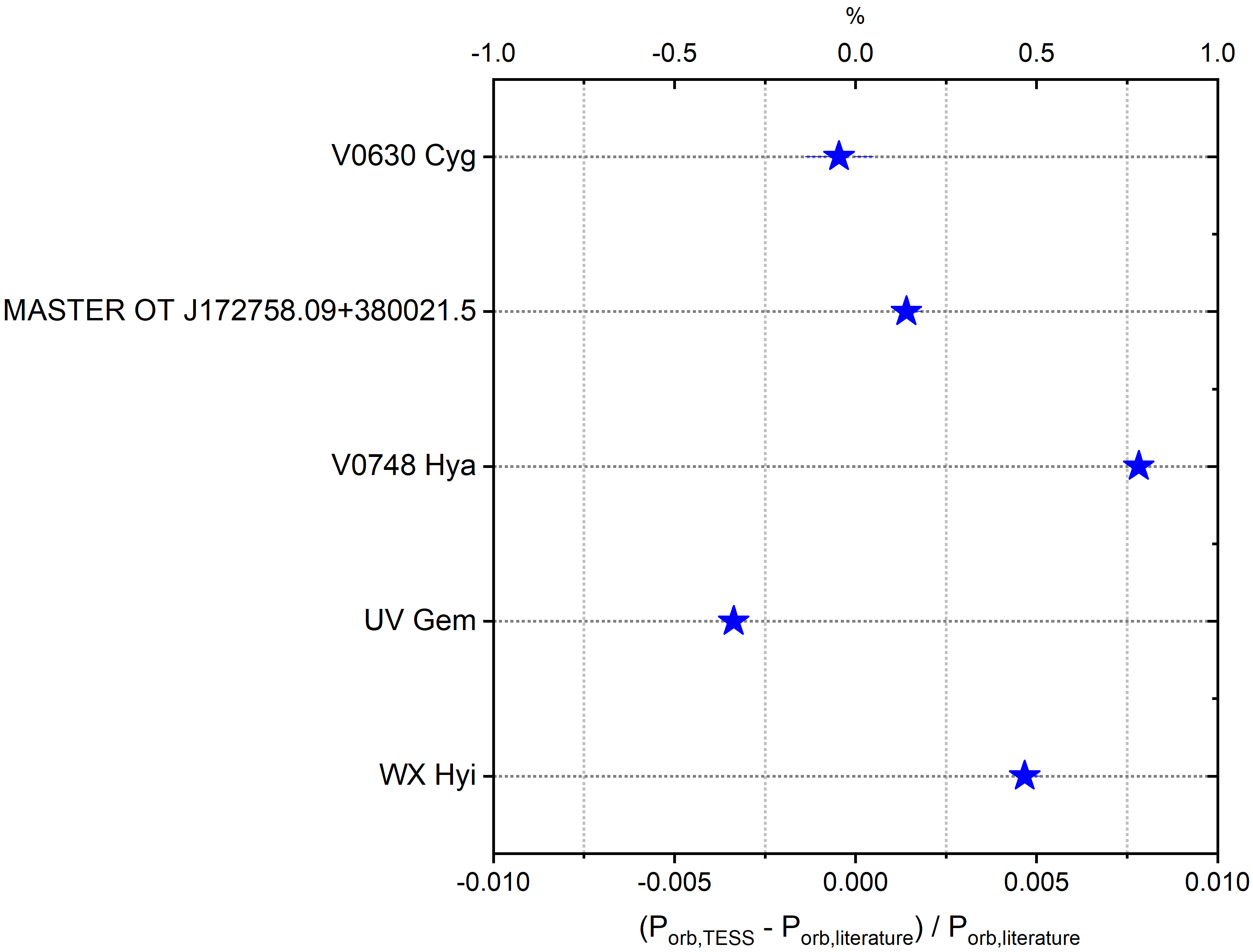}
	\includegraphics[width=0.45\columnwidth]{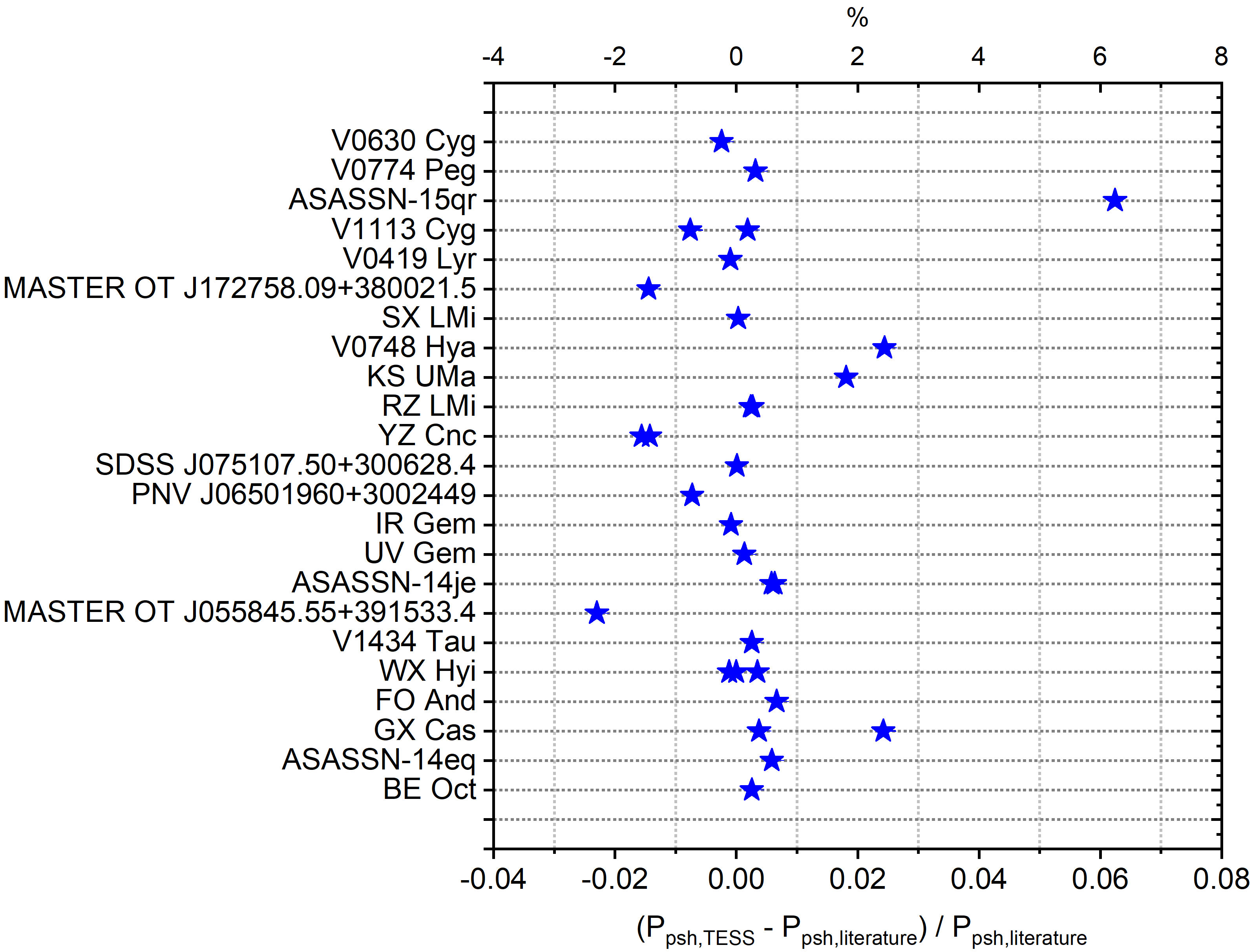}
	\caption{The left panel shows the fractional differences between the TESS orbital periods and literature orbital periods for the five targets. The right panel presents the fractional differences between the TESS PSH periods and literature PSH periods for 23 targets.}
	\label{fig:VS}
\end{figure}

We derived orbital periods for 13 targets from TESS data. Only five of these objects have published literature orbital periods; the absolute differences between our measurements and the literature values range from 0.000035 to 0.00048 (all deviations are less than 1\%, see Fig.~\ref{fig:VS}), indicating excellent agreement between our results and prior values. The main discrepancies arise from the periods of PSHs, as PSHs measured at different superoutburst stages differ. For TESS PSH measurements compared with literature values, 22 detections show deviations within 1\%, and 7 detections fall in the 1\%--3\% range (see Fig.~\ref{fig:VS}). Notably, GX~Cas exhibits two measurements: one below 0.5\% and another above 2\%. The light curve reveals that the 2\% deviation arises because TESS only captured the early outburst phase (S85), which has longer PSH periods (see Figs.~\ref{fig:1-6}(c)A and B). This further implies that PSHs measured at different evolutionary stages show substantial differences. We find that the PSH period of ASASSN-15qr is 6\% larger than the literature value (see Fig.~\ref{fig:VS}). The published PSH period originates from a communication by Vanmunster (2020, vsnet-alert~24642), who reported clear superhump modulations and preliminary parameters of $P_{\rm sh}=0.075$~d without releasing the periodogram, suggesting a large uncertainty for the literature value. We argue that our TESS PSH period of $0.079681(10)$~d for ASASSN-15qr is more reliable.

\subsection{Mass Ratio Determination from Stage~A Superhumps}
\label{sec:mass_ratio}

The mass ratio $q=M_2/M_1$ is a fundamental parameter governing the tidal structure and evolutionary state of CVs, yet it has historically been difficult to measure: traditional determinations require either quiescent eclipse photometry or spectroscopy of the faint secondary, both inaccessible for short-period and WZ~Sge-type systems. The Stage~A superhump method \citep{2013PASJ...65..115K,2022arXiv220102945K} circumvents these limitations by deriving $q$ purely dynamically --- at the onset of a superoutburst, the eccentric wave is confined to the 3:1 resonance radius where pressure effects are negligible, so the observed superhump period directly yields $q$ with accuracy comparable to eclipse modeling, without requiring eclipses or empirical calibration. Superhumps in SU~UMa-type dwarf novae arise from the apsidal precession of an eccentric accretion disk at the 3:1 tidal resonance \citep{whitehurst1988numerical, 1990PASJ...42..135H, lubow1991model}. The normalized precession rate is
\begin{equation}
	\epsilon^* \equiv \frac{\omega_{\rm pr}}{\omega_{\rm orb}} = 1 - \frac{P_{\rm orb}}{P_{\rm SH}} = \frac{\epsilon}{1+\epsilon},
	\label{eq:epsilon_star}
\end{equation}
where $\epsilon \equiv P_{\rm SH}/P_{\rm orb} - 1$ and $\omega_{\rm orb} = 2\pi/P_{\rm orb}$.

In general, the disk precession rate receives contributions from the dynamical tidal torque, the retrograde pressure effect, and a minor stress term \citep{1992ApJ...401..317L, 2006MNRAS.371..235P}. However, during Stage~A --- the initial growing phase of superhumps --- the eccentric wave is confined to the vicinity of the 3:1 resonance radius, and the pressure effect is negligible \citep{2013PASJ...65..115K}. The observed $\epsilon^*_{\rm A}$ therefore equals the dynamical precession rate evaluated at $r_{3:1}$ \citep{2013PASJ...65..115K,2022arXiv220102945K}:
\begin{equation}
	\epsilon^*_{\rm A} \simeq \left.\frac{\omega_{\rm dyn}}{\omega_{\rm orb}}\right|_{r\,=\,r_{3:1}} = \frac{q}{\sqrt{1+q}}\left[\frac{1}{4}\sqrt{r_{3:1}}\;b_{3/2}^{(1)}(r_{3:1})\right],
	\label{eq:stageA_q}
\end{equation}
where $r_{3:1} = 3^{-2/3}(1+q)^{-1/3}$ is the 3:1 resonance radius \citep{1990PASJ...42..135H} and $b_{3/2}^{(1)}$ is the Laplace coefficient. Because $r_{3:1}$ is itself a function of $q$, the right-hand side of Equation~(\ref{eq:stageA_q}) depends on $q$ alone, enabling a direct determination of $q$ from the observed $\epsilon^*_{\rm A}$ without any free parameter or empirical coefficient. The reliability of this method has been extensively validated against quiescent eclipse modeling across ${>}100$ objects spanning $0.04 \lesssim q \lesssim 0.35$ \citep{2013PASJ...65..115K,2022arXiv220102945K}, with both methods yielding mutually consistent $P_{\rm orb}$--$q$ relations that follow the CV evolutionary tracks of \citet{2011ApJS..194...28K}. 

\begin{figure}
	\centering
	\includegraphics[width=0.45\columnwidth]{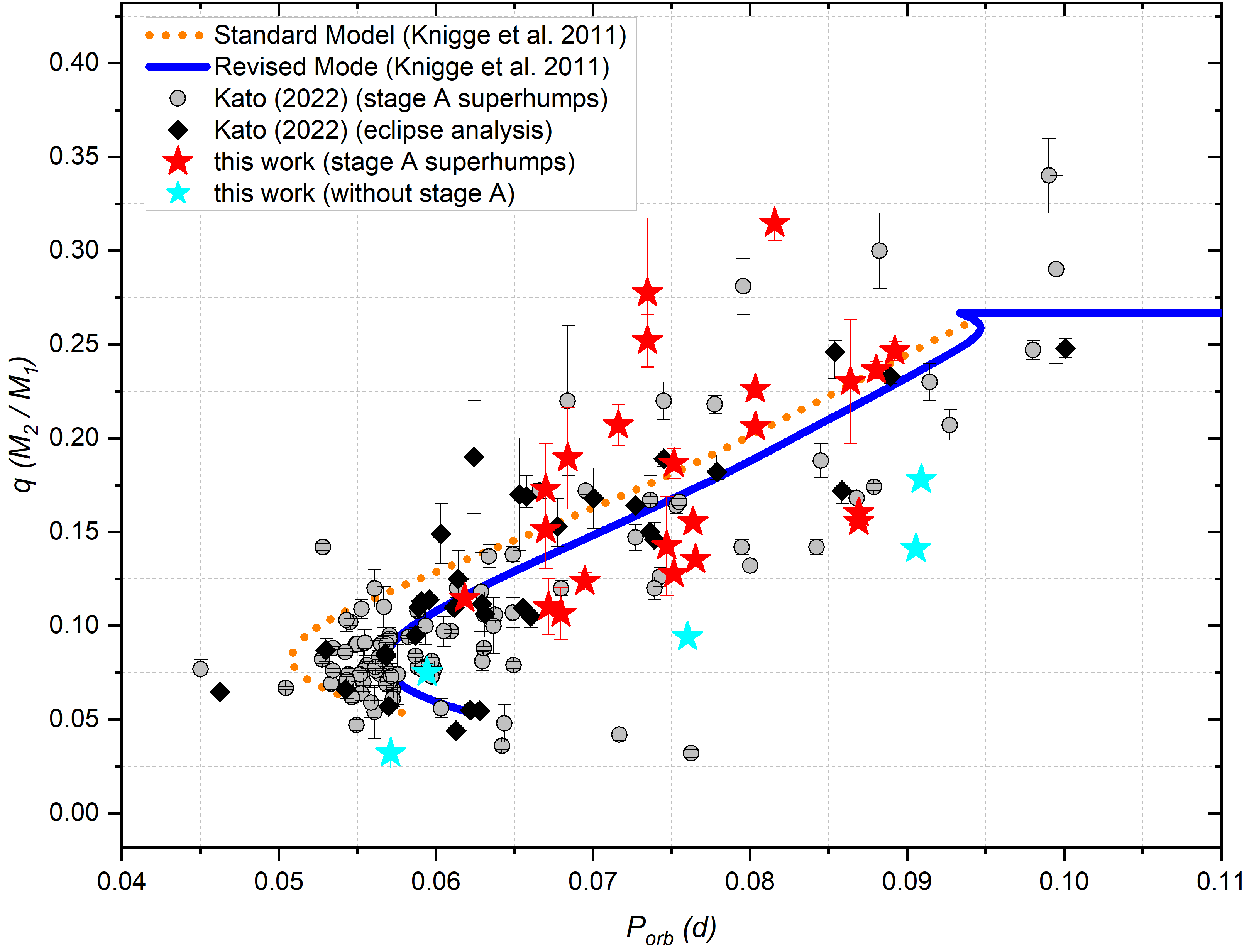}
	\includegraphics[width=0.45\columnwidth]{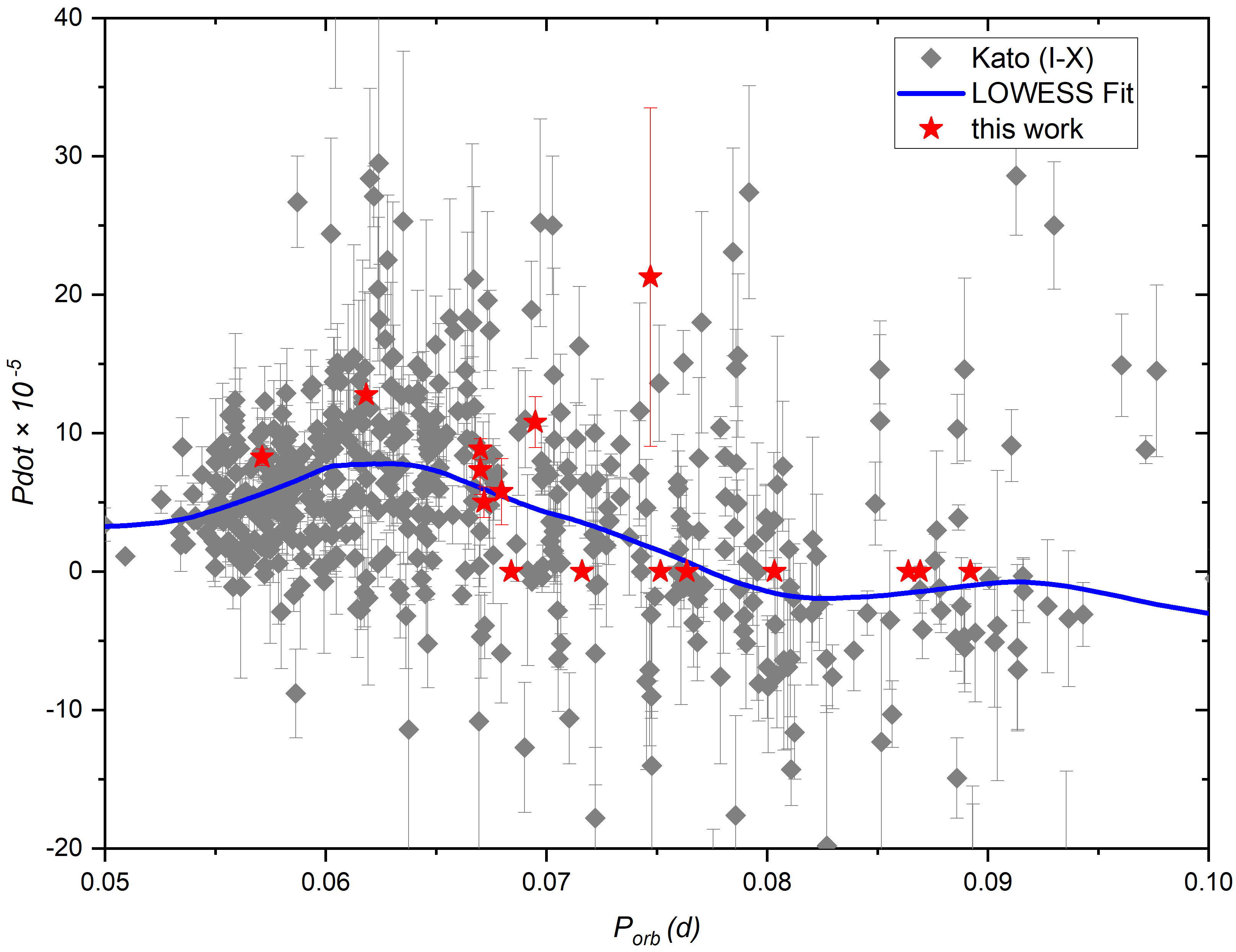}
	\caption{(left) Mass ratio ($q$) versus orbital period ($P_{\rm orb}$). Gray circles and black diamonds denote $q$ values derived by \cite{2022arXiv220102945K} using Stage~A PSHs and eclipse measurements, respectively. The red dashed and blue solid lines represent the standard and best-fit evolutionary tracks from \citet{2011ApJS..194...28K}. The red and bright-yellow stars mark mass ratios determined in this work using the Stage~A PSH method and PSH periods from Fourier transforms (where no Stage~A is present). (right) Period versus the Stage~B period derivative. Gray diamonds are taken from the Kato sequence, red stars show our new measurements, and the blue solid line is the LOWESS smoothed fit.}
	\label{fig:q}
\end{figure}

Our full sample comprises 30 objects. Superhump signals were detected in 29 of them, orbital periods are available for 24, and 23 display clear Stage~A evolution in the $O-C$ diagrams of superhump maxima during their superoutbursts. For the 18 objects that possess both a known orbital period and a well-measured Stage~A superhump period, we computed $\epsilon^*_{\rm A}$ via Equation~(\ref{eq:epsilon_star}) and derived $q$ by numerically inverting Equation~(\ref{eq:stageA_q}). The resulting $q$ are shown in Figure~\ref{fig:q} (red stars in Figure~\ref{fig:q}): they closely follow the distribution established by the Kato sequence and lie along the standard CV evolutionary tracks, independently confirming the reliability of both our $O-C$ timing analysis and our Fourier frequency analysis. Six of these objects have previously published $q$ (GX~Cas, FO~And, IR~Gem, YZ~Cnc, KS~UMa, SX~LMi; see Table~\ref{tab:qcomp}). Earlier mass-ratio estimates were obtained in \citet{2005PASP..117.1204P} via the fitting relation $\varepsilon = 0.18 q + 0.29 q^2$, and in \citet{2022arXiv220102945K} using the Stage~A superhump method. \citet{2022arXiv220102945K} compared their technique against the formula from \citet{2005PASP..117.1204P} and argued that the Stage~A superhump method is more physically justified. Our $q$ generally show substantial differences from the values of \citet{2005PASP..117.1204P} (see Table~\ref{tab:qcomp}), likely arising from differences in calculation methodology. Our results are broadly consistent with those of \citet{2022arXiv220102945K}. We adopt the same Stage~A superhump method as \citet{2022arXiv220102945K}; the only difference lies in the measured Stage~A PSH periods. We suggest that the high-cadence TESS photometry yields $q$ estimates with improved reliability.

Stage~A superhumps were not recorded in five objects (ASASSN-14kj, RZ~LMi, MASTER~OT~J172758.09+380021.5, MGAB-V233, and V630~Cyg), so that $\epsilon^*_{\rm A}$ could not be determined. For these systems we instead adopted the mean superhump period from the Fourier analysis to calculate $\epsilon^*$ and hence $q$; the resulting values lie systematically and significantly below the theoretical expectations (yellow stars in Figure~\ref{fig:q}), consistent with the known retrograde pressure contribution that biases Stage~B-based estimates toward low $q$.

\begin{deluxetable}{lccc}
	\tablecaption{Mass ratios derived from Stage~A PSH periods for the six targets with previous determinations.\label{tab:qcomp}}
	\tablewidth{1pt}
	\tablehead{	
		\colhead{Object} & \colhead{$q$ (this work)} & \colhead{$q$ \citep{2005PASP..117.1204P}} & \colhead{$q$ \citep{2022arXiv220102945K}}}
	\startdata
	GX~Cas   & $0.246(5)$                & $0.191(11)$ & --- \\
	FO~And   & $0.207(11)$               & $0.151(16)$ & --- \\
	IR~Gem   & $0.189(27)$               & $0.152(26)$ & $0.22(4)$ \\
	YZ~Cnc   & $0.160(3),\,0.156(3)$     & $0.236(12)$ & $0.168(5)$ \\
	KS~UMa   & $0.107(14)$               & $0.110(11)$ & $0.120(4)$ \\
	SX~LMi   & $0.110(15)$               & $0.150(10)$ & --- \\
	\enddata 
\end{deluxetable}

\subsection{Thermal–Tidal Decoupling of Superhump Activity}

The classical TTI model describes synchronized thermal and tidal cycles in dwarf novae: superoutburst-driven disk expansion crosses the 3:1 Lindblad resonance, generating an eccentric, progradely precessing disk, and subsequent disk contraction causes the hot accretion state and eccentric disk to decay in tandem \citep{1989PASJ...41.1005O,1996PASP..108...39O}. In this framework, PSHs arise exclusively from tidally excited eccentric disks during superoutbursts \citep{1988MNRAS.232...35W,2000NewAR..44...51M,2005PJABOsaki}. Observationally, PSHs appear 1–2 days after superoutburst maximum, a delay attributed to the competition between resonant eccentricity growth and post-outburst disk contraction \citep{1996ASSL..208...45S,2000NewAR..44..171S,2005PJABOsaki}. For low-mass-ratio systems, the 3:1 resonance lies well inside the tidal truncation radius, allowing eccentric disk structures to persist into quiescence and decouple tidal eccentricity evolution from thermal outburst cycles, violating the strict phase locking of standard TTI theory.

Our TESS observations confirm this thermal–tidal decoupling in three SU\,UMa-type dwarf novae, where clear PSH signals are detected in sectors dominated by normal outbursts, with no concurrent superoutburst. In WX\,Hyi, five consecutive normal outbursts following a superoutburst still host detectable PSHs at $P_{\rm SH}=0.077519(13)$\,d (see Fig. \ref{fig:1-6}(e)D), 0.47\% shorter than the superoutburst-period value. EC\,05200$-$5205 exhibits similar behavior in Sector\,87: two normal outbursts produce PSHs at $0.082864(15)$\,d (see Fig. \ref{fig:1-6}(f)D), 0.91\% shorter than the superoutburst counterpart. V1113\,Cyg also shows persistent PSHs during a single normal outburst in Sector\,82, with $P_{\rm SH}=0.078309(9)$\,d (0.95\% lower than its superoutburst period; see Fig. \ref{fig:19-24}(e)D).

The 0.5–1\% PSH period decrease in these systems matches the canonical Stage\,B–Stage\,C period drop in the Kato sequence. We interpret these normal-outburst PSHs as residual eccentric disk signatures from prior superoutbursts, rather than newly excited resonant features. After superoutburst termination, the disk cools, and tidal/viscous torques deplete outer-disk angular momentum. The eccentric region retreats inward, lowering the precession rate and shortening the PSH period \citep{2019MNRAS.488.4149C}. While normal outbursts temporarily inflate and replenish the disk, their weak, short-lived expansion cannot re-establish 3:1 resonant eccentricity excitation. Eccentricity damping thus dominates over regeneration, driving continuous reductions in PSH period and amplitude until the eccentric disk fully dissipates.

Persistent PSHs across multiple normal outbursts without superoutburst activity definitively prove decoupled thermal accretion and tidal eccentricity cycles. This behavior is consistent with recent findings for low-$q$ cataclysmic variables. The low-mass-ratio AM\,CVn system V803\,Cen hosts persistent PSHs across diverse disk states, with period evolution driven by decaying eccentric resonance structures \citep{2026ApJV803}, consistent with early detections of state-independent persistent superhumps \citep{2000PASP..112..625P}. Low-$q$ systems possess weaker tidal torques \citep{Hellier2001Echo}, resulting in slower eccentricity damping that allows eccentric disk modes to persist long after the parent superoutburst ends.
 TESS high-precision photometry continuously tracks the gradual decay of eccentric precession, providing robust quantitative constraints on eccentricity dissipation following the breakdown of thermal–tidal synchronization.

\subsection{Variety and repeatability of stage evolution}
\label{sec:stage_diversity}

Five systems (ASASSN-14je, WX~Hyi, EC~05200$-$5205, ASASSN-18hr, YZ~Cnc) were observed during two or more superoutbursts (Table~\ref{tab:repeat}). Their comparison reveals a tiered pattern rather than strict repeatability. The Stage~B and Stage~C properties repeat well. ASASSN-14je shows the closest match (Fig. \ref{fig:7-12}(c)D): its two events agree in the Stage~B derivative at $\sim1\sigma$ ($(8.84\pm0.76)$ vs. $(7.35\pm1.80)\times10^{-5}$~d~d$^{-1}$) and in the Stage~C period to $0.05\%$ (Table~\ref{tab:repeat}). The two fully covered superoutbursts of EC~05200$-$5205 (Sectors~67, 97) reproduce the same A1--A2--B--C structure within $0.1\%$ (Fig. \ref{fig:1-6}(f)E); its third sector (94) looks different only because the early stages fell outside the sector, its first segment coinciding with the Stage~B period of the complete events. ASASSN-18hr likewise repeats in its resolved stages ($0.1\%$ in the second segment), although early coverage is incomplete in both of its events (Fig. \ref{fig:19-24}(d)D). YZ~Cnc matches in Stage~A and Stage~B ($\lesssim0.15\%$) between its two complete superoutbursts, but Stage~C appears only in Sector~72.

WX~Hyi shows the clearest event-to-event differences. Both superoutbursts have three segments with linear Stage~B evolution; their Stage~B and Stage~C periods agree within $0.08\%$ and $0.55\%$, respectively. However, the mean Stage~A period differs by $1.7\%$: $0.078665$~d in Sector~3 (no precursor) versus $0.080026$~d in Sector~28, which features a prominent precursor normal outburst before the main superoutburst (Fig. \ref{fig:1-6-R}(e)).
	The WX~Hyi result is not unique. \citet{2005A&A...432..261U} reported stark contrasts in superhump period derivatives between precursor and precursor-free superoutbursts of TV~Crv. They attributed this difference to the amount of gas around and beyond the 3:1 resonance radius, which differs between precursor and no-precursor events in the modified thermal-tidal instability model of \citet{2003A&A...401..325O}. Kepler observations of V1504~Cyg also reveal strong event-to-event variations in Stage~A behaviour across its 12 superoutbursts \citet{Osaki2013a,Osaki2013b}, including long quiescent gaps between precursor and main outburst and transient "failed superhumps" that fail to initiate a superoutburst. Together, WX~Hyi, TV~Crv and V1504~Cyg demonstrate that Stage~A parameters depend on the disk state at outburst onset, and a $\sim$1--2\% event-to-event scatter of Stage~A quantities is intrinsic to SU~UMa systems.

As summarized in Table~\ref{tab:repeat}, among these five targets with repeated superoutbursts, four systems show consistent Stage~A periods across different outburst events; only WX~Hyi exhibits a significant difference in its Stage~A period. Stage~B parameters agree well for all five systems between repeated outbursts. Stage~C is measurable in four of the five targets, and these four systems display good consistency of Stage~C periods across outbursts. Repeated superoutbursts exhibit coherent evolutionary patterns, demonstrating that superhump morphology may represent an intrinsic, repeatable property for some systems.

Two objects depart from the canonical sequence altogether. In RZ~LMi (S48) the first $O$--$C$ segment has a negative slope and a mean period (0.058315~d) shorter than the following segment (0.059429~d) (see Fig.~\ref{fig:13-18}(e)D); with $P_{\rm orb}=0.05792$~d, this yields a period excess $\epsilon\simeq0.69\%$ in the first segment but $\epsilon\simeq2.6\%$ in the second. The conflict with standard precession theory is direct: Stage~A is defined as the phase in which the eccentric disturbance is confined to the 3:1 Lindblad resonance, where the retrograde gas-pressure contribution is negligible and the apsidal precession rate is purely dynamical, so for $q=0.105$ the expected Stage~A excess is $\sim$2--3\% \citep{Kato2013PASJ,2022arXiv220102945K}---exactly the value measured in the second segment, not in the first. The first segment is therefore too short to represent a disk at the 3:1 resonance, yet it precedes the dynamically consistent one, inverting the standard A-before-B ordering. RZ~LMi is an extreme ER~UMa system with an exceptionally high mass-transfer rate ($\dot{M}$ some $\sim$100 times that of WZ~Sge); strong gas pressure inside the disk can add a substantial retrograde contribution to the precession rate \citep{1992ApJ...401..317L,2006MNRAS.371..235P}, reducing the net prograde rate and thereby shrinking $\epsilon^*$ during an early, inner-disk eccentric phase before the 3:1-resonant Stage~A sets in. Second, RZ~LMi has been proposed as a stripped-secondary system evolving toward the AM~CVn sequence \citep{2025MUPB...80S.236V}; a peculiar disk structure or a rapidly changing $\dot{M}$ during the superoutburst rise could transiently alter the resonance structure and produce a non-standard early signal. 

In ASASSN-14kj, the 51 measured maxima show no obvious stage evolution (see Fig.~\ref{fig:13-18}(b)D), but instead carry a coherent 1.00(2)-d modulation---about an order of magnitude longer than both the superhump [$0.095167(54)$~d] and orbital [$0.09056(3)$~d; \citep{2016AJ....152..226T}] periods. From the beat between the orbital and superhump periods, the prograde apsidal precession period of the eccentric disk is $P_{\rm prec}\simeq1.87$~d, which is of the same order as the observed 1.00(2)-d signal but roughly a factor of two longer; the modulation may therefore be related to disk precession, but the frequency offset rules out a straightforward identification. Alternatively, the 51 maxima could be interpreted as three parabolic segments, i.e.\ three successive stage transitions, rather than a single coherent modulation---the present sampling does not decisively distinguish these two possibilities. In either case, the physical origin cannot be established from the current data alone, and an accretion-disk precession origin cannot be ruled out; the discovery of additional objects exhibiting similar long-period modulations, especially in systems with independently known precession periods, would help determine whether the signal traces disk precession or a distinct physical mechanism. It is likewise excluded from the dynamical mass-ratio determination.

The A--B--C template thus describes the majority of our sample but is not universal: Stage~B may be parabolic or linear, Stage~A single- or two-segmented, and the staged evolution can be inverted (RZ~LMi) or absent altogether (ASASSN-14kj), with each system faithfully reproducing its own $O$--$C$ morphology in independent superoutbursts.

\begin{deluxetable}{lcccccc}
	\tablecaption{Event-to-event comparison of stage evolution for systems with multiple TESS superoutbursts.\label{tab:repeat}}
	\tablewidth{1pt}
	\tabletypesize{\scriptsize}
	\tablehead{
		\colhead{Name} & \colhead{Sector} & \colhead{$P_{\rm A}$} & \colhead{$P_{\rm B}$} & \colhead{$P_{\rm C}$} & \colhead{$\dot{P}_{\rm B}$} & \colhead{$P_{\rm SH}$}\\
		\colhead{} & \colhead{} & \colhead{(d)} & \colhead{(d)} & \colhead{(d)} & \colhead{($10^{-5}$~d~d$^{-1}$)} & \colhead{(d)}}
	\startdata
	ASASSN-14je & S89  & 0.07062 & 0.06926 & 0.06920 &  $+8.84\pm0.76$ & 0.069507(2) \\
	& S96  & 0.07107 & 0.06934 & 0.06917 &  $+7.35\pm1.80$ & 0.069473(2) \\
	WX~Hyi      & S3   & 0.07867 & 0.07776 & 0.07733 & 0       & 0.077610(2) \\
	& S28  & \textbf{0.08003} & 0.07782 & 0.07691 & 0       & 0.077882(3) \\
	EC~05200$-$5205 & S67  & 0.08643 (A1+A2) & 0.08358 & 0.08319 & $0$ & 0.083591(3) \\
	& S94* & 0.08356 (incomplete) & --- & --- & --- & 0.082864(15) \\
	& S97  & 0.08599 (A1+A2) & 0.08361 & 0.08319 & $0$ & 0.083626(4) \\
	ASASSN-18hr & S74  & 0.07955 & 0.07833 & --- & $0$ & 0.078450(7) \\
	& S80  & 0.08003 & 0.07841 & --- & $0$ & 0.078462(10) \\
	YZ~Cnc      & S47  & 0.09187 & 0.09052 & --- & $0$ & 0.090728(3) \\
	& S72  & 0.09174 & 0.09049 & 0.09023 & $0$ & 0.090604(4) \\
	\enddata
\begin{threeparttable}
	\small
	\begin{tablenotes}
		\item [*] Sector~94 of EC~05200$-$5205 only observed the late superoutburst; its first-segment mean \\period matches the Stage~B period obtained from complete coverage in Sectors~67 and 97.
	\end{tablenotes}
\end{threeparttable}
\end{deluxetable}

\subsection{Stage~B period derivatives}
\label{sec:pdot}

After the A-to-B transition, the eccentric mode is fully established and the superhump period settles to its shorter, systematically evolving Stage~B value. During this phase the disk is in the hot state, and the apsidal precession rate of the eccentric region receives contributions from the dynamical tidal torque and from the retrograde gas-pressure term \citep{1992ApJ...401..317L,2006MNRAS.371..235P}. The Stage~B period derivative therefore traces the redistribution of eccentricity within the disk and the changing balance between tidal excitation and pressure support as the superoutburst proceeds \citep{Osaki2013b}. Empirically, the ground-based survey series found the period derivative to be tightly correlated with the superhump period excess, and hence with the mass ratio, making it a sensitive diagnostic of the outbursting disk \citep{2009PASJ...61S.395K}.

Of the 29 objects in which superhumps were detected, twelve have a parabolic Stage~B segment with a positive coefficient. The measured derivatives span $\dot{P}_{\rm SH}=(1.19\pm1.36)\times10^{-5}$ to $(85.09\pm17.83)\times10^{-5}$~d~d$^{-1}$ (Table~\ref{tab:stage_param}); the two smallest values (BE~Oct and ASASSN-14eq; Figs.\ref{fig:1-6}(a)D and (b) D) remain only marginally detected, whereas the long, well-sampled Stage~B intervals (e.g.\ ASASSN-14je, Fig.\ref{fig:7-12}(c)D; V0748~Hya, Fig. \ref{fig:19-24}(a)D; PNV~J06501960+3002449, Fig.\ref{fig:7-12}(f)D MASTER~OT~J172758.09+380021.5; Fig.\ref{fig:7-12}(b)D) yield strongly significant positive coefficient. The overall range agrees with that established by the Kato series. In a further ten objects the Stage~B $O$--$C$ segments are linear within the measurement uncertainties, and we adopt $\dot{P}_{\rm SH}=0$; no object exhibits a significantly negative derivative.

The concentration of the strongest positive derivatives among short-period systems is consistent with the standard picture: in low-$q$ disks the tidal forcing is weak, the eccentric disturbance spreads outward only slowly through a long-lived Stage~B, and the superhump period lengthens secularly as the eccentric region expands and its mean precession rate drops \citep{2009PASJ...61S.395K,Osaki2013b}. Negative derivatives, by contrast, are confined to the long-period, high-$q$ end of the sequence, where the stronger tidal torque establishes the eccentric configuration rapidly and Stage~B is intrinsically short---typically only a few days. Such brief episodes are readily missed or only partially sampled by a single $\sim$27-d TESS sector, and several of our long-period objects were indeed observed only during the onset or the decline of their superoutbursts; their linear Stage~B segments are consistent with the curvature being unresolved rather than absent. Notably, the prototypical negative-derivative system V0419~Lyr ($\dot{P}=-32.4(2.4)\times10^{-5}$; \citealp{2009PASJ...61S.395K}) here shows a purely linear Stage~B, while UV~Gem ($\dot{P}=-5.34\times10^{-4}$; \citealp{2009PASJ...61S.395K})---a long-period, high-$q$ object ($q=0.236$)---displays the largest derivative in our sample, $(85.09\pm17.84)\times10^{-5}$~d~d$^{-1}$, and also the largest Stage~B period uncertainty (Table~\ref{tab:stage_param}); this measurement is discrepant with its population statistics and should be verified against additional superoutbursts. Taken together, the measured derivatives follow the same $\dot{P}_{\rm SH}$--$P_{\rm orb}$ sequence established from the ground (Fig. \ref{fig:q}), independently corroborating the stage identification from our $O$--$C$ and frequency analyses. Following the empirical trend of \citet{2009PASJ...61S.395K}, the short-period objects with linear or only marginally positive Stage~B segments are the most promising period-minimum or post-period-minimum candidates, where the weak tidal forcing produces little detectable period drift, and deserve dedicated monitoring during future superoutbursts.

\section{Conclusions} \label{sec:Conclusions}
We analysed superoutbursts and superhump evolution for 30 SU~UMa-type dwarf novae using continuous, high-cadence TESS light curves. Our key findings are summarised below.

\begin{itemize}
	\item \textbf{Superhump detection and timing measurements.}
	We recovered coherent positive superhump signals in 29 out of 30 targets, including six (EC 05200-5205, ASASSN-14kj, SDSS J080303.90+251627.0, ASASSN-18hr, ASASSN-18xz, and MGAB-V233) sources where superhump periods had not been reported before. We revised orbital periods for 13 systems, eight (GX Cas, EC 05200-5205, V1434 Tau, ASASSN-18hr, Gaia17cuh, ASASSN-15qr, V0774 Peg, and MGAB-V233) of which are new measurements presented in this work. One target, Gaia17cuh, was only observed in TESS Sector~81, capturing a partially covered, long outburst of ambiguous classification (Fig. \ref{fig:25-30}(a)A). Its detected periodic signal lies outside the expected SU~UMa period range, indicating it may not be a true SU~UMa system, and we interpret the measured period as its orbital period. Continuous TESS coverage yields dozens to hundreds of superhump maxima per outburst, enabling reliable $O$--$C$ diagrams that resolve the stepwise period evolution across different stages. 
	
	\item \textbf{Stage~A mass ratios.}
	We applied the dynamical 3:1 tidal resonance method to Stage~A superhumps and derived mass ratios for 18 sources with known orbital periods. These $q$ values follow the observed $P_{\rm orb}$–$q$ distribution and lie on the CV evolutionary tracks of \citet{2011ApJS..194...28K}. This consistency supports both our timing analysis and the validity of the Stage~A dynamical method. Five sources without identifiable Stage~A segments give systematically lower mass ratios when estimated from mean superhump periods
	
	\item \textbf{Stage~B period derivatives.}
	Our measured Stage~B $\dot{P}_{\rm SH}$ range from $(1.19\pm1.36)$ to $(85.09\pm17.83)\times10^{-5}$~d~d$^{-1}$, covering the full parameter range reported in ground-based surveys. Twelve systems show parabolic Stage~B evolution with positive $\dot{P}_{\rm SH}$; these are mostly short-period, low-$q$ systems where weak tidal torque allows slow outward propagation of the eccentric disturbance wave. Ten systems show linear Stage~B behaviour with $\dot{P}_{\rm SH}\approx0$. This set includes V0419~Lyr, a classic candidate for negative derivatives, for which the Stage~B interval is too short within a single TESS sector to resolve curvature. We find no systems with statistically significant negative derivatives. This selection effect arises because long-period systems have intrinsically brief Stage~B phases and are often only observed near outburst rise or decline.
	
	\item \textbf{Variety and repeatability of stage evolution.}
	
Five systems with two or more observed superoutbursts demonstrate that Stage~B/C morphology is broadly repeatable, tracing the mass ratio and the 3:1 resonance template (e.g., ASASSN-14je: Stage~B derivatives agree within $\sim1\sigma$, and Stage~C periods match to $0.05\%$). In contrast, Stage~A properties can vary across outburst events: for WX~Hyi, the Stage~A period differs by $1.7\%$ between a superoutburst with a prominent precursor and one without. Although YZ~Cnc exhibits consistent Stage~A/B periods in its two events, Stage~C is detected in only one of them.
 EC~05200$-$5205 additionally shows a two-part Stage~A (A1–A2), implying the eccentricity growth phase can be more complex than the simple standard picture. Two sources deviate strongly from the standard A–B–C sequence. RZ~LMi has reversed ordering: its Stage~A-like segment has a shorter period than Stage~B. ASASSN-14kj shows no recognisable staged evolution, only a coherent 1.00(2)-d modulation whose physical origin remains unclear.

	\item \textbf{Thermal and tidal decoupling.}
	We detected persistent positive superhumps during normal outbursts after superoutbursts in WX~Hyi, EC~05200$-$5205 and V1113~Cyg. Their periods are 0.5–1\% shorter than those measured during the preceding superoutburst. This directly demonstrates eccentric disk structures can survive after the superoutburst and become decoupled from the thermal accretion cycle. Low-$q$ systems have weaker tidal torques, which slow eccentricity damping and allow the eccentric mode to persist through multiple normal outbursts. This extends the scenario of long-lived eccentric disks, previously found in AM~CVn systems such as V803~Cen, to classical hydrogen-rich SU~UMa dwarf novae.
\end{itemize}

Continuous TESS photometry resolves superhump stage evolution with unprecedented continuity and precision unavailable from ground observations. The diverse $O$--$C$ behaviours identified in our sample, including inverted stage sequences, missing stage evolution, and multi-step eccentricity growth, place tight constraints on precession models incorporating gas pressure, eccentricity transport, and thermal-tidal coupling. Future studies should enlarge the sample of high-$q$ and long-period systems, combine TESS data with multi-wavelength observations to constrain disk structures, and explore anomalous stage evolution to clarify the boundaries of the canonical A–B–C framework. The detection of persistent superhumps during normal outbursts offers new insights into long-term eccentric disk evolution and thermal-tidal decoupling, encouraging systematic searches for similar phenomena across cataclysmic variable populations.

\section*{Acknowledgments}
This research received support from the following sources: the National Natural Science Foundation of China (Grant Nos. 12503040, 11933008, and 12303040), the National Key R\&D Program of China (Grant No. 2022YFE0116800), the Yunnan Fundamental Research Projects (Grant Nos. 202601AT070222, 202501AS070055, 202503AP140013, 202201AT070092, and 202401AT070143), the China Manned Space Program (Grant No. CMS-CSST-2025-A16), the International Partnership Program of the Chinese Academy of Sciences (Grant No. 020GJHZ2023030GC), and the 2022 CAS ``Light of West China'' Program.
All observational data employed in this work are publicly available. TESS photometry can be accessed via the Mikulski Archive for Space Telescopes (MAST) at \dataset[doi:10.17909/m88m-dk35]{https://doi.org/10.17909/m88m-dk35}. 

\section*{Appendix}\label{sec:Appendix}
This appendix contains Figures~\ref{fig:1-6}-\ref{fig:max-example} and Table~\ref{tab:stage_param}.
Figures~\ref{fig:1-6}--\ref{fig:25-30} present the light curves, periodograms, $O$--$C$ diagrams, and folded light curves of the corresponding signals for the 30 targets analyzed in this work.
 Figure~\ref{fig:out-cal-examp1} gives examples of confirmed outburst epochs. Figure~\ref{fig:max-example} shows an example of the linear-plus-Gaussian fitting. Table~\ref{tab:stage_param} compiles superhump stage periods, Stage-B period derivatives, mass ratios and orbital periods for all targets in our sample.

\bibliography{sample7}{}
\bibliographystyle{aasjournalv7}

\renewcommand\thefigure{A1}
\begin{figure}
	\centering
	\begin{subfigure}{0.45\columnwidth}
		\includegraphics[width=\linewidth]{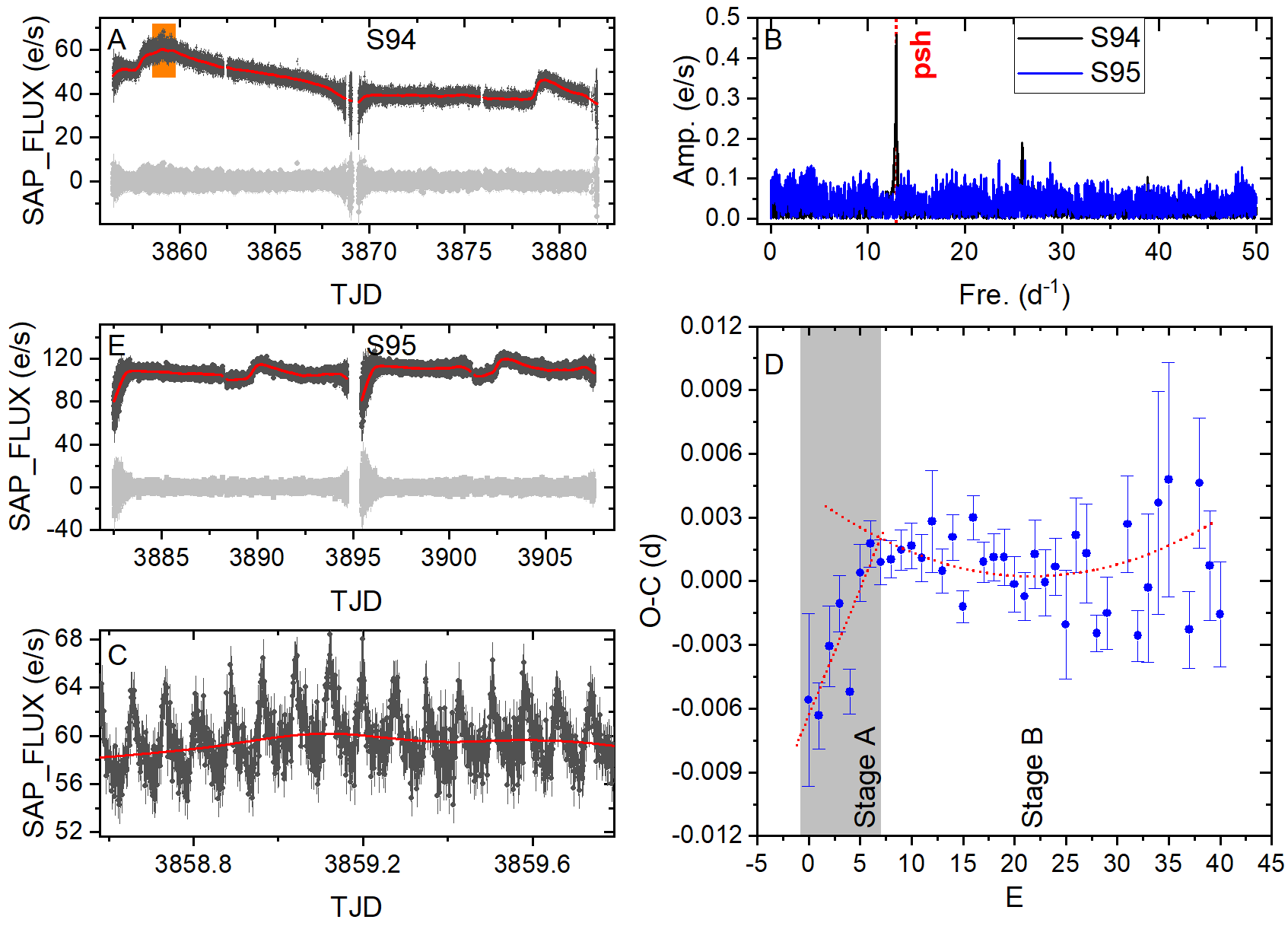}
		\caption{BE Oct}
	\end{subfigure}
	\begin{subfigure}{0.45\columnwidth}
		\includegraphics[width=\linewidth]{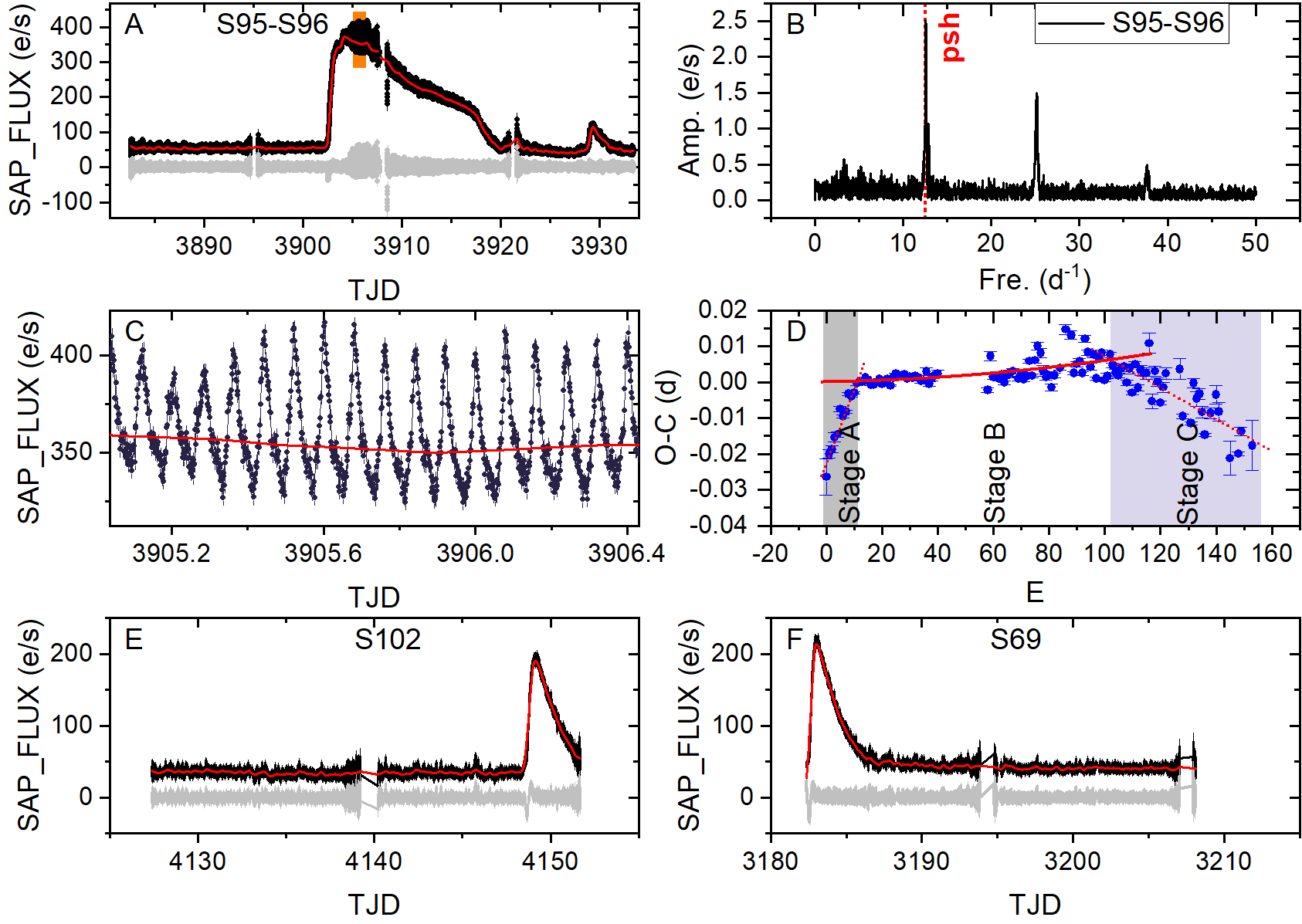}
		\caption{ASASSN-14eq}
	\end{subfigure}
	\begin{subfigure}{0.45\columnwidth}
		\includegraphics[width=\linewidth]{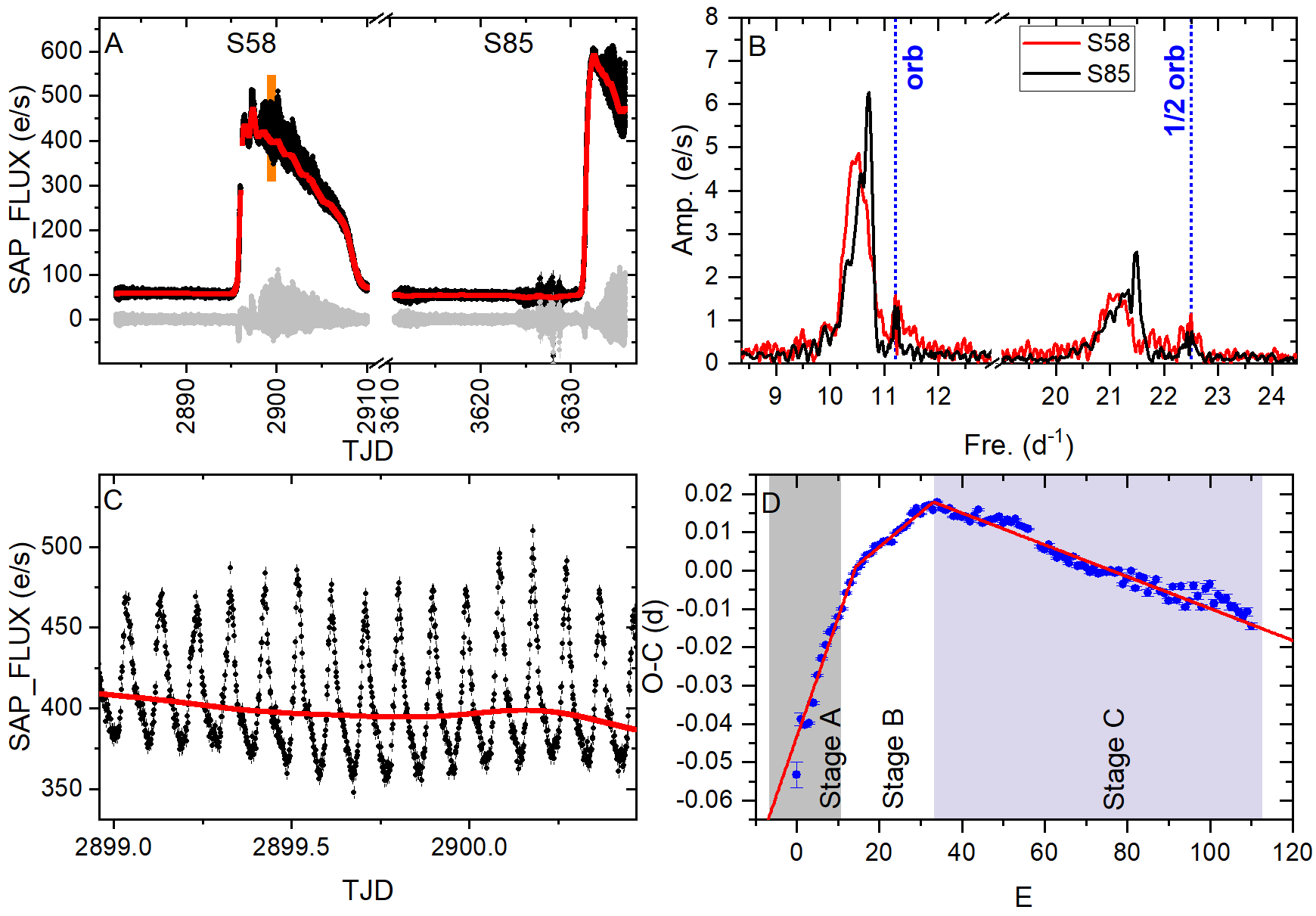}
		\caption{GX~Cas}
	\end{subfigure}
	\begin{subfigure}{0.45\columnwidth}
		\includegraphics[width=\linewidth]{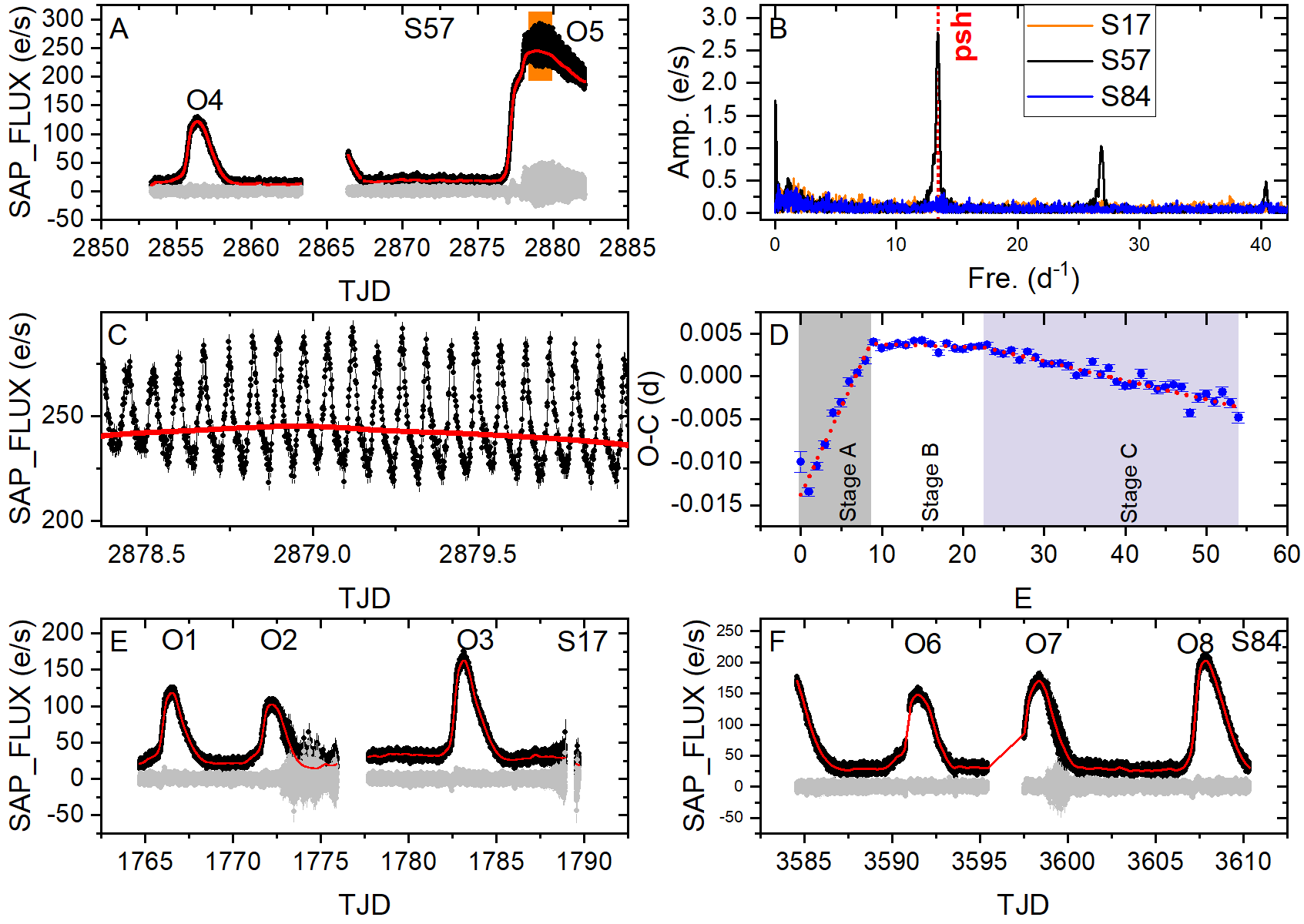}
		\caption{FO And}
	\end{subfigure}
	\begin{subfigure}{0.45\columnwidth}
		\includegraphics[width=\linewidth]{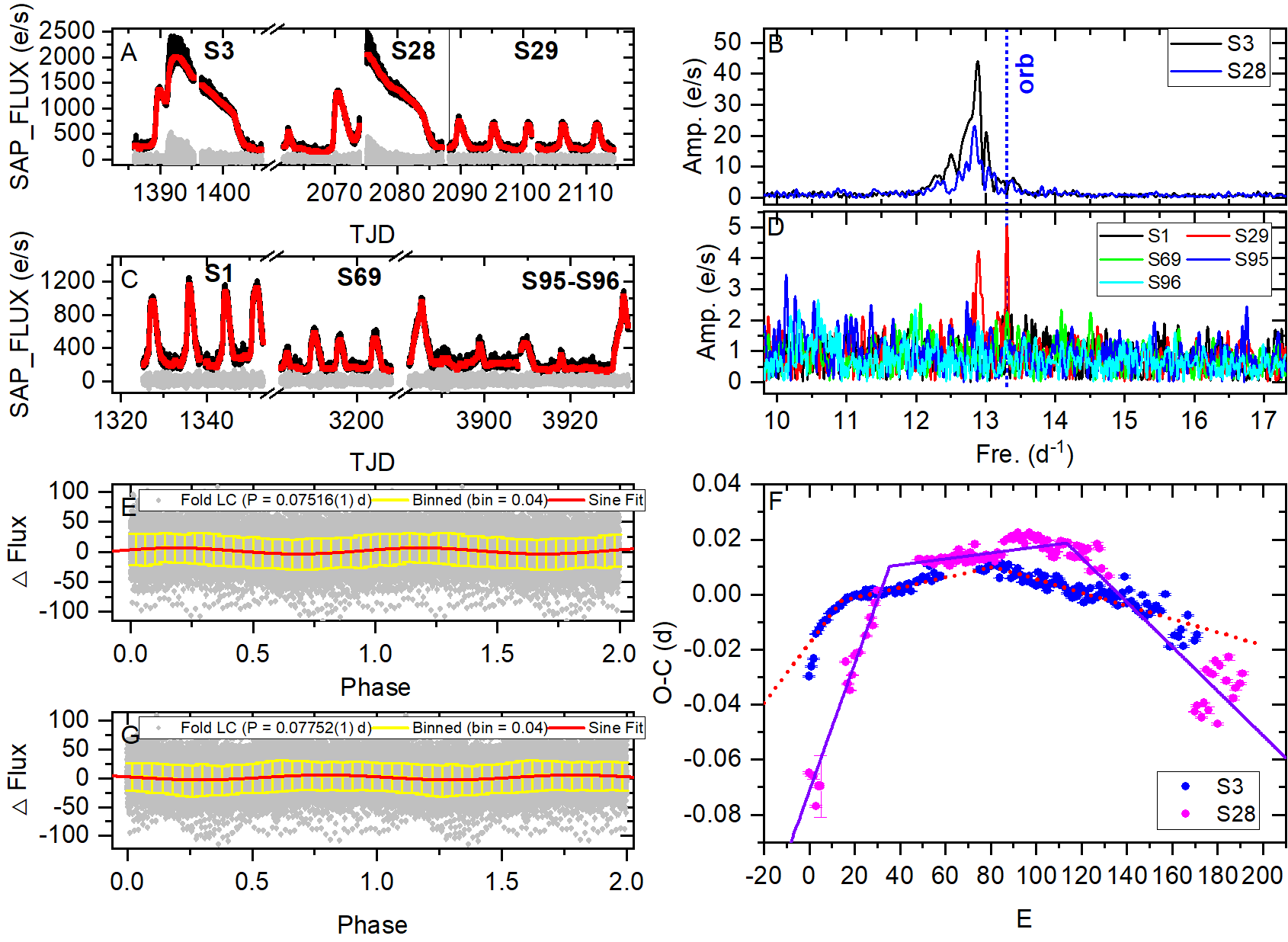}
		\caption{WX~Hyi}
	\end{subfigure}
	\begin{subfigure}{0.45\columnwidth}
		\includegraphics[width=\linewidth]{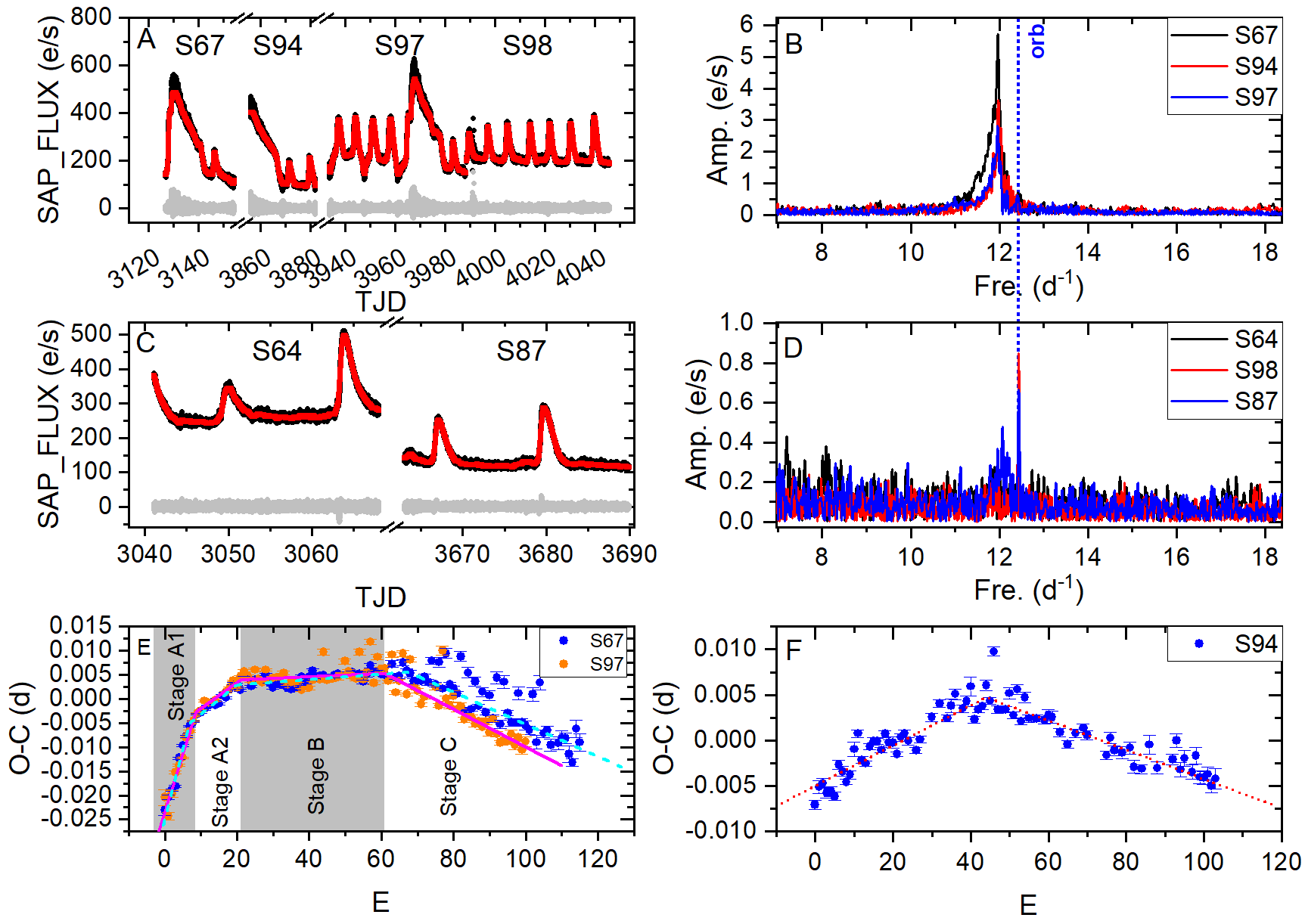}
		\caption{EC 05200-5205}
	\end{subfigure}
	\caption{Light curves and analysis results of BE\,Oct, ASASSN-14eq, GX\,Cas, FO\,And, WX\,Hyi, and EC\,05200-5205.
		In panel~A of each subfigure, the black, red, and grey lines represent the TESS light curve, LOWESS fit, and fit residuals, respectively; this convention is adopted for all similar figures in this work.
		Panel~B of each subfigure shows the frequency spectrum derived from the Fourier transform of the corresponding TESS data.
		Except for (e) WX\,Hyi and (f) EC\,05200-5205, panel~C in all other subpanels is an enlargement of the orange region marked in panel~A to visually illustrate the PSHs.
		(a) For BE\,Oct, the $O$--$C$ interval shown in panel~D was fitted with stage~A (linear fit) and stage~B (parabolic fit).
		(b) For ASASSN-14eq, panel~D includes fitted stage~C (linear fit) in addition to stage~A and stage~B.
		(c) GX\,Cas, (d) FO\,And, and (e) WX\,Hyi have their $O$--$C$ diagrams in panel~D modelled using a three-segment piecewise-linear function.
		(f) For EC\,05200-5205, panel~E presents the $O$--$C$ curves from different sectors fitted with a four-segment piecewise-linear function, while panel~F shows the two-segment piecewise-linear fit only.}
	\label{fig:1-6}
\end{figure}

\renewcommand\thefigure{A2}
\begin{figure}
	\centering
	\begin{subfigure}{0.45\columnwidth}
		\includegraphics[width=\linewidth]{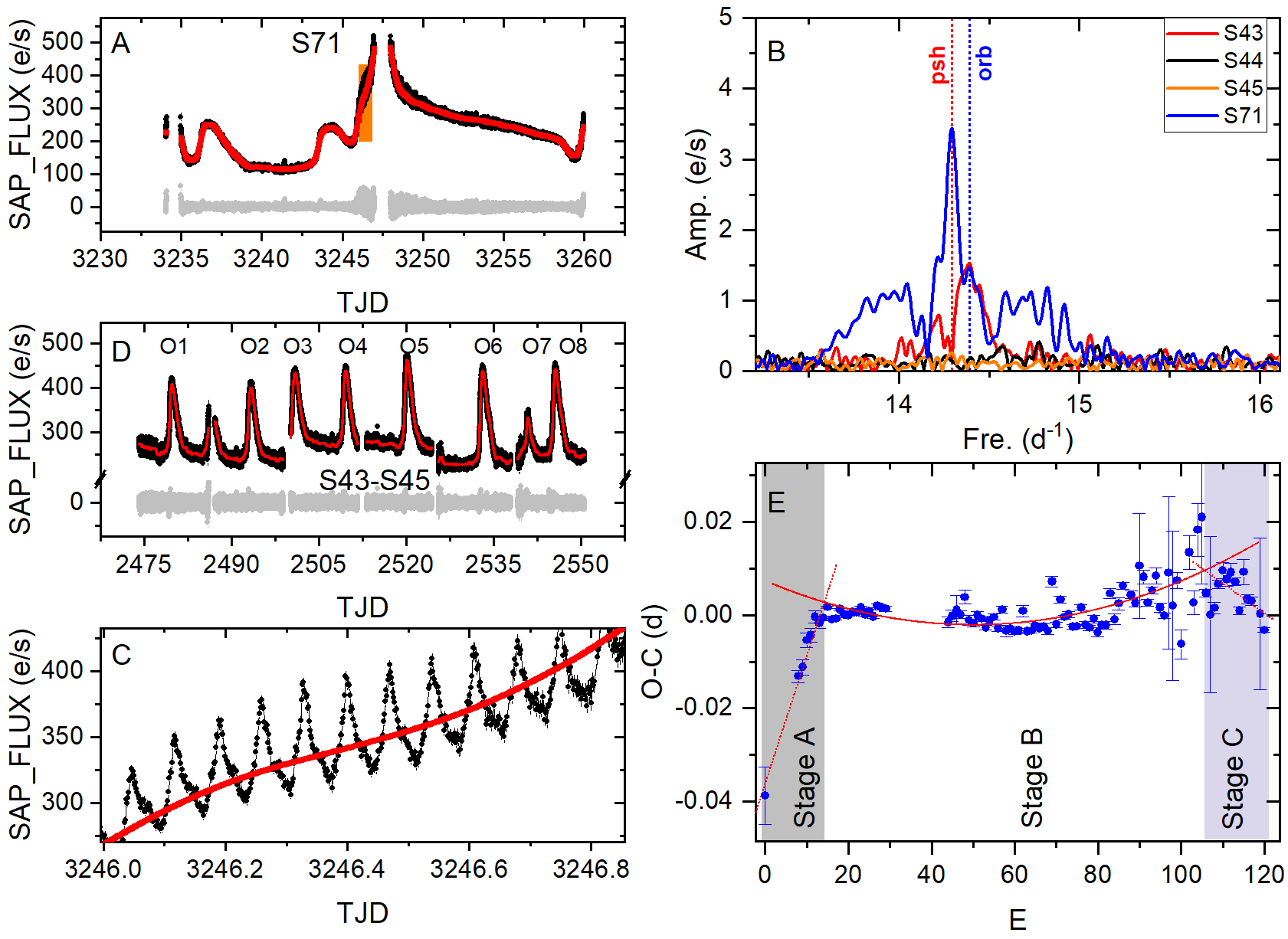}
		\caption{V1434 Tau}
	\end{subfigure}	
	\begin{subfigure}{0.45\columnwidth}
		\includegraphics[width=\linewidth]{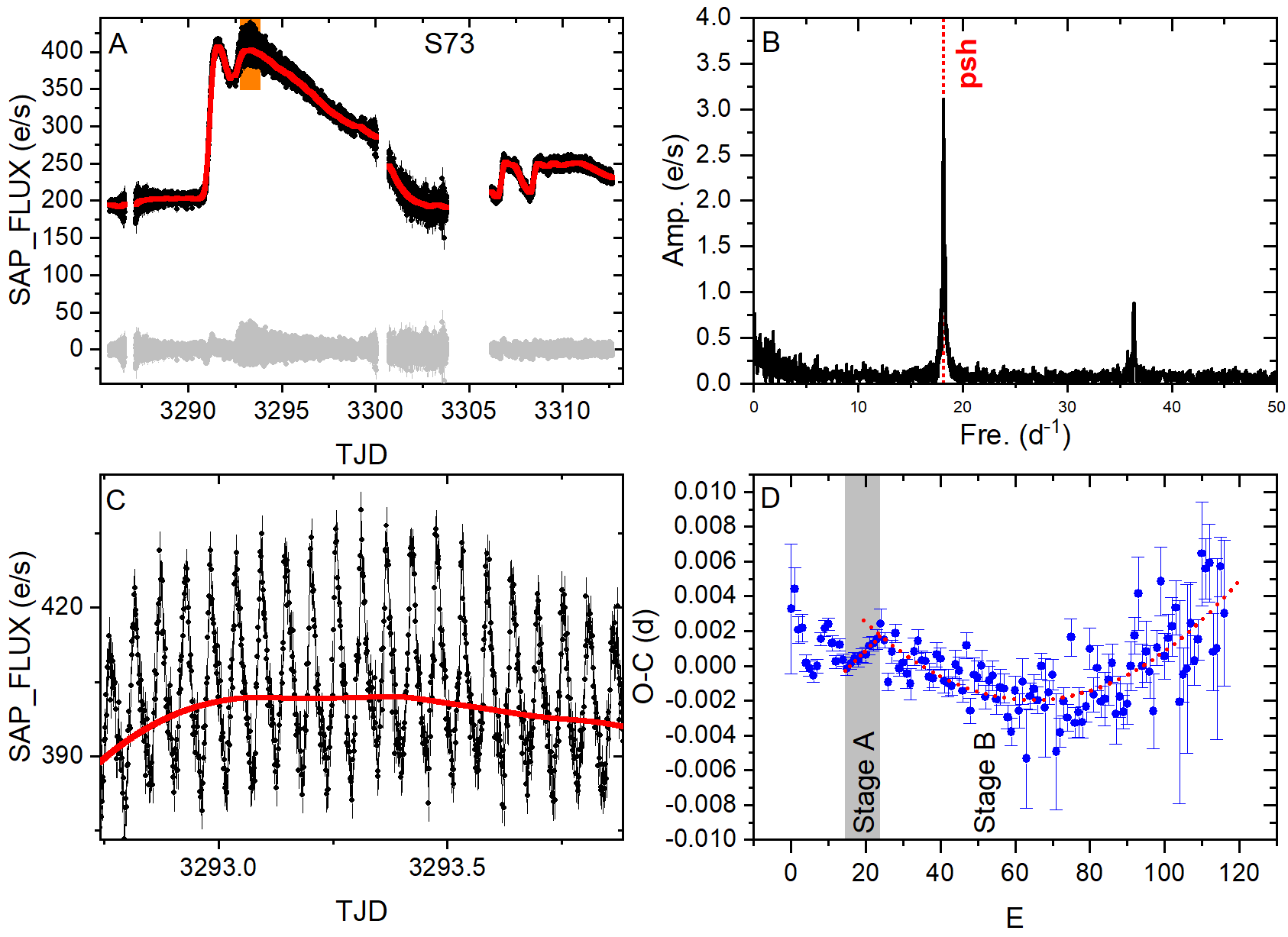}
		\caption{MASTER OT J055845.55+391533.4}
	\end{subfigure}

	\begin{subfigure}{0.45\columnwidth}
		\includegraphics[width=\linewidth]{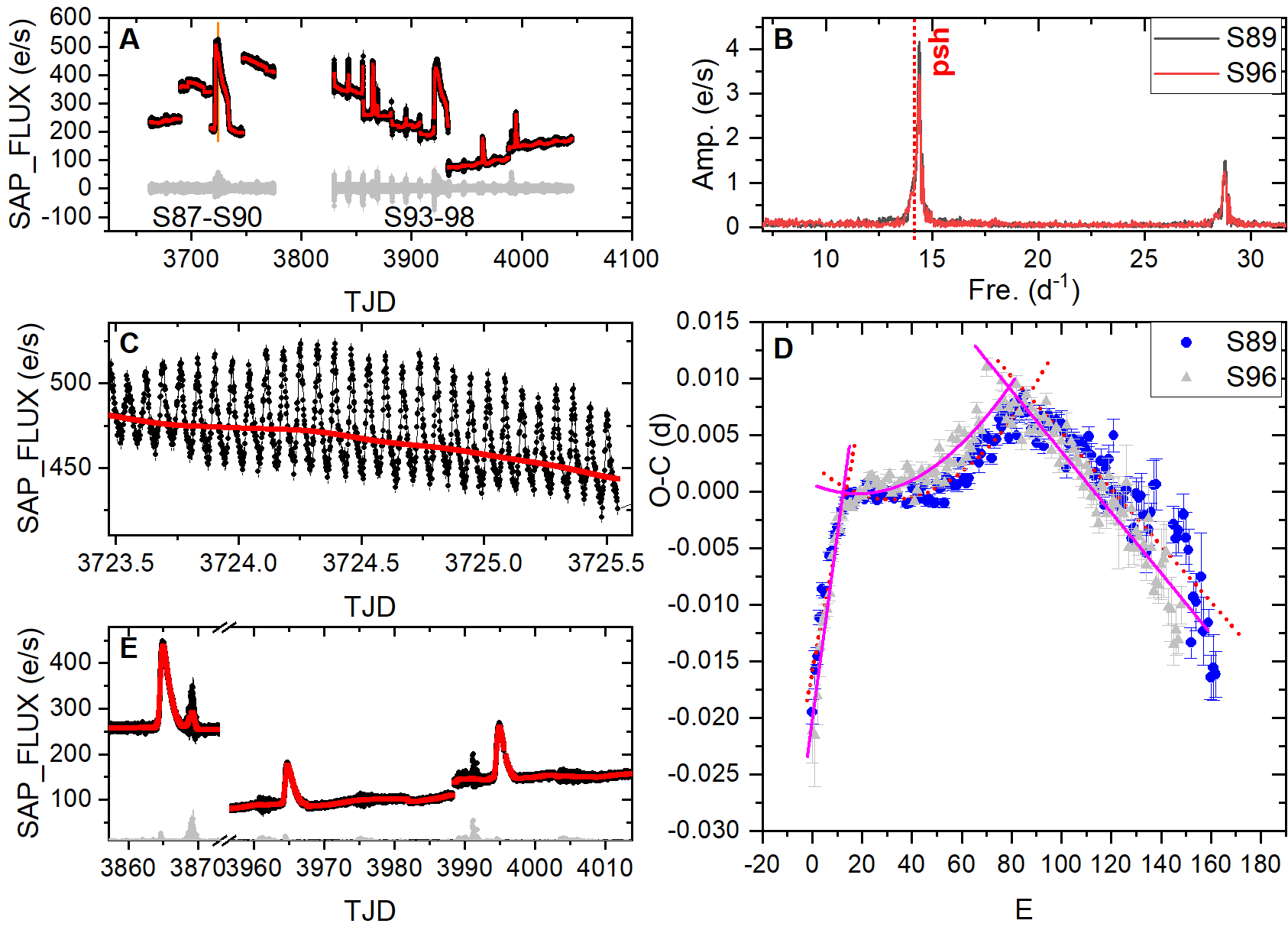}
		\caption{ASASSN-14je}
	\end{subfigure}
	\begin{subfigure}{0.45\columnwidth}
		\includegraphics[width=\linewidth]{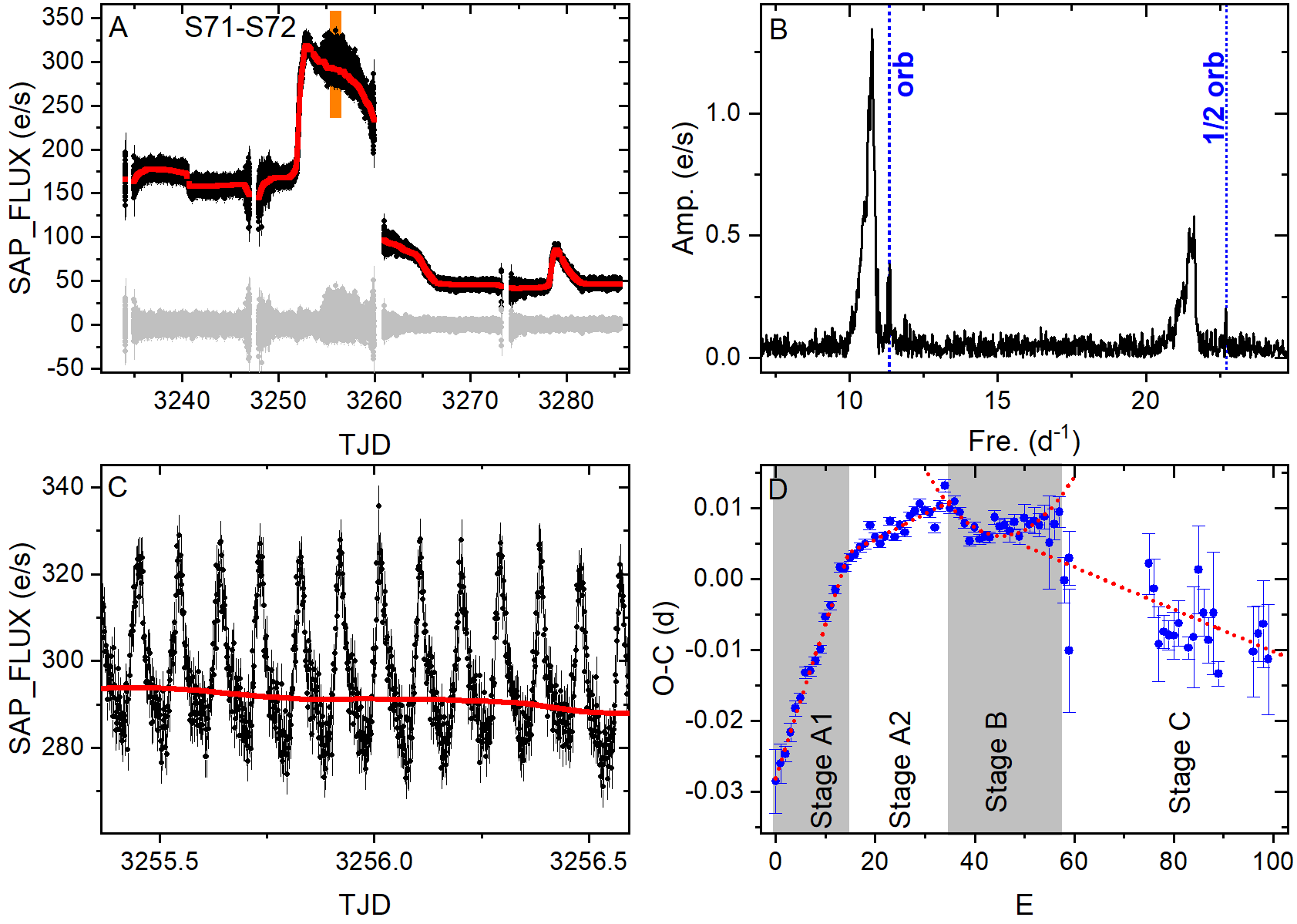}
		\caption{UV Gem}
	\end{subfigure}

	\begin{subfigure}{0.45\columnwidth}
		\includegraphics[width=\linewidth]{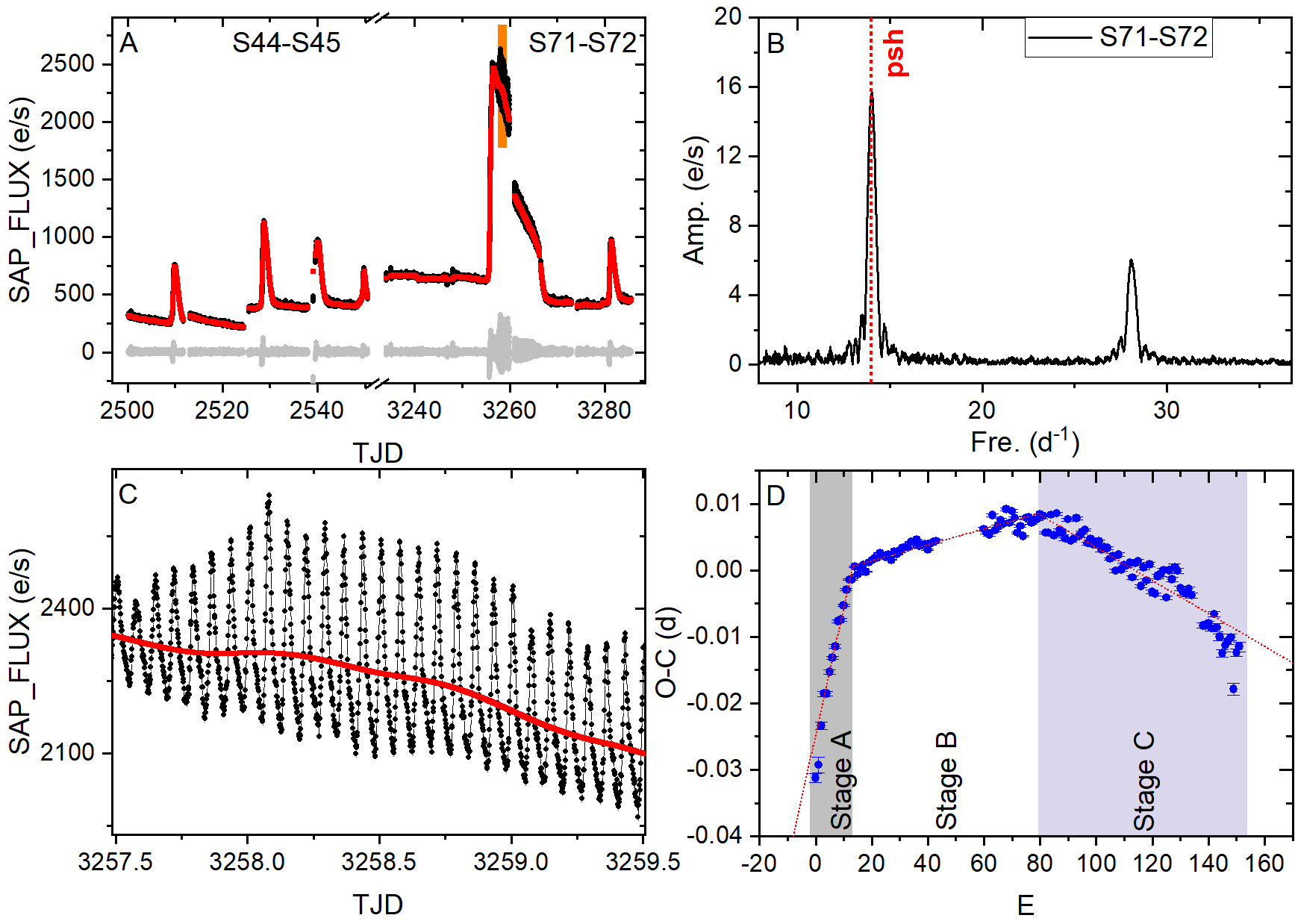}
		\caption{IR Gem}
	\end{subfigure}
	\begin{subfigure}{0.45\columnwidth}
		\includegraphics[width=\linewidth]{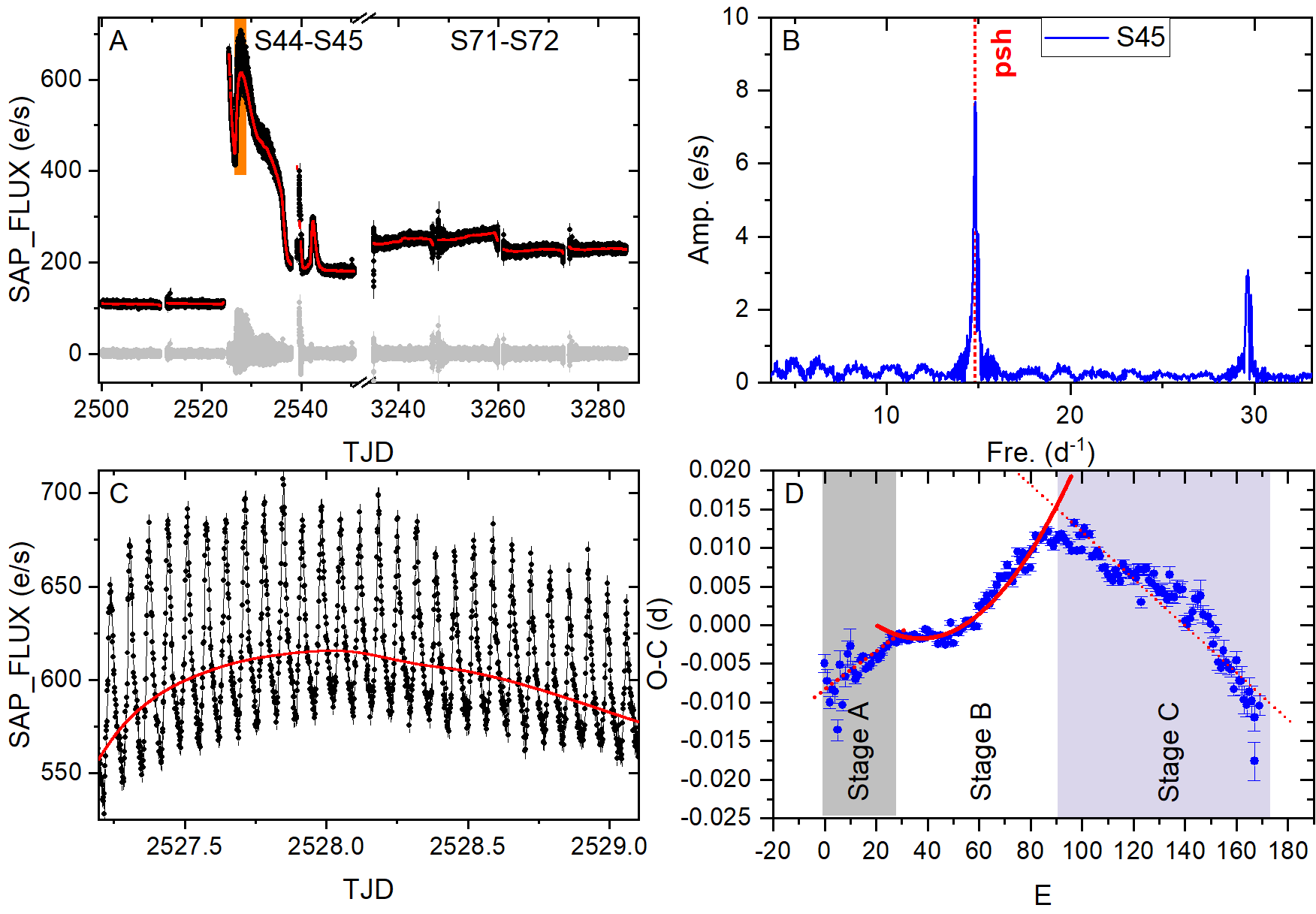}
		\caption{PNV J06501960+3002449}
	\end{subfigure}
	\caption{Panels A share identical color schemes with panel A in Fig.~\ref{fig:1-6} but correspond to different samples and different TESS sectors. Panels B are all Fourier transform results. All panels C are consistent with Fig.~\ref{fig:1-6}, which are magnified views of the orange curves in panels A.
		(a) For V1434 Tau, the $O$--$C$ interval shown in panel~E was fitted with stage~A (linear fit), stage~B (parabolic fit), and stage~C (linear fit).
		(b) For MASTER OT J055845.55+391533.4, panel~D includes fitted stage~A and stage~B.
		(c) ASASSN-14je, (d) UV Gem, and (f) PNV J06501960+3002449 have their $O$--$C$ diagrams in panel~D containing stage~A, stage~B and stage~C.
		(e) For IR Gem, panel~D presents the $O$--$C$ curves fitted with a three-segment piecewise-linear function.}
	\label{fig:7-12}
\end{figure}

\renewcommand\thefigure{A3}
\begin{figure}
	\centering
	\begin{subfigure}{0.45\columnwidth}
		\includegraphics[width=\linewidth]{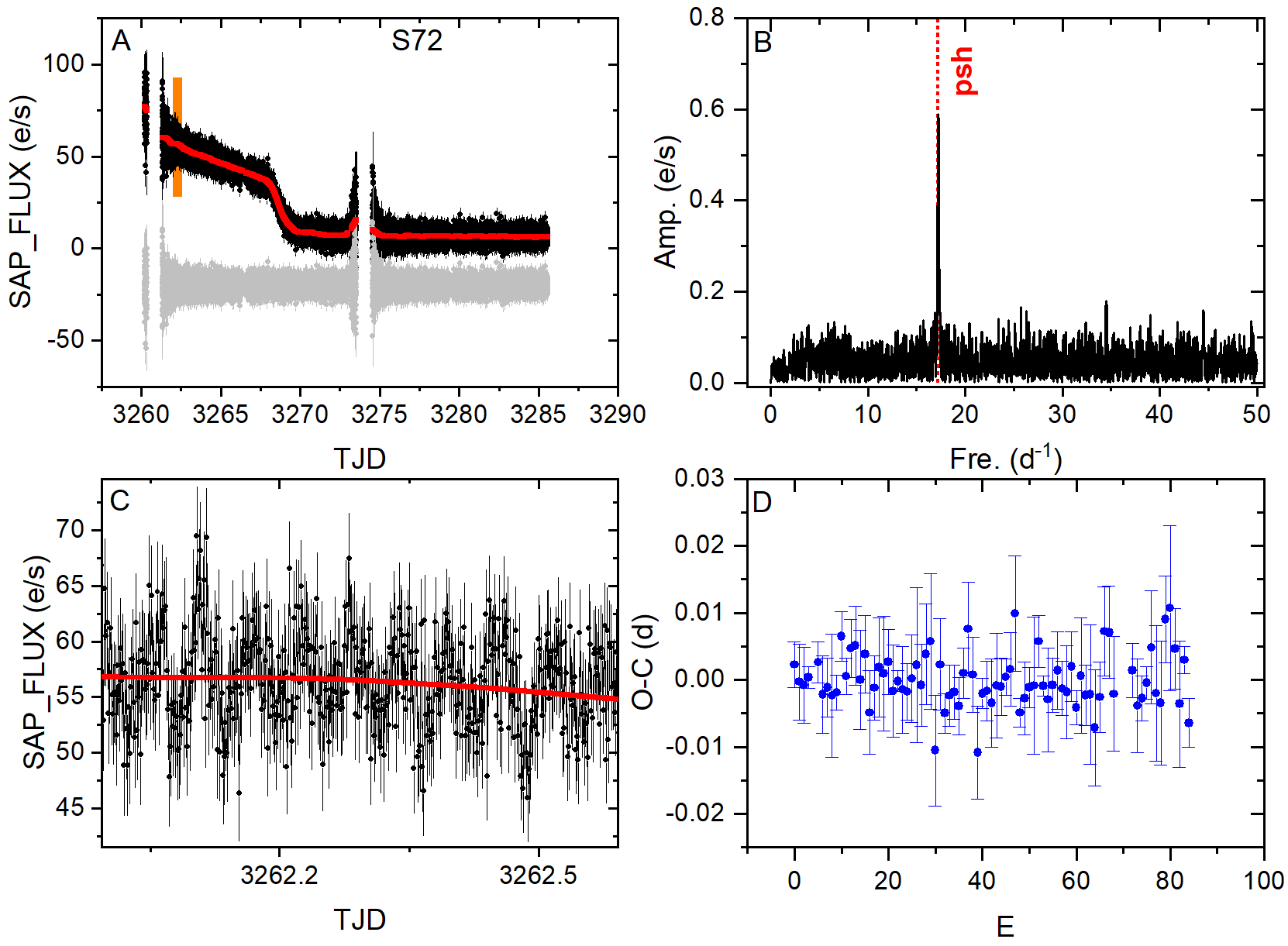}
		\caption{SDSS\,J075107.50+300628.4}
	\end{subfigure}	
	\begin{subfigure}{0.45\columnwidth}
		\includegraphics[width=\linewidth]{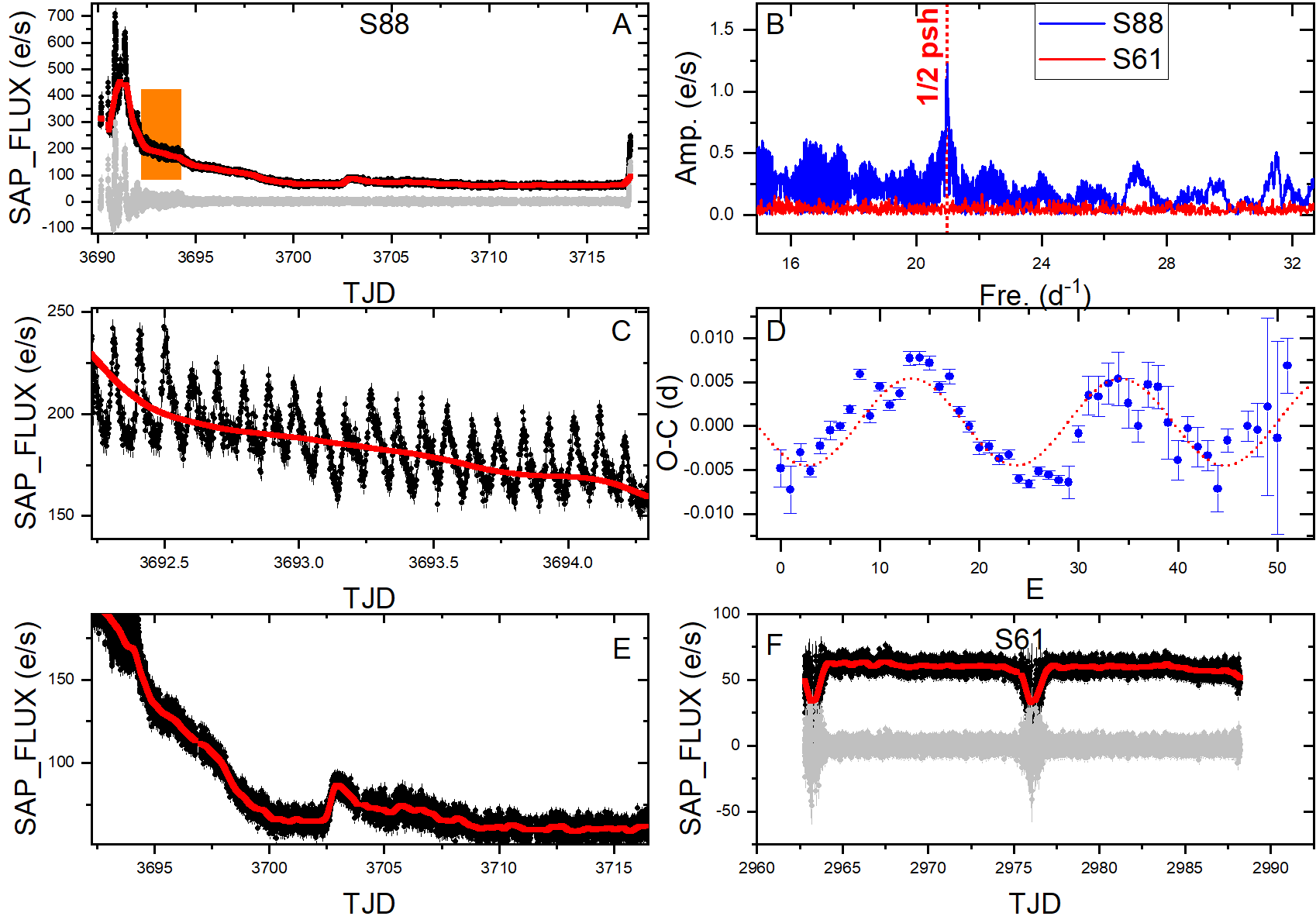}
		\caption{ASASSN-14kj}
	\end{subfigure}

	\begin{subfigure}{0.45\columnwidth}
		\includegraphics[width=\linewidth]{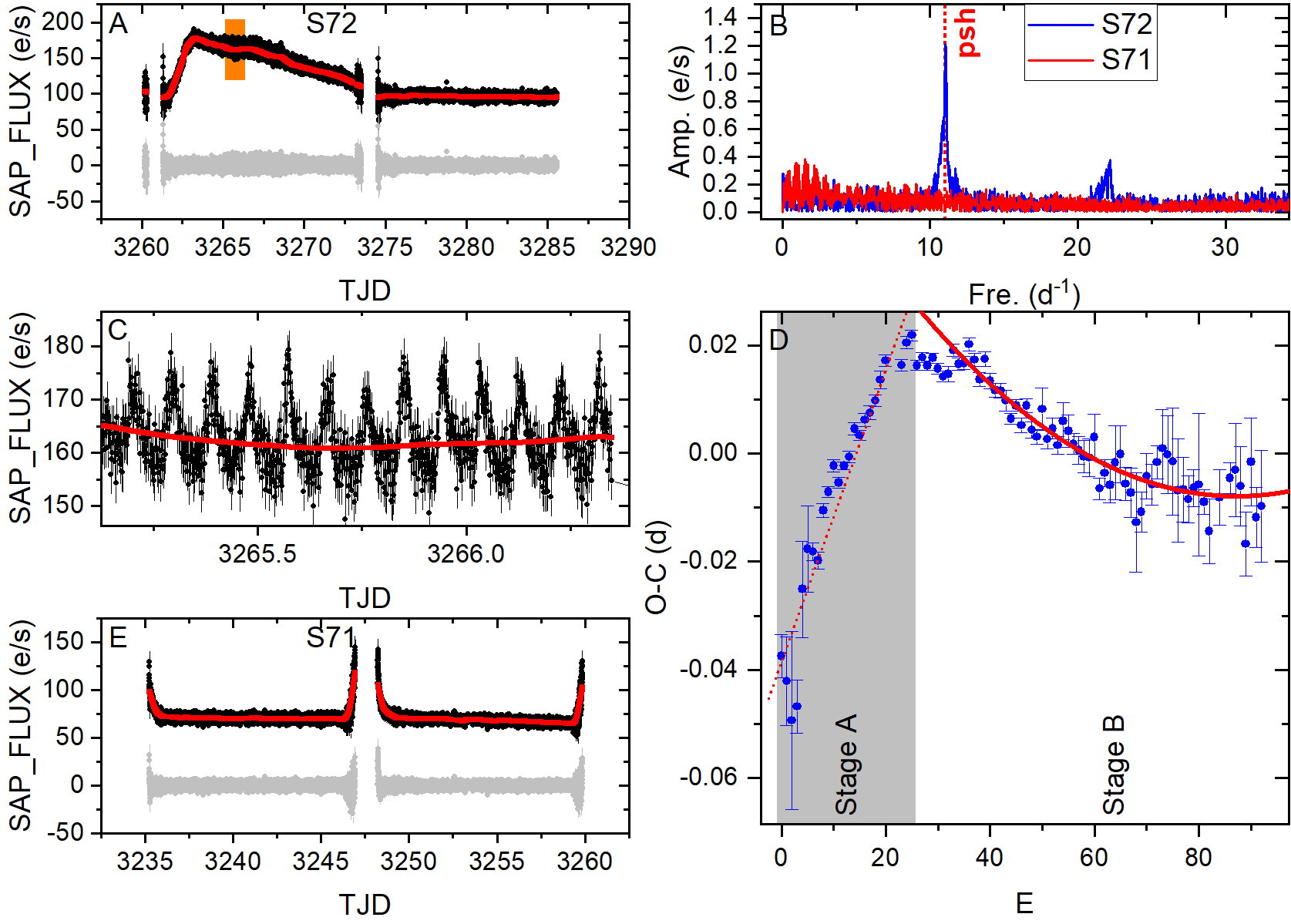}
		\caption{SDSS J080303.90+251627.0}
	\end{subfigure}
	\begin{subfigure}{0.45\columnwidth}
		\includegraphics[width=\linewidth]{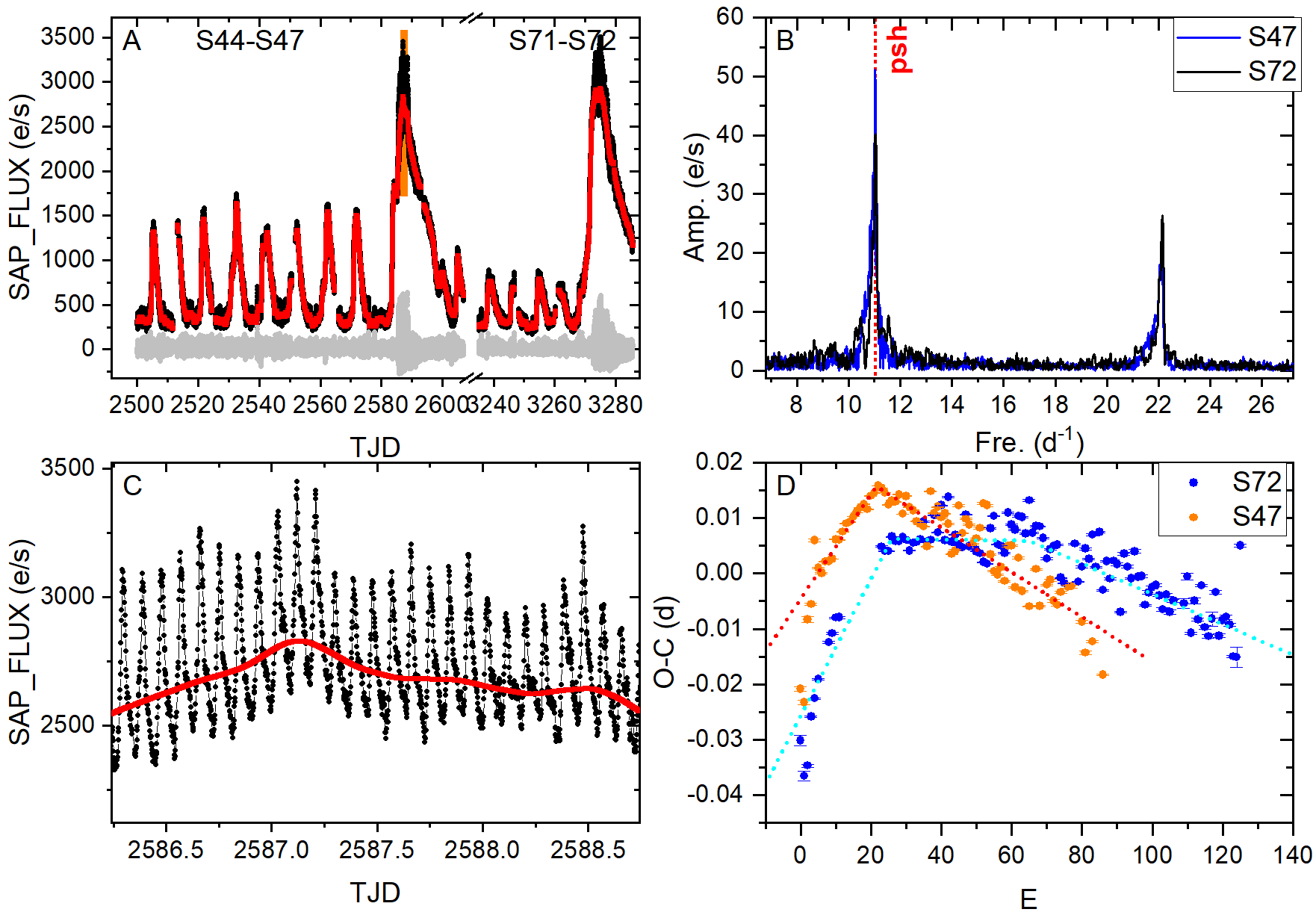}
		\caption{YZ Cnc}
	\end{subfigure}

	\begin{subfigure}{0.45\columnwidth}
		\includegraphics[width=\linewidth]{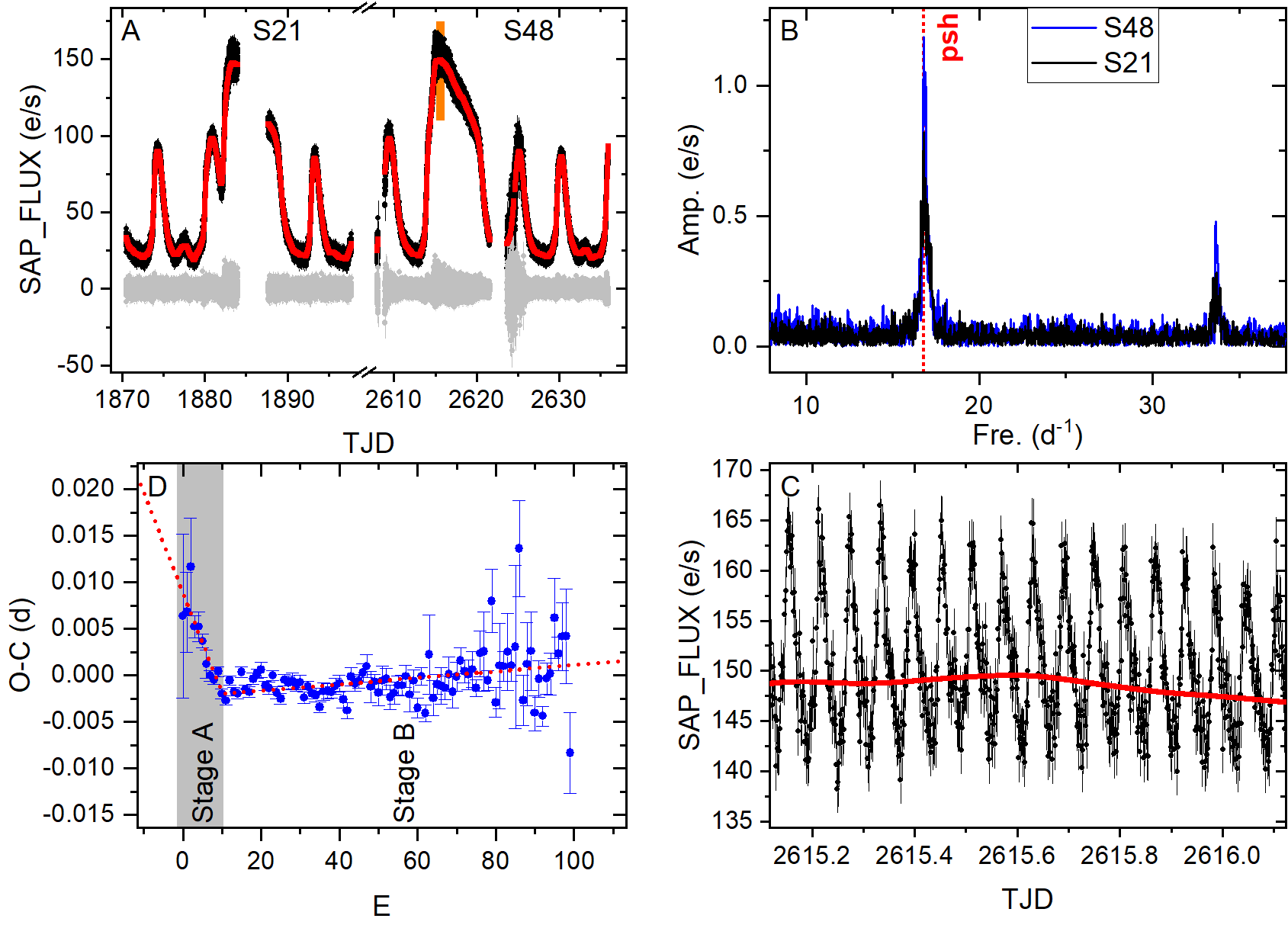}
		\caption{RZ LMi}
	\end{subfigure}
	\begin{subfigure}{0.45\columnwidth}
		\includegraphics[width=\linewidth]{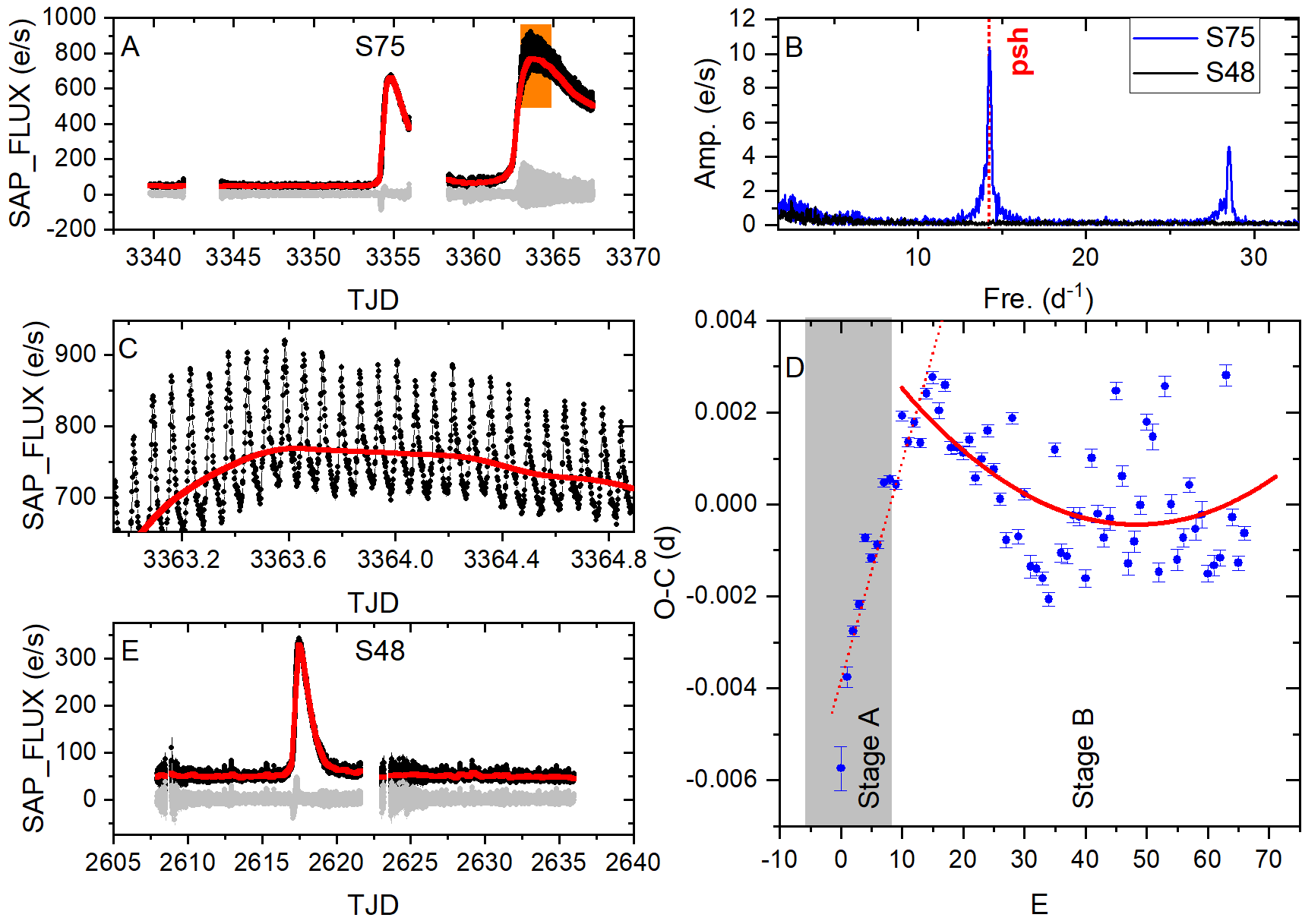}
		\caption{KS UMa}
	\end{subfigure}
	\caption{Panels A, B, C and D of the subpanels in this figure follow the same layout as Figs.~\ref{fig:1-6} and \ref{fig:7-12}, showing the light curve, frequency spectrum, zoomed-in view, and $O$--$C$ diagram, respectively.
		(a) For SDSS\,J075107.50+300628.4, no fit was performed on the $O$--$C$ diagram.
		(b) For ASASSN-14kj, a sinusoidal fit was applied to the $O$--$C$ curve, revealing a periodic variation with a period of approximately one day.
		(c) The $O$--$C$ diagrams of SDSS\,J080303.90+251627.0 and (f) KS UMa were fitted for stage~A and stage~B.
		(d) Two sets of $O$--$C$ fits were carried out for YZ Cnc, corresponding to a 2-segment piecewise-linear function and a 3-segment piecewise-linear function.
		(e) The $O$--$C$ curve of RZ LMi was fitted using a 2-segment piecewise-linear function.}
	\label{fig:13-18}
\end{figure}

\renewcommand\thefigure{A4}
\begin{figure}
	\centering
	\begin{subfigure}{0.45\columnwidth}
		\includegraphics[width=\linewidth]{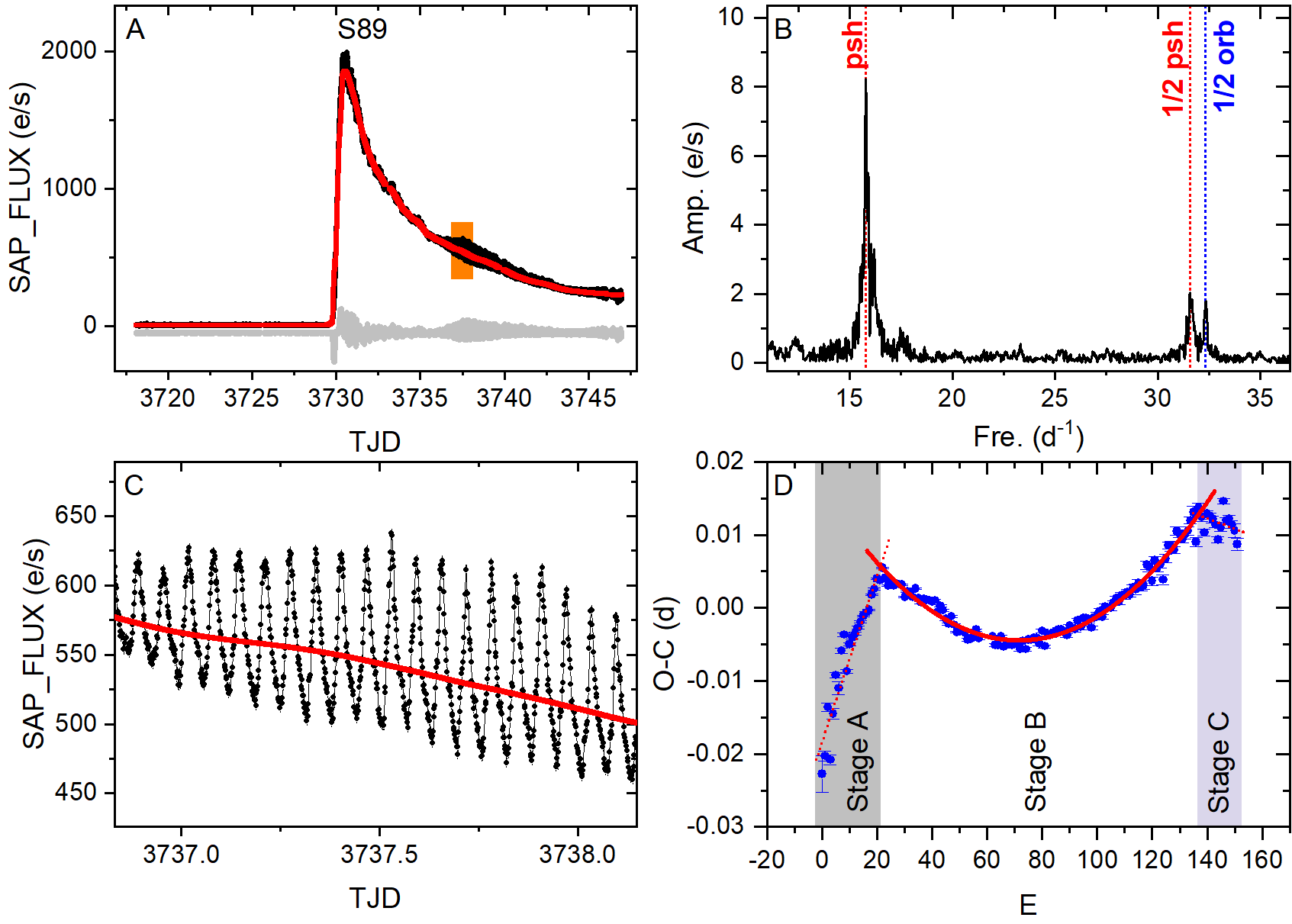}
		\caption{V0748 Hya}
	\end{subfigure}	
	\begin{subfigure}{0.45\columnwidth}
		\includegraphics[width=\linewidth]{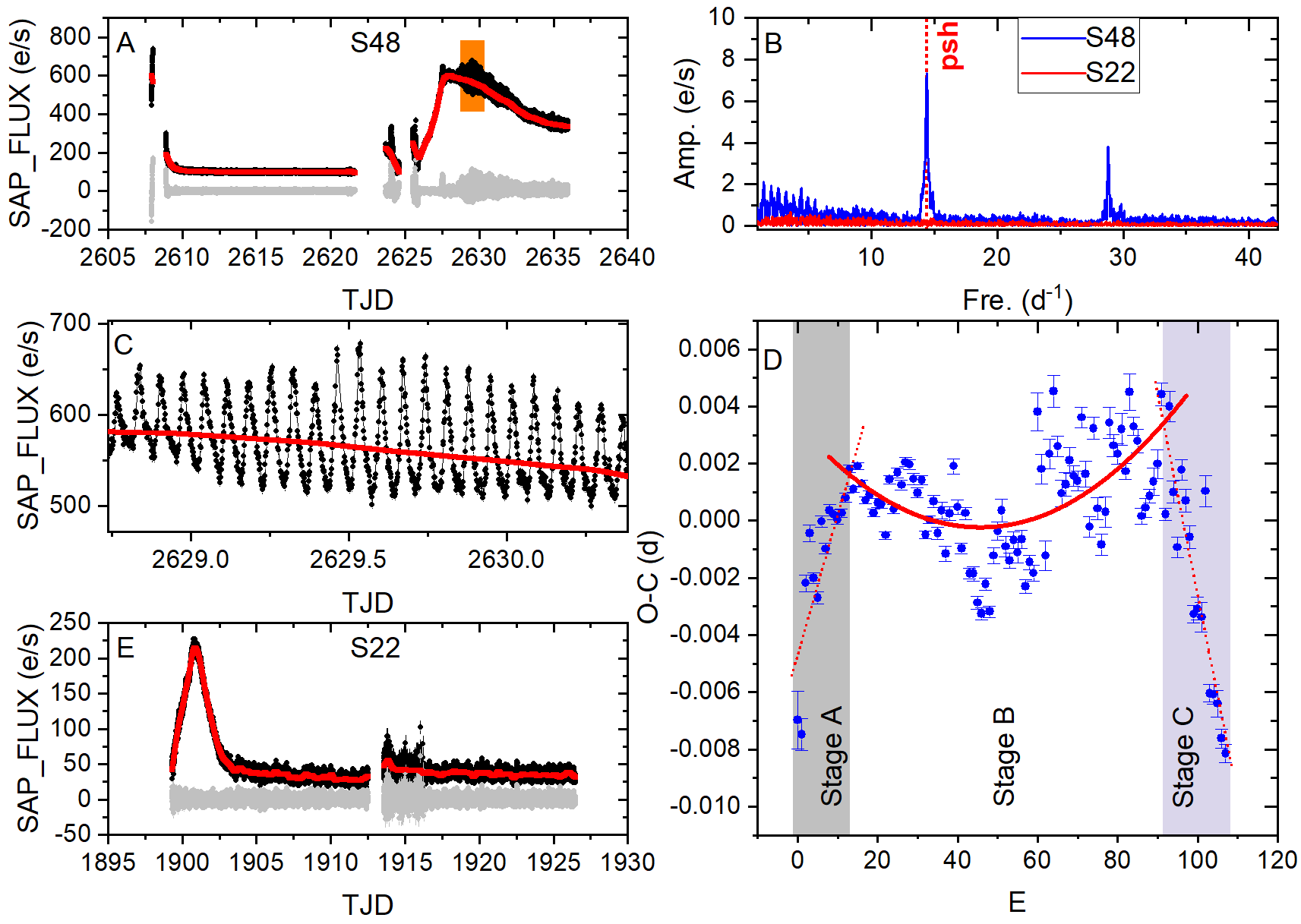}
		\caption{SX LMi}
	\end{subfigure}

	\begin{subfigure}{0.45\columnwidth}
		\includegraphics[width=\linewidth]{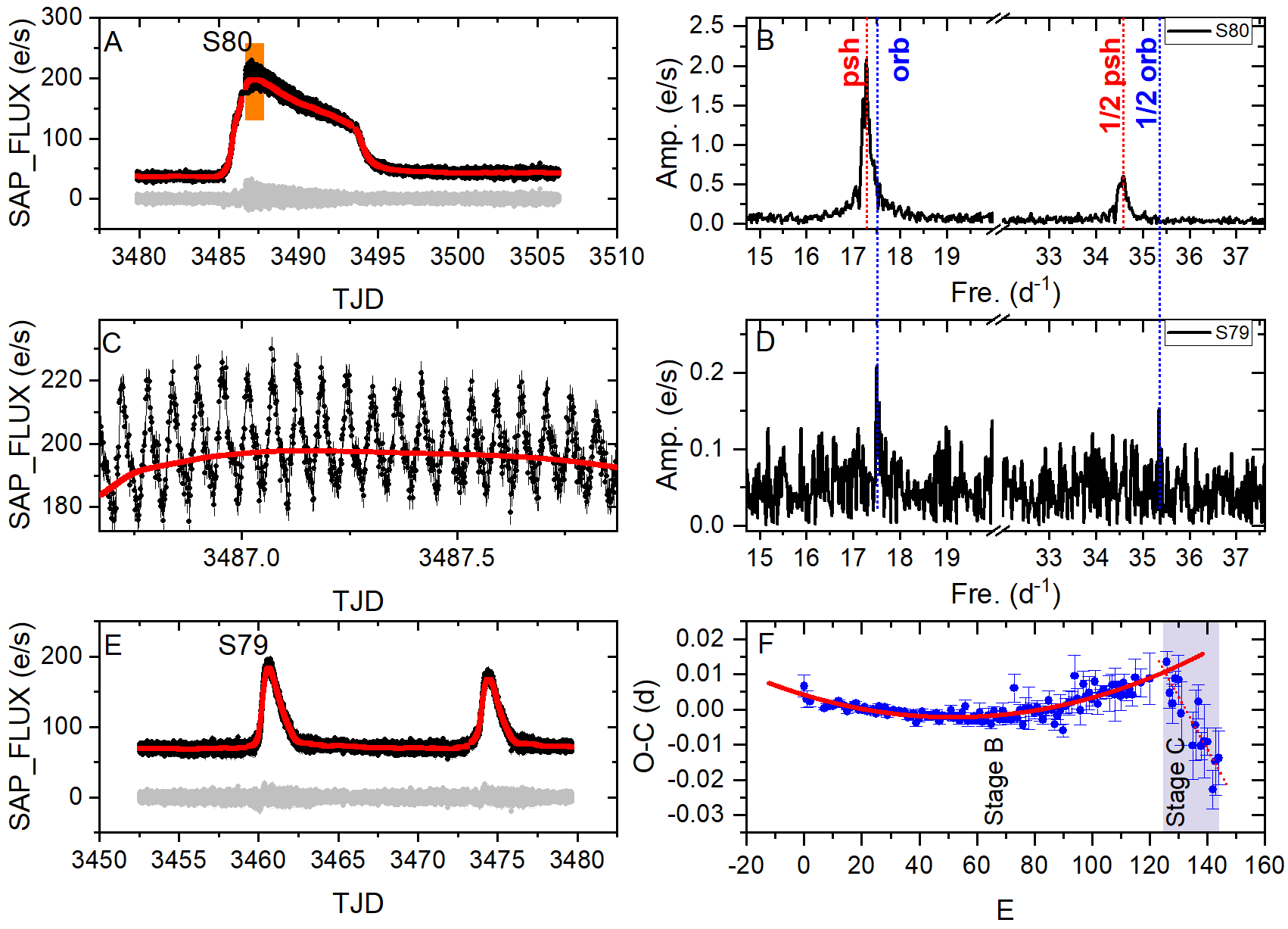}
		\caption{MASTER OT J172758.09+380021.5}
	\end{subfigure}
	\begin{subfigure}{0.45\columnwidth}
		\includegraphics[width=\linewidth]{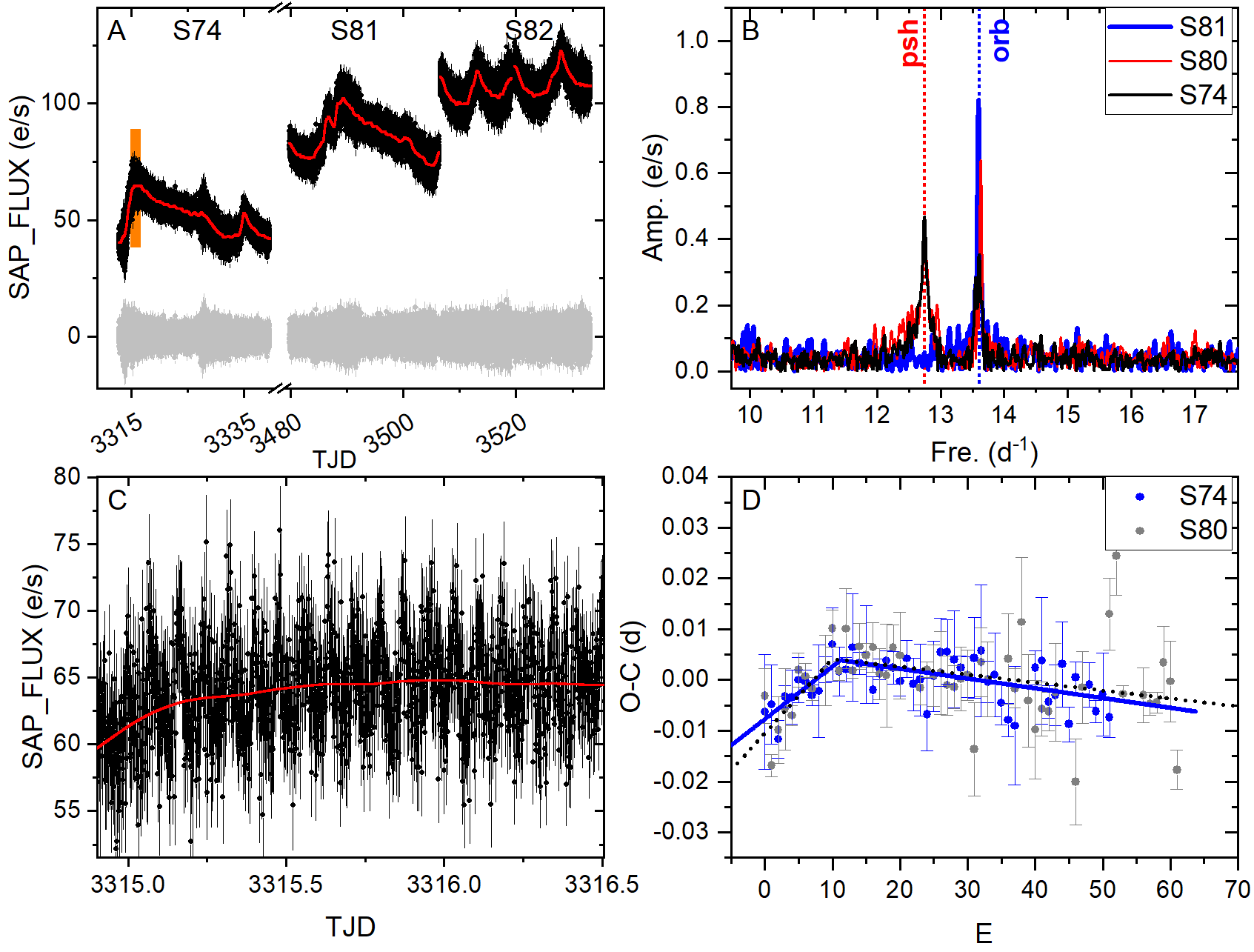}
		\caption{ASASSN-18hr}
	\end{subfigure}

	\begin{subfigure}{0.45\columnwidth}
		\includegraphics[width=\linewidth]{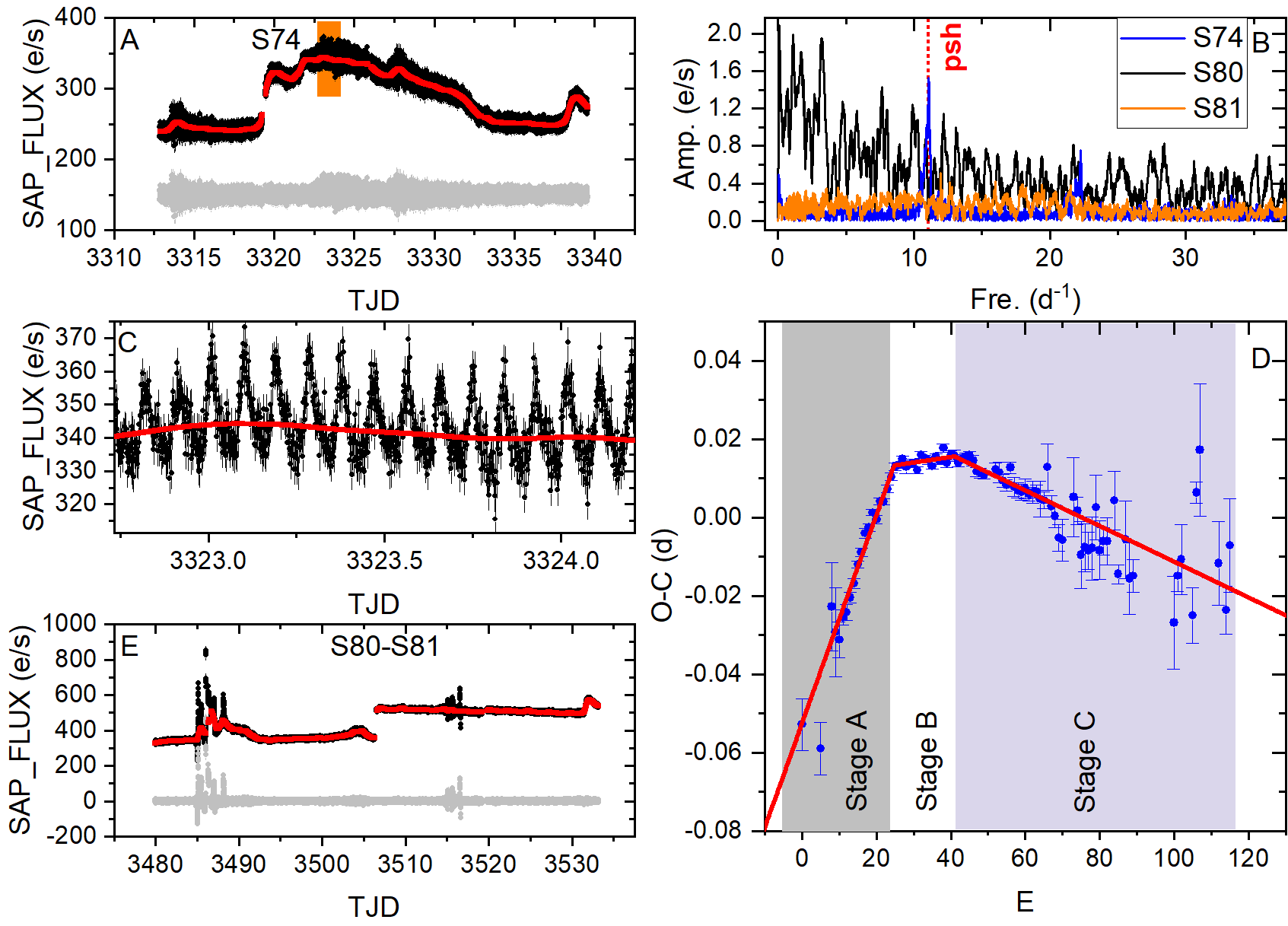}
		\caption{V0419 Lyr}
	\end{subfigure}
	\begin{subfigure}{0.45\columnwidth}
		\includegraphics[width=\linewidth]{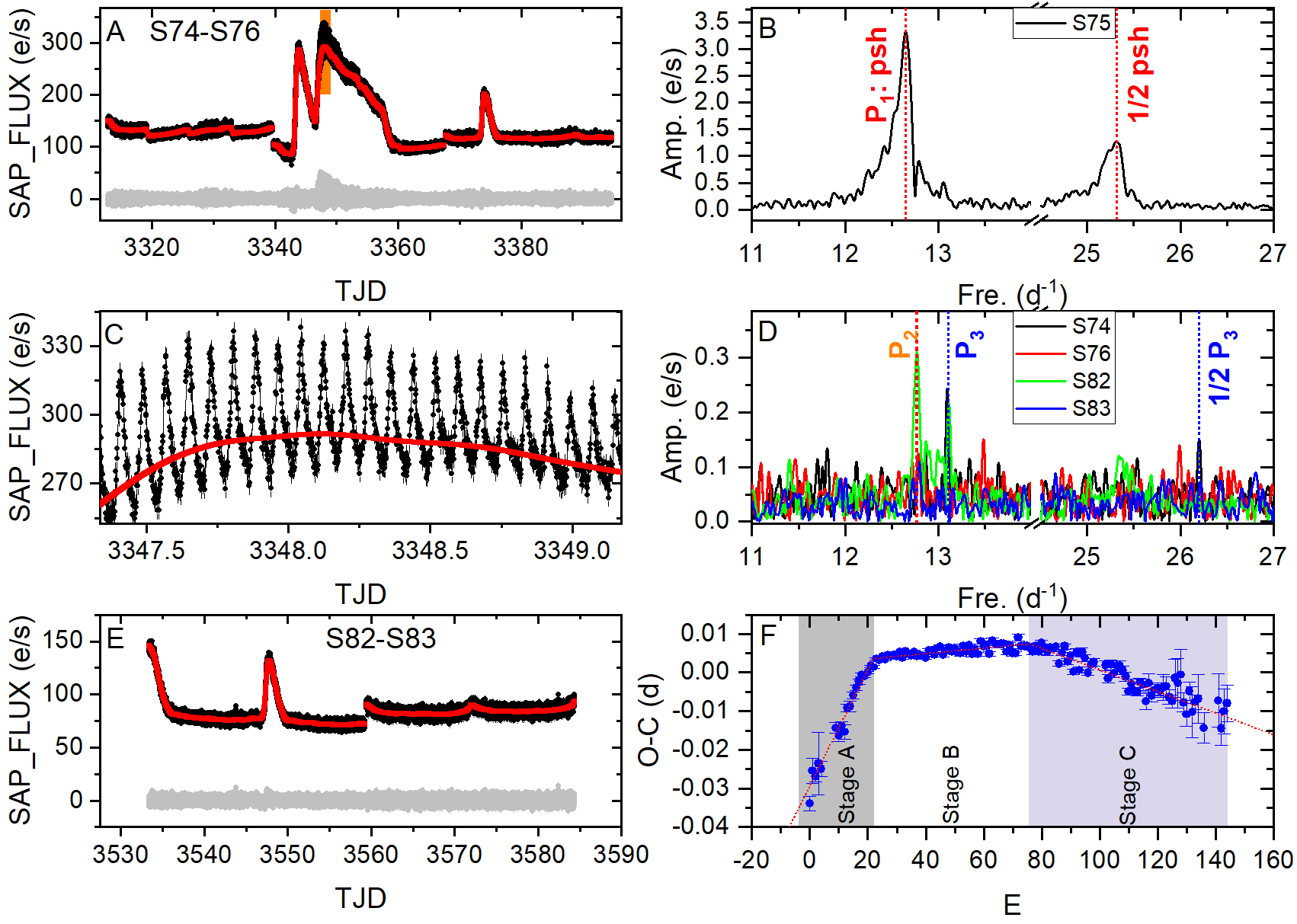}
		\caption{V1113 Cyg}
	\end{subfigure}
	\caption{Light curves, frequency spectra, zoomed-in insets, $O$--$C$ curves and their fits (including linear fits, parabolic fits, and piecewise-linear function fits) for (a) V0748 Hya, (b) SX LMi, (c) MASTER\,OT\,J172758.09+380021.5, (d) ASASSN-18hr, (e) V0419 Lyr, and (f) V1113 Cyg. The symbols in this figure are generally consistent with those in Figs.~\ref{fig:1-6}--\ref{fig:13-18}.}
	\label{fig:19-24}
\end{figure}

\renewcommand\thefigure{A5}

\begin{figure}
	\centering
	\begin{subfigure}{0.45\columnwidth}
		\includegraphics[width=\linewidth]{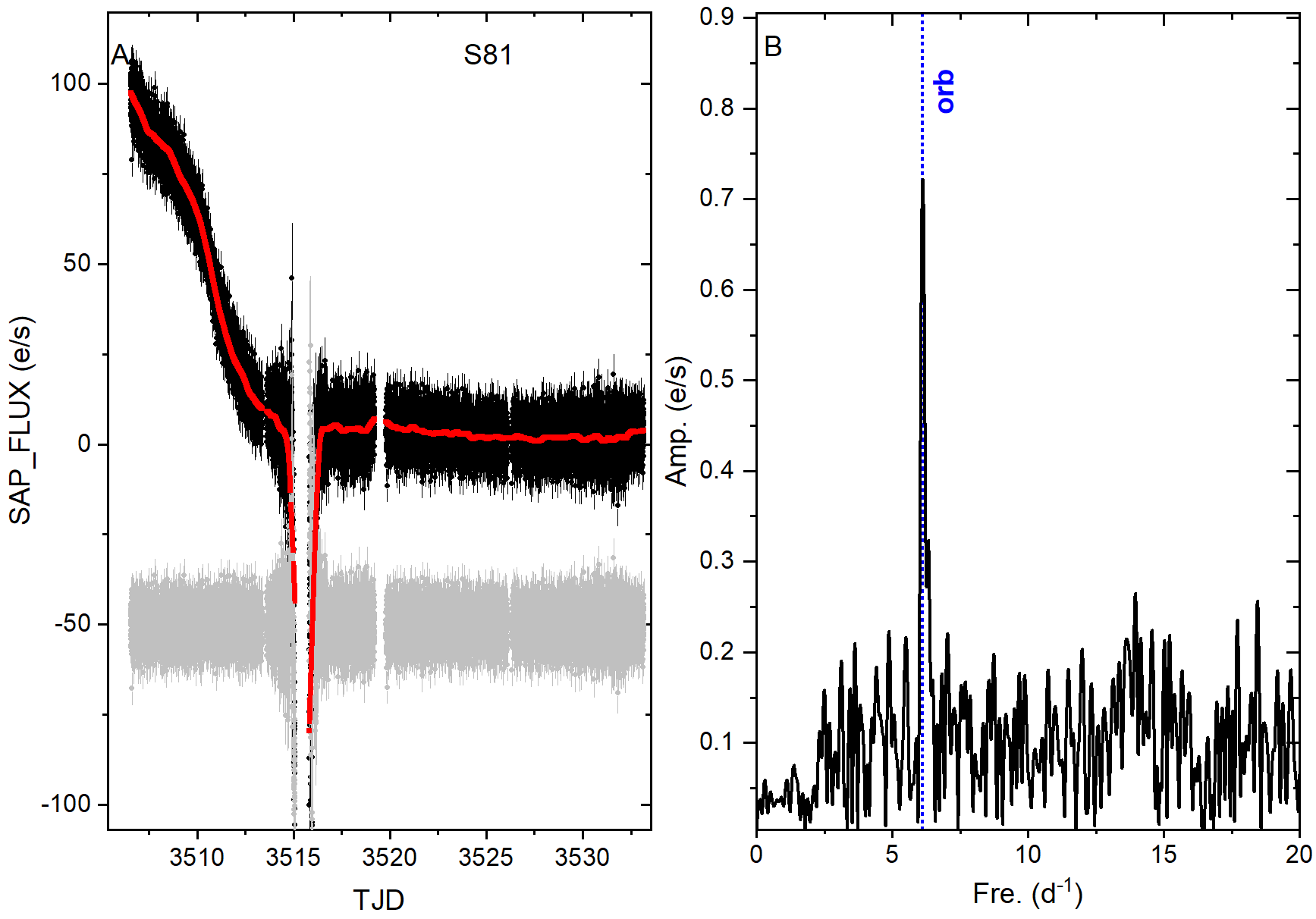}
		\caption{Gaia17cuh}
	\end{subfigure}	
	\begin{subfigure}{0.45\columnwidth}
		\includegraphics[width=\linewidth]{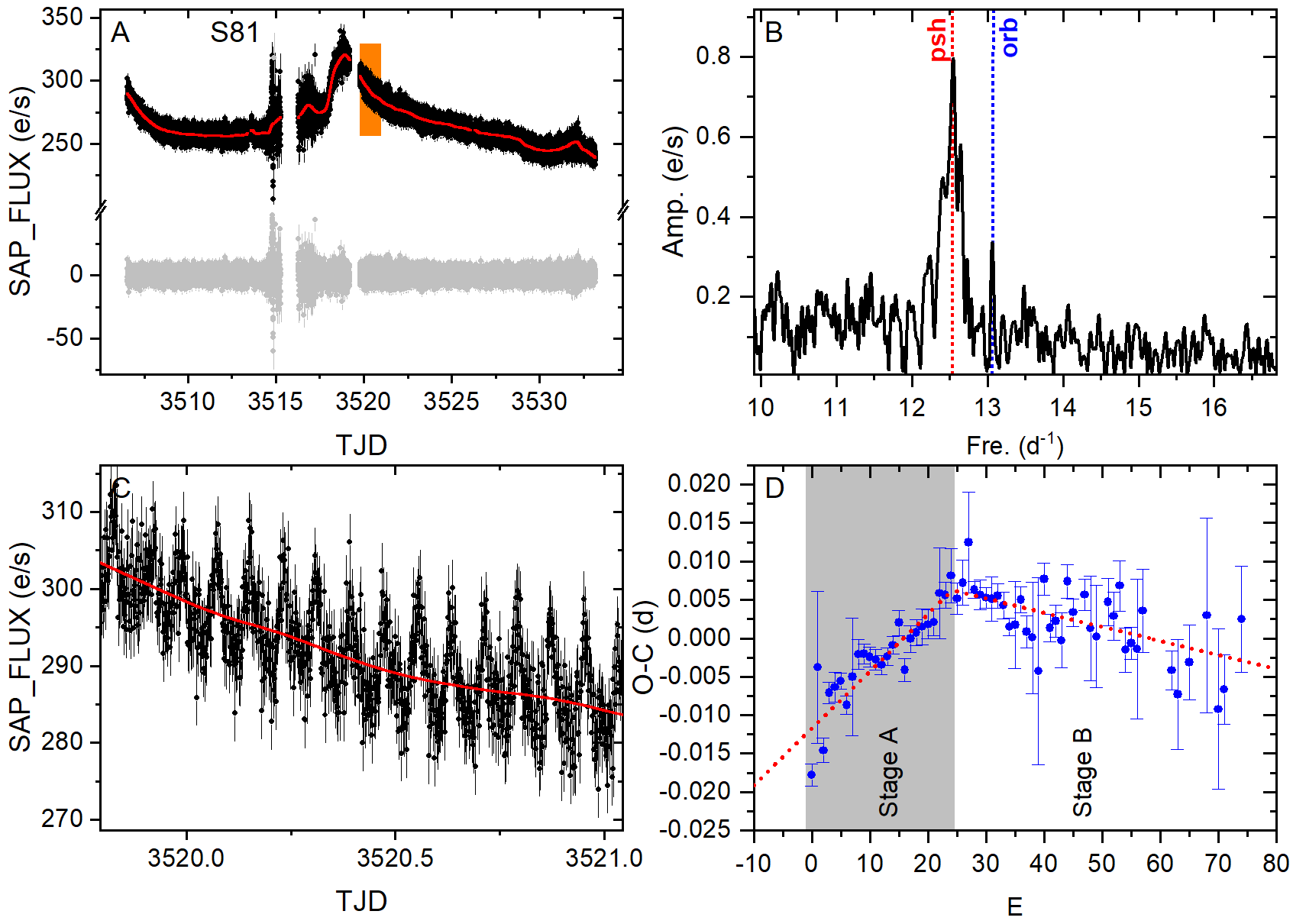}
		\caption{ASASSN-15qr}
	\end{subfigure}

	\begin{subfigure}{0.45\columnwidth}
		\includegraphics[width=\linewidth]{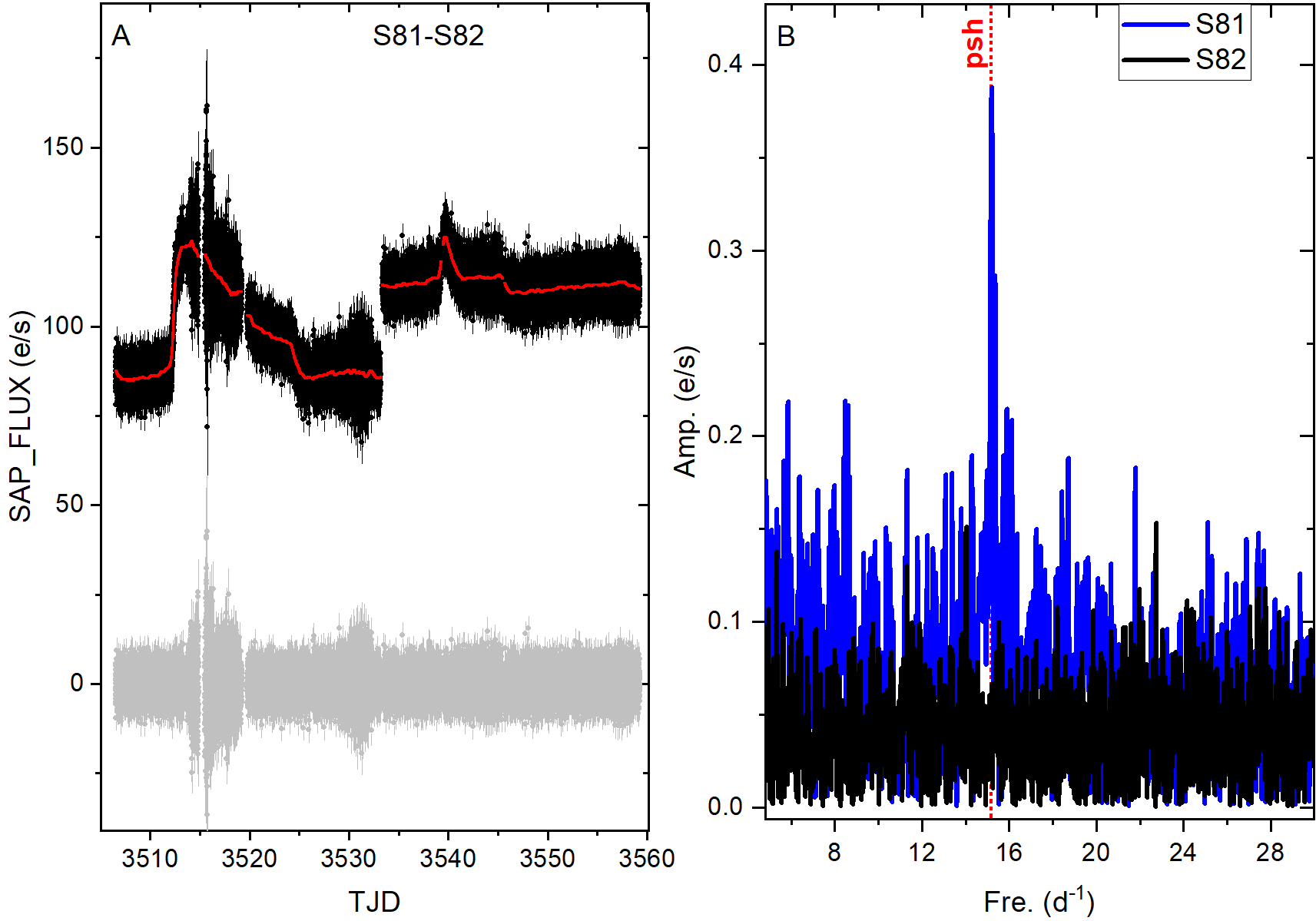}
		\caption{ASASSN-18xz}
	\end{subfigure}
	\begin{subfigure}{0.45\columnwidth}
		\includegraphics[width=\linewidth]{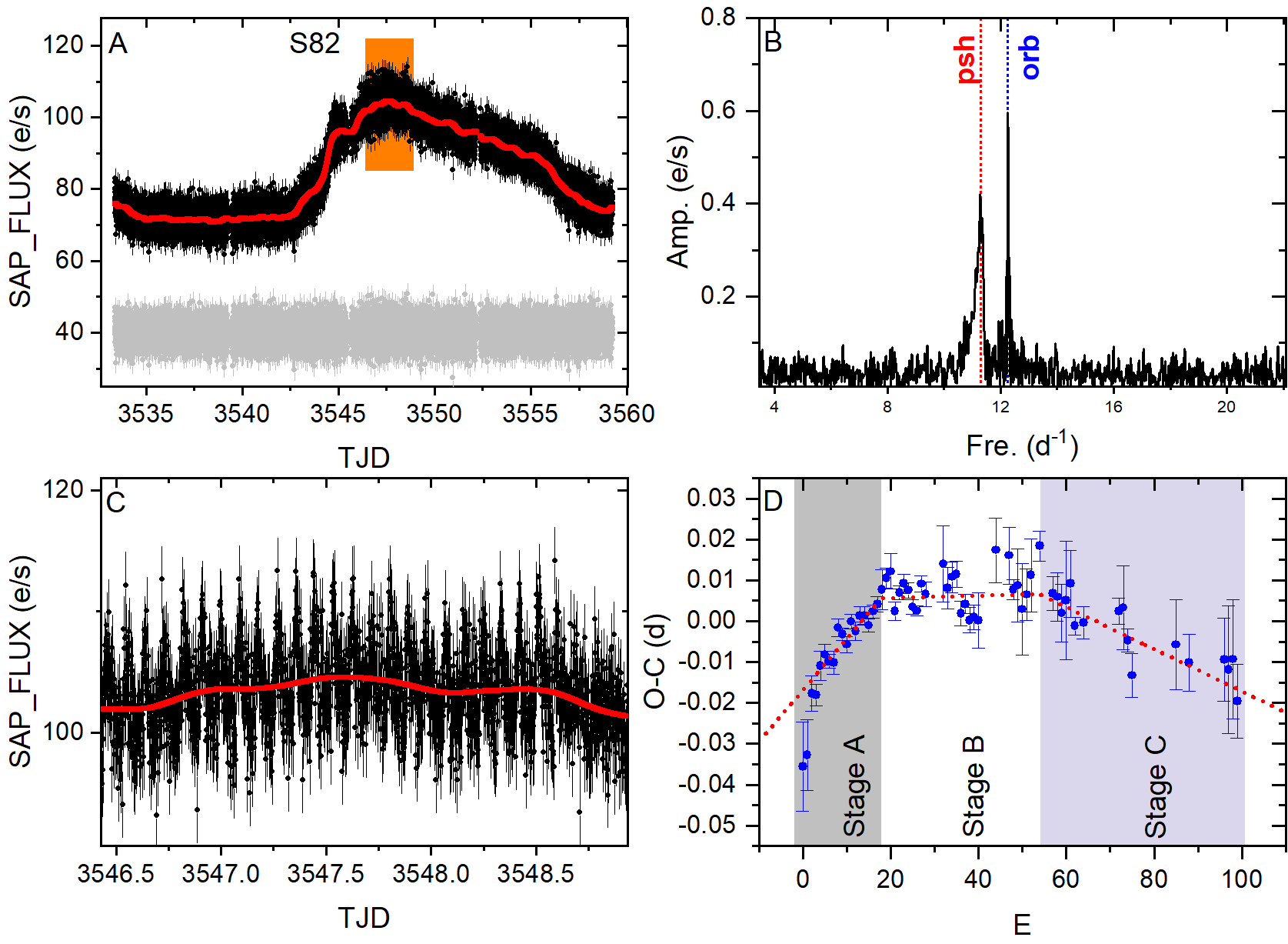}
		\caption{V0774 Peg}
	\end{subfigure}

	\begin{subfigure}{0.45\columnwidth}
		\includegraphics[width=\linewidth]{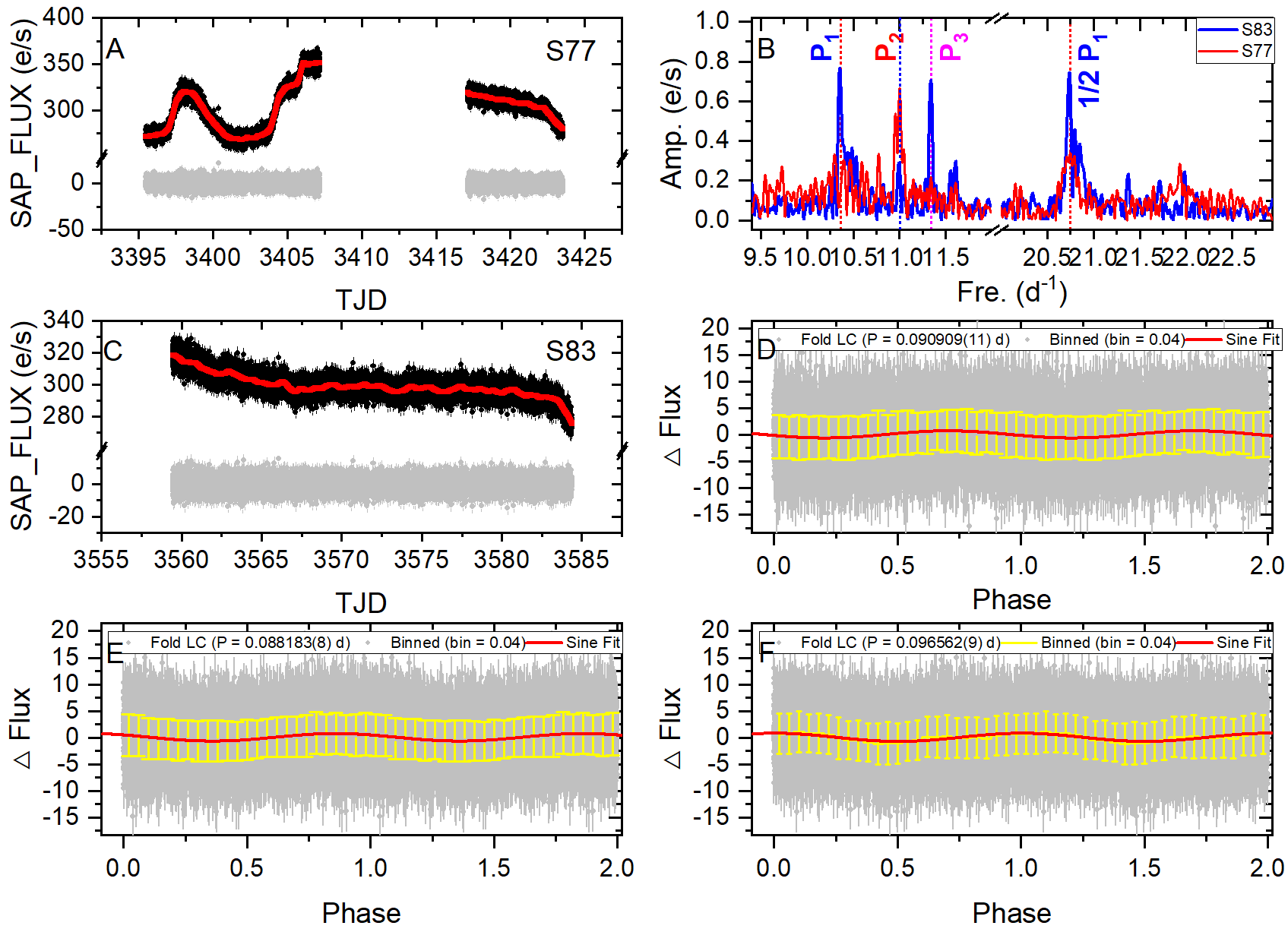}
		\caption{MGAB-V233}
	\end{subfigure}
	\begin{subfigure}{0.45\columnwidth}
		\includegraphics[width=\linewidth]{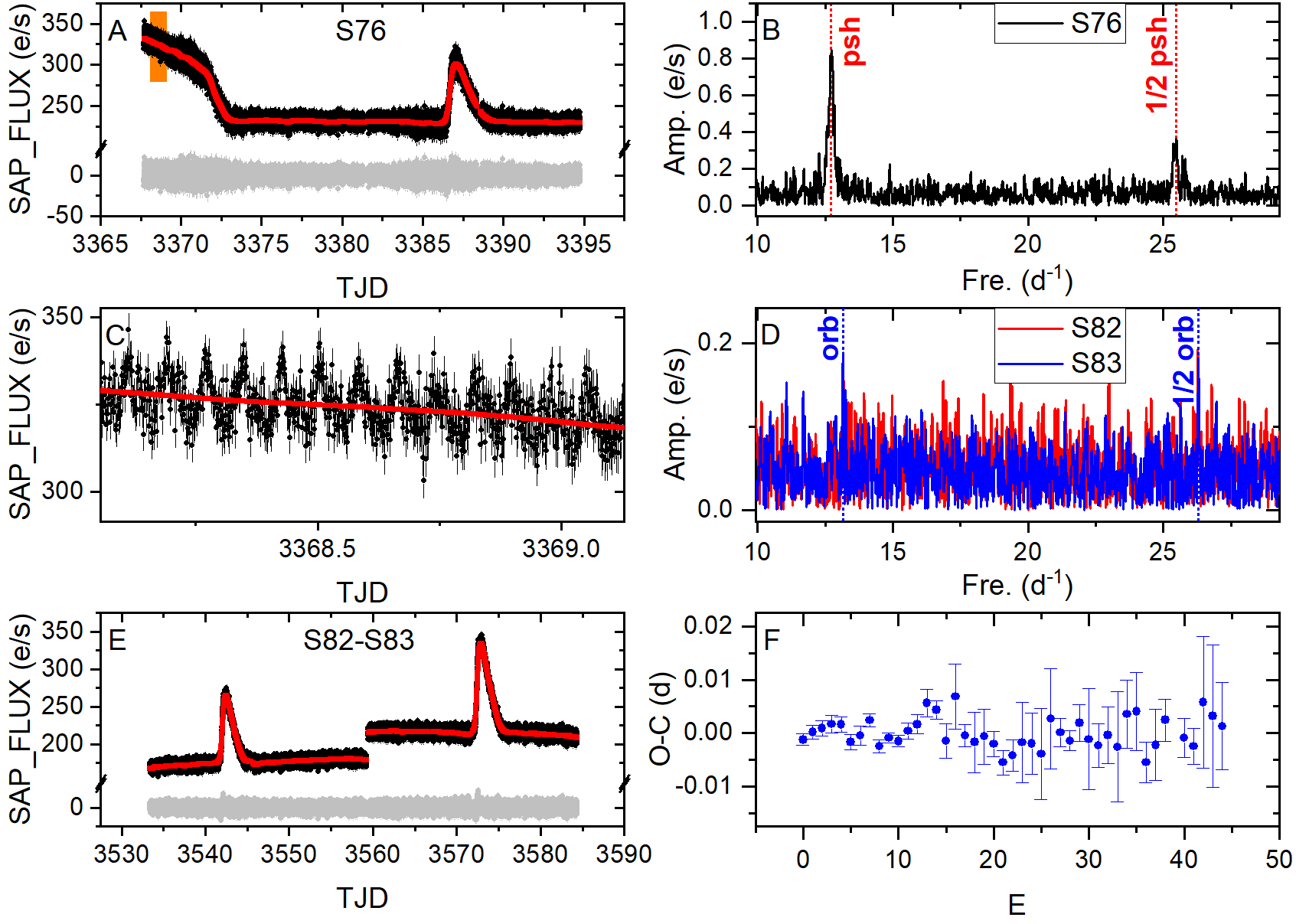}
		\caption{V0630 Cyg}
	\end{subfigure}
	\caption{(a) For Gaia17cuh, panel~C shows the folded light curve corresponding to the prominent period identified in the frequency spectrum.
		(e) For MGAB-V233, panels D--F present folded curves for periods P2, P1, and P3. The remaining panels follow the layout shown in Figs.~\ref{fig:1-6}--\ref{fig:19-24}.}
	\label{fig:25-30}
\end{figure}

\renewcommand\thefigure{A6}
\begin{figure}
	\centering
	\includegraphics[width=0.7\columnwidth]{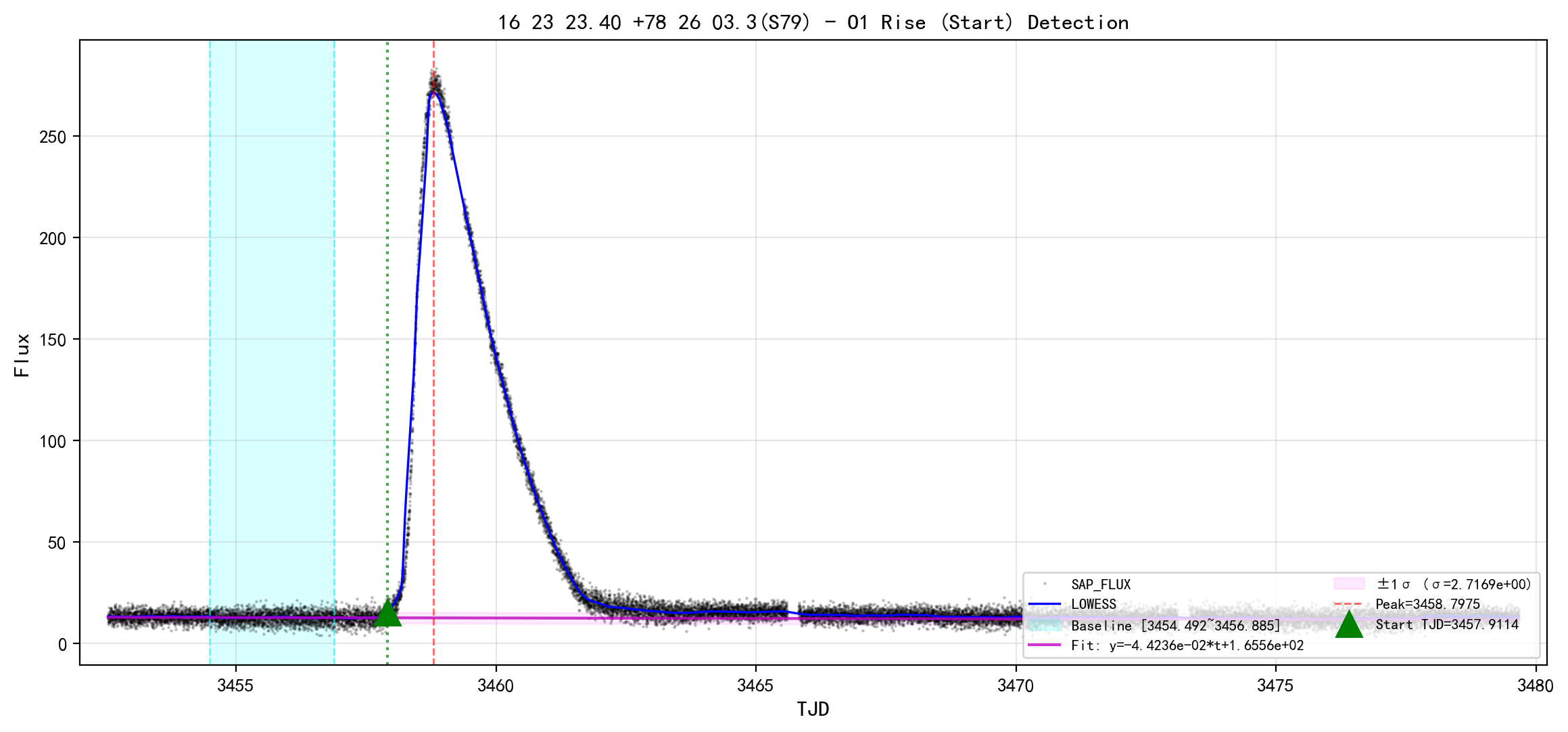}
	\includegraphics[width=0.7\columnwidth]{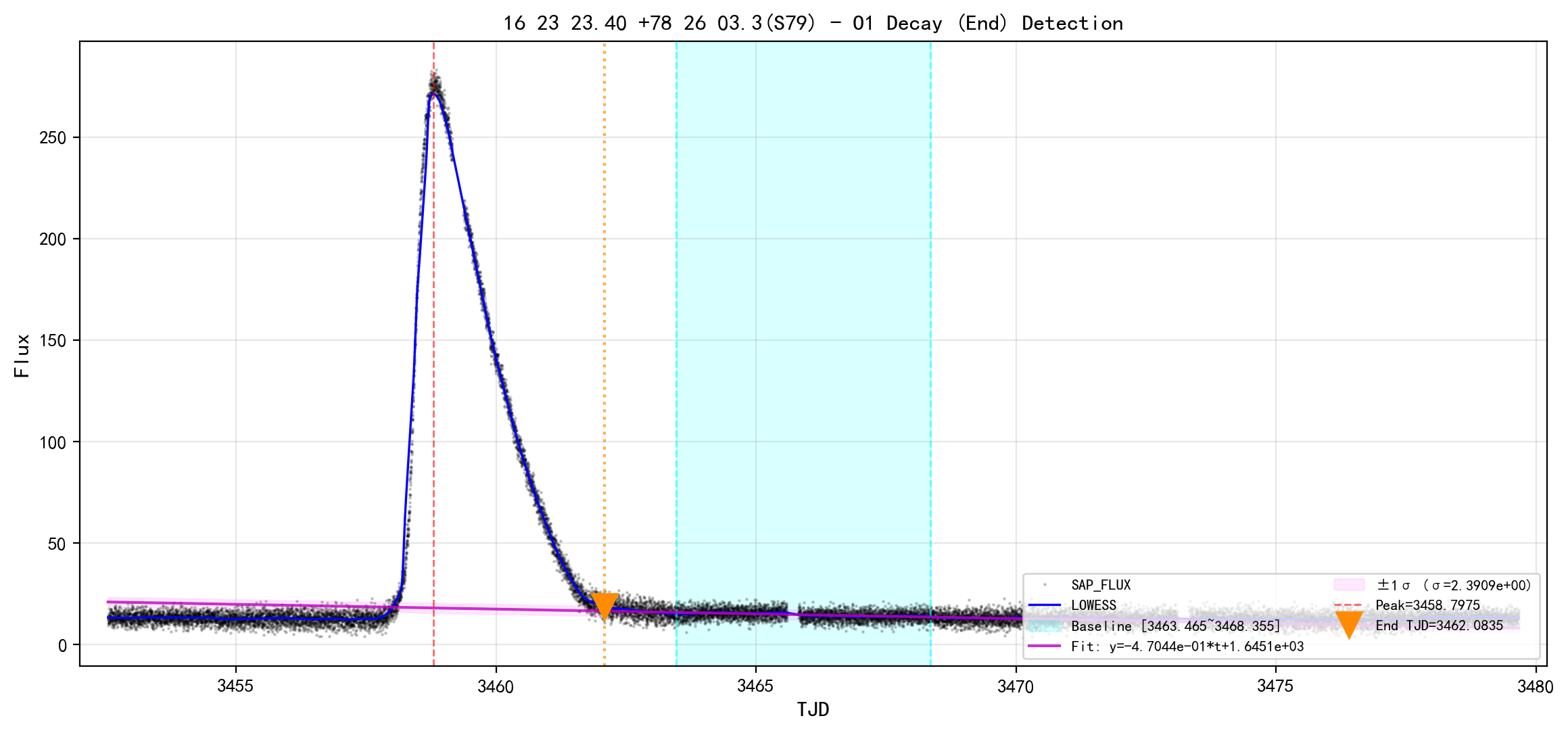}
	\caption{Examples of confirmed outburst epochs. The upper panel shows the detection of the rise (start) of the outburst, and the lower panel shows the detection of the decay (end) of the same outburst. The black dots represent SAP\_FLUX photometric data, and the solid blue curve is the LOWESS smoothed light curve. The magenta solid line denotes the linear fit to the baseline region, with the magenta shaded region indicating the $\pm1\sigma$ uncertainty band. The cyan shaded region marks the selected baseline interval. The red dashed vertical line indicates the outburst peak. The green upward triangle marks the determined outburst start epoch, and the dark-orange downward triangle marks the determined outburst end epoch. The time uncertainty is defined as the time difference between the point where the LOWESS curve crosses $\mathrm{fit}+1\sigma$ and the point where the LOWESS curve crosses the fitted baseline.}
	\label{fig:out-cal-examp1}
\end{figure}

\renewcommand\thefigure{A7}

\begin{figure}
	\centering

	\begin{subfigure}{\columnwidth}
		\centering
		\includegraphics[width=0.8\linewidth]{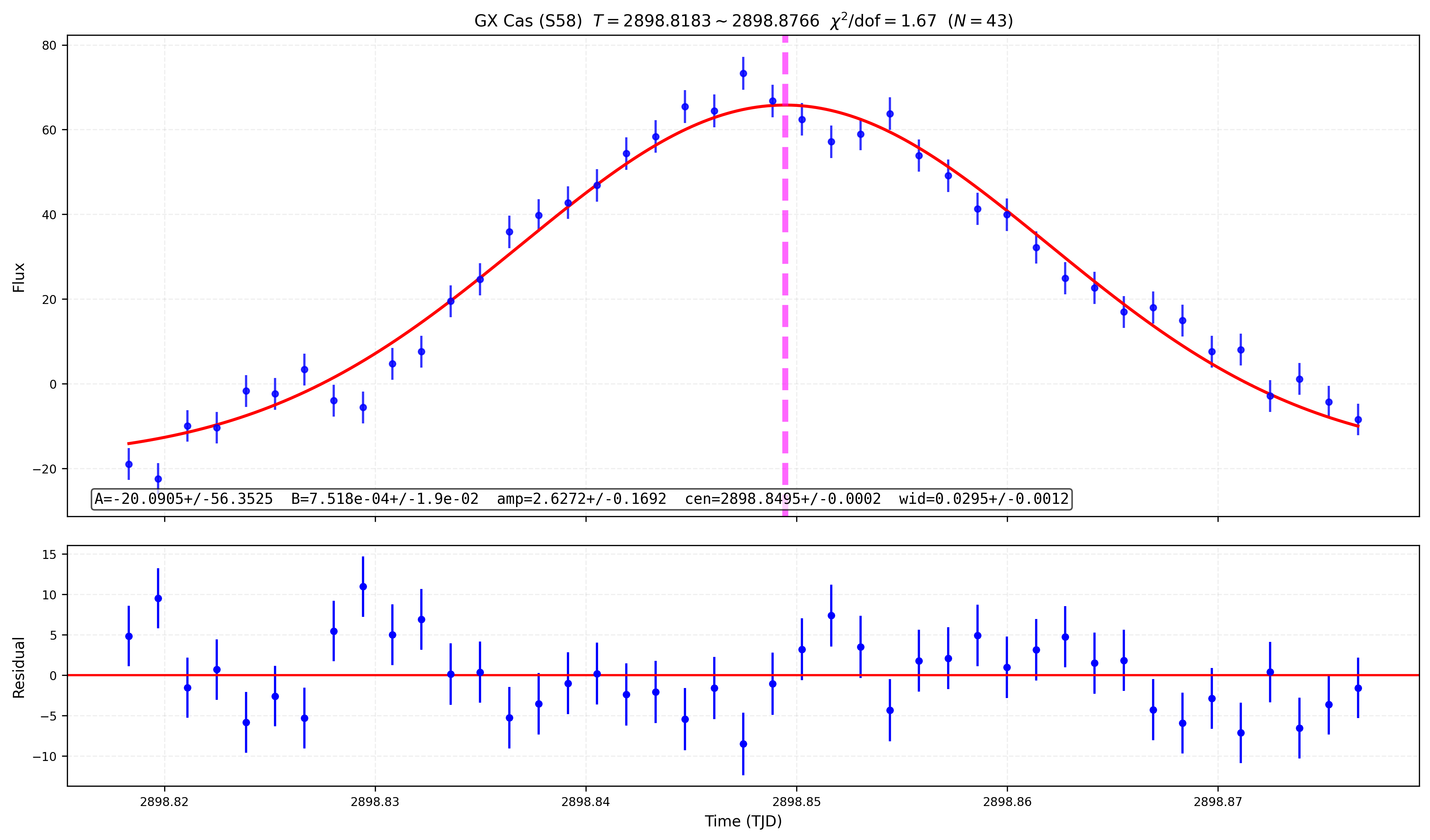}
		\caption{GX Cas}
	\end{subfigure}
	\begin{subfigure}{\columnwidth}
		\centering
		\includegraphics[width=0.8\linewidth]{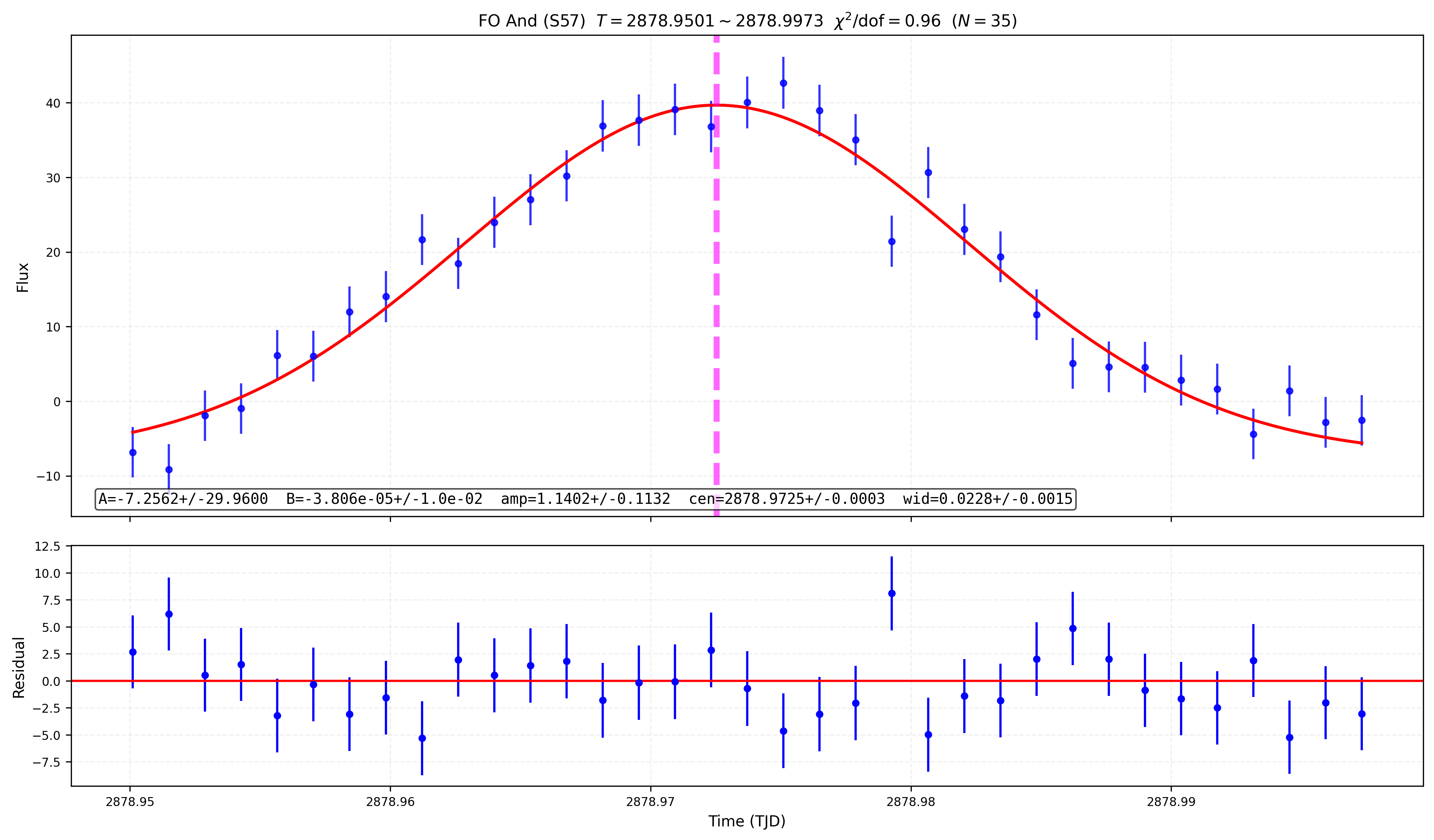}
		\caption{FO And}
	\end{subfigure}
	\caption{Example of the linear-plus-Gaussian fit. The red curve in the figure represents the linear-plus-Gaussian model, and the magenta vertical line marks the peak maxima in TJD. The definitions of the model parameters are given in Equation~\ref{eq:1}.}
	\label{fig:max-example}
\end{figure}

\newpage
\clearpage
\renewcommand\thetable{A1}
\begin{deluxetable}{lccccccccccccc}
	\tablecaption{Superhump stage periods, stage-B period derivatives, mass ratios, and orbital periods of the sample.\label{tab:stage_param}}
	\tablewidth{1pt}
	\tabletypesize{\scriptsize}
	\tablehead{
		\colhead{Name} & \colhead{$P_{\rm A}$} & \colhead{err} & \colhead{$P_{\rm B}$} & \colhead{err} & \colhead{$P_{\rm C}$} & \colhead{err} & \colhead{$\dot{P}_{\rm SH}$} & \colhead{err} & \colhead{$q$} & \colhead{err} & \colhead{$P_{\rm orb}$} & \colhead{err} & \colhead{Ref.\tablenotemark{c}}\\
		\colhead{} & \colhead{(d)} & \colhead{(d)} & \colhead{(d)} & \colhead{(d)} & \colhead{(d)} & \colhead{(d)} & \colhead{($10^{-5}$)} & \colhead{($10^{-5}$)} & \colhead{} & \colhead{} & \colhead{(d)} & \colhead{(d)} & \colhead{}
	}
	\startdata
	BE~Oct                          & 0.07854 & 0.00042 & 0.07699 & 0.00024 & \nodata & \nodata & 21.264 & 12.219 & 0.143 & 0.0263 & 0.074700 & \nodata & VSX \\
	ASASSN-14eq                     & 0.08108 & 0.00018 & 0.07903 & 0.00008 & 0.07858 & 0.00007 & 1.187  & 1.359  & \nodata & \nodata & \nodata & \nodata & \nodata \\
	GX~Cas                          & 0.09646 & 0.00015 & 0.09426 & 0.00009 & 0.09295 & 0.00006 & 0.000  & 0.000  & 0.246 & 0.0052 & 0.089206 & 0.000017 & this work \\
	FO~And                          & 0.07667 & 0.00014 & 0.07459 & 0.00007 & 0.07440 & 0.00005 & 0.000  & 0.000  & 0.207 & 0.0109 & 0.071610 & 0.000180 & \citet{1996PASP..108...73T} \\
	WX~Hyi                          & 0.07866 & 0.00009 & 0.07776 & 0.00003 & 0.07733 & 0.00003 & 0.000  & 0.000  & 0.128 & 0.0021 & 0.075160 & 0.000011 & this work \\
	& 0.08003 & 0.00027 & 0.07782 & 0.00007 & 0.07691 & 0.00009 & 0.000  & 0.000  & 0.187 & 0.0079 & \nodata & \nodata & \nodata \\
	EC~05200$-$5205                 & 0.08643 & 0.00013 & 0.08358 & 0.00003 & 0.08319 & 0.00003 & 0.000  & 0.000  & 0.226 & 0.0047 & 0.080334 & 0.000015 & this work \\
	& \nodata & \nodata & 0.08356 & 0.00003 & 0.08317 & 0.00003 & \nodata & \nodata & \nodata & \nodata & \nodata & \nodata & \nodata \\
	& 0.08599 & 0.00008 & 0.08361 & 0.00003 & 0.08319 & 0.00004 & 0.000  & 0.000  & 0.206 & 0.0027 & \nodata & \nodata & \nodata \\
	V1434~Tau                       & 0.07264 & 0.00022 & 0.06952 & 0.00009 & 0.06925 & 0.00030 & 10.787 & 1.837  & 0.124 & 0.0047 & 0.069488 & 0.000006 & this work \\
	MASTER~OT~J055845.55+391533.4   & 0.05520 & 0.00003 & 0.05470 & 0.00004 & \nodata & \nodata & 8.147  & 0.959  & \nodata & \nodata & \nodata & \nodata & \nodata \\
	ASASSN-14je                     & 0.07062 & 0.00010 & 0.06926 & 0.00004 & 0.06920 & 0.00003 & 8.837  & 0.762  & 0.151 & 0.0208 & 0.067000 & \nodata & VSX \\
	& 0.07107 & 0.00012 & 0.06934 & 0.00007 & 0.06917 & 0.00002 & 7.347  & 1.798  & 0.173 & 0.0244 & \nodata & \nodata & \nodata \\
	UV~Gem                          & 0.09497 & 0.00014 & 0.08921 & 0.00077 & 0.09249 & 0.00012 & 85.086 & 17.838 & 0.236 & 0.0046 & 0.088045 & 0.000009 & this work \\
	IR~Gem                          & 0.07288 & 0.00014 & 0.07109 & 0.00002 & 0.07073 & 0.00003 & 0.000  & 0.000  & 0.189 & 0.0271 & 0.068400 & 0.000684 & \cite{1984ApJ...282..236S} \\
	PNV~J06501960+3002449           & 0.06761 & 0.00010 & 0.06691 & 0.00006 & 0.06705 & 0.00003 & 18.125 & 1.287  & \nodata & \nodata & \nodata & \nodata & \nodata \\
	SDSS~J075107.50+300628.4        & \nodata & \nodata & \nodata & \nodata & \nodata & \nodata & \nodata & \nodata & \nodata & \nodata & \nodata & \nodata & \nodata \\
	ASASSN-14kj\tablenotemark{a}    & \nodata & \nodata & \nodata & \nodata & \nodata & \nodata & \nodata & \nodata & 0.141 & 0.0015 & 0.090560 & 0.000030 & \cite{2016AJ....152..226T} \\
	SDSS~J080303.90+251627.0        & 0.09355 & 0.00028 & 0.09044 & 0.00014 & 0.09017 & 0.00036 & 8.302  & 1.419  & \nodata & \nodata & \nodata & \nodata & \nodata \\
	YZ~Cnc                          & 0.09187 & 0.00013 & \nodata & \nodata & 0.09052 & 0.00007 & 0.000  & 0.000  & 0.160 & 0.0028 & 0.086924 & 0.000007 & \cite{1994MNRAS.267..465V} \\
	& 0.09174 & 0.00013 & 0.09049 & 0.00006 & 0.09023 & 0.00007 & \nodata & \nodata & 0.156 & 0.0028 & \nodata & \nodata & \nodata \\
	RZ~LMi\tablenotemark{a}         & 0.05831 & 0.00025 & 0.05943 & 0.00002 & \nodata & \nodata & \nodata & \nodata & 0.075	 & 0.0041 & 0.05792 & \nodata & \cite{2016PASJ...68...59K} \\
	KS~UMa                         & 0.07065 & 0.00006 & 0.06998 & 0.00008 & \nodata & \nodata & 5.772  & 2.388  & 0.107 & 0.0138 & 0.067960 & 0.000680 & \cite{2009MNRAS.397.2170G} \\
	V0748~Hya                       & 0.06445 & 0.00010 & 0.06273 & 0.00003 & 0.06317 & 0.00010 & 12.762 & 0.252  & 0.115 & 0.0022 & 0.061838 & 0.000005 & this work \\
	SX~LMi                          & 0.06991 & 0.00009 & 0.06926 & 0.00005 & 0.06872 & 0.00010 & 5.013  & 1.103  & 0.110 & 0.0150 & 0.067170 & 0.000672 & \citet{2009PASJ...61S.395K} \\
	MASTER~OT~J172758.09+380021.5\tablenotemark{a} & \nodata & \nodata & 0.05763 & 0.00003 & 0.05639 & 0.00023 & 8.254 & 0.590 & 0.032 & 0.0001 & 0.057117 & 0.000011 & this work \\
	ASASSN-18hr                     & 0.07955 & 0.00036 & 0.07833 & 0.00011 & \nodata & \nodata & 0.000  & 0.000  &0.252
	 & 0.0140 & 0.073466 & 0.000006 & this work \\
	& 0.08003 & 0.00095 & 0.07841 & 0.00015 & \nodata & \nodata & 0.000  & 0.000  & 0.278& 0.0398 & \nodata & \nodata & \nodata \\
	V0419~Lyr                       & 0.09305 & 0.00019 & 0.09054 & 0.00015 & 0.08994 & 0.00010 & 0.000  & 0.000  & 0.230 & 0.0332 & 0.086400 & \nodata & VSX \\
	V1113~Cyg                       & 0.08059 & 0.00009 & 0.07916 & 0.00003 & 0.07881 & 0.00004 & 0.000  & 0.000  & 0.155 & 0.0025 & 0.076365 & 0.000015 & this work \\
	Gaia17cuh\tablenotemark{b}      & \nodata & \nodata & \nodata & \nodata & \nodata & \nodata & 0.000  & 0.000  & \nodata & \nodata & 0.163312 & 0.000253 & this work \\
	ASASSN-15qr                     & 0.08029 & 0.00011 & 0.07936 & 0.00009 & \nodata & \nodata & \nodata & \nodata & 0.135 & 0.0026 & 0.076533 & 0.000021 & this work \\
	V0774~Peg                       & 0.08959 & 0.00021 & 0.08835 & 0.00014 & 0.08781 & 0.00020 & \nodata & \nodata & 0.315 & 0.0091 & 0.081566 & 0.000006 & this work \\
	MGAB-V233\tablenotemark{a}      & \nodata & \nodata & \nodata & \nodata & \nodata & \nodata & \nodata & \nodata & 0.178 & 0.0005 & 0.090909 & 0.000011 & this work \\
	V0630~Cyg\tablenotemark{a}      & \nodata & \nodata & \nodata & \nodata & \nodata & \nodata & \nodata & \nodata & 0.094 & 0.0004 & 0.076025 & 0.000022 & this work \\
	\enddata
	\begin{threeparttable}
		\begin{tablenotes}
			\item [a] Stage~A was not detected or its evolution was anomalous; the mass ratio was not derived from stage-A superhump timing and is taken from the PSH period derived from the Fourier transform.
			\item [b] No significant superhump signal was detected; only the orbital period is reported.
			\item [] $P_{\rm A}$, $P_{\rm B}$, and $P_{\rm C}$ are the mean superhump periods of stages~A, B, and C, respectively, in days. $\dot{P}_{\rm SH}$ is the stage-B period derivative in units of $10^{-5}$; entries of 0.000 indicate a stage~B $O$--$C$ diagram consistent with a constant period (linear $O$--$C$).
			\item [c] ``this work'' denotes orbital periods derived from the TESS data in this study and adopted for our $q$ calculations. We primarily adopt our newly derived orbital periods. For targets without TESS orbital-period detections, we use literature values instead.
		\end{tablenotes}
	\end{threeparttable}
\end{deluxetable}

\end{document}